\documentclass[journal]{IEEEtran}
\usepackage{cite}
\usepackage{algorithm,algorithmicx}
\usepackage{algpseudocode}
\usepackage[small,bf]{caption}
\usepackage[cmex10]{amsmath}
\usepackage{amssymb,latexsym,epsfig,subfigure,epic,amscd,mathrsfs,euscript,eufrak,bm}
\usepackage{color}

\usepackage{amsfonts}
\usepackage{amsopn}
\usepackage{graphicx}
\usepackage{url} 
\newtheorem{myprop}{\bf{Proposition}}

\newtheorem{remark}{\bf{Remark}}
\newtheorem{lemma}{\bf{Lemma}}
\newcommand{\argmin}{\operatornamewithlimits{arg\,min}}

\DeclareMathOperator{\tr}{tr}
\DeclareMathOperator{\card}{card}

\DeclareMathOperator{\diag}{diag}
\DeclareMathOperator*{\minimize}{\text{minimize}}
\DeclareMathOperator*{\maximize}{\text{maximize}}

\DeclareMathOperator*{\Csj}{\text{\tiny{C}}}
\DeclareMathOperator*{\Gsj}{\text{\tiny{G}}}
\DeclareMathOperator*{\Tsj}{\text{\tiny{T}}}
\DeclareMathOperator*{\JDsj}{\text{\tiny{JD}}}
\DeclareMathOperator*{\JNsj}{\text{\tiny{JN}}}
\DeclareMathOperator*{\st}{\text{subject to}}
\DeclareMathAlphabet\mathbfcal{OMS}{cmsy}{b}{n}
\newcommand{\Def}[0]{\mathrel{\mathop:}=}
\renewcommand{\baselinestretch}{0.973}

\begin{document}
\title{Sparsity-Aware Sensor Collaboration \\for Linear Coherent Estimation}

\author{Sijia~Liu~\IEEEmembership{Student Member,~IEEE,}
        Swarnendu~Kar,~\IEEEmembership{Member,~IEEE,}
		Makan~Fardad,~\IEEEmembership{Member,~IEEE,}
        and~Pramod~K.~Varshney~\IEEEmembership{Fellow,~IEEE}% <-this % stops a space
        
\thanks{A preliminary version of this paper appears in the 2014 IEEE
International Symposium on Information Theory.}            
\thanks{S. Liu, M. Fardad and P. K. Varshney are with the Department
of Electrical Engineering and Computer Science, Syracuse University, Syracuse,
NY, 13244 USA e-mail: \{sliu17, makan, varshney\}@syr.edu.}% <-this % stops a space
\thanks{S. Kar is with 
New Devices Group, Intel Corporation, Hillsboro, Oregon, 97124 USA email: swarnendu.kar@intel.com.}
\thanks{The work of S. Liu and P. K. Varshney was supported by the U.S. Air Force Office of Scientific Research (AFOSR) under Grants FA9550-10-1-0263 and FA9550-10-1-0458. The work of M. Fardad was supported by the National Science Foundation under awards CNS-1329885 and  CMMI-0927509. }
% <-this % stops a space
%%Important
%\thanks{Manuscript received April 19, 2005; revised December 27, 2012.}

}



\maketitle

%%%%%%%%%%%%%%%%%%%%%%%ABSTRACT%%%%%%%%%%%%%%%%%%%%%%%%%%%%
\begin{abstract}
In the context of distributed estimation,
we consider the problem of sensor collaboration, which refers
to the act of sharing measurements with neighboring sensors
prior to transmission to a fusion center. While incorporating
the cost of sensor collaboration, we aim to find optimal sparse
collaboration schemes subject to a certain information or
energy constraint. Two types of sensor collaboration problems are studied:   minimum energy with an information constraint; and maximum information with an energy constraint.
To solve the resulting sensor collaboration problems, 
we present tractable
optimization formulations and propose efficient methods which render  near-optimal solutions in numerical experiments.
{We also explore the situation in which there is a cost associated with the involvement of each sensor in the estimation scheme. In such situations, the participating sensors must be chosen judiciously.  We introduce a unified framework to jointly design the optimal sensor selection and collaboration schemes.} For a given estimation performance, we empirically show  that there exists  a trade-off between sensor selection and sensor collaboration.
\end{abstract}


\IEEEpeerreviewmaketitle

%%%%%%%%%%%%%%%%%%%%%%%INTRODUCTION%%%%%%%%%%%%%%%%%%%%%%%%%%%%

\section{Introduction}

Wireless sensor networks, consisting of a large
number of spatially distributed sensors, have been widely used for applications such as environment monitoring, target tracking and optimal control \cite{olirod11,hevicyan06,caochegao09}. 
In this paper, we study the problem of distributed estimation, where each sensor  reports its local observation of the phenomenon of interest and transmits a \textit{processed} message (after inter-sensor communication) to a fusion center (FC) that determines the global estimate. 
The act of inter-sensor communication is referred to as sensor collaboration  \cite{fanli09}, where 
sensors are allowed to share their raw observations with a set of neighboring nodes prior to transmission to the FC.

In the absence of collaboration, the estimation architecture reduces to a classical distributed estimation network, where \textit{scaled} versions of sensor measurements are transmitted using an %analog 
amplify-and-forward strategy \cite{cuixiagolluopoo07}. In this setting, one of the key problems is to design the optimal power amplifying factors (i.e., scaling laws) %for each sensor 
to reach certain design criteria for performance measures, such as estimation distortion and energy cost. Several variations of the conventional distributed estimation problem have been addressed in the literature depending on the quantity to be estimated (random parameter or process) \cite{MAH11,jiacheswi13},
the type of communication (analog-based or quantization-based) \cite{gasrimvet03,ribgia06}, nature of transmission channels (coherent or orthogonal) \cite{xiacuiluogol08,luogiazha05} and energy constraints \cite{leodeyeva11}. 

In the aforementioned literature \cite{cuixiagolluopoo07,MAH11,jiacheswi13,gasrimvet03,ribgia06,xiacuiluogol08,luogiazha05,leodeyeva11}, it is assumed that 
there is no inter-sensor collaboration. In contrast, here we study the problem of sensor collaboration, which is motivated by a significant improvement of estimation performance resulting from collaboration \cite{fanli09}.
Furthermore, in the collaborative estimation system, we consider the energy cost for sensor activation in order to determine the optimal subset of sensors that communicate with the FC.  
%To the best of our knowledge, the problem of sensor selection in the collaborative estimation system has not been investigated in the literature. Here,
We will derive optimal schemes for sensor selection and collaboration simultaneously.

The problem of sensor collaboration was first proposed in \cite{fanli09} by assuming an orthogonal multiple access channel (MAC) setting with a fully connected network, where all the sensors are allowed to collaborate.
It was shown that the optimal strategy is 
to transmit the processed signal (after collaboration) over the best available channels with power levels consistent with the channel qualities. 
%{\color{blue} 
In \cite{thamit08,thamit06_asilomar}, a related problem on power allocation
was studied for distributed estimation in a sensor network with fixed topologies. This can be interpreted as the problem of sensor collaboration with given network topologies.
Recently, the problem of sensor collaboration over a coherent MAC was studied in \cite{karvar12isit,karvar12allerton}, where it
was observed that even a partially connected network can yield performance close to that of a fully connected network, and the problem of sensor collaboration for a family of sparsely connected networks was investigated.
Further, the problem of sensor collaboration for estimating a vector of random parameters is studied in \cite{fanvaljamsch_14}. The works \cite{fanli09,karvar12isit, karvar12allerton,thamit08,thamit06_asilomar,fanvaljamsch_14}
assumed that there is no cost associated with collaboration, the collaboration topologies are fixed and given in advance, and the only unknowns are the collaboration weights used to combine sensor observations.

A more relevant reference to this work is  our earlier work \cite{karvar13}, where 
the nonzero collaboration cost was taken into account for linear coherent estimation, and a greedy algorithm was developed for seeking the optimal collaboration topology in energy constrained sensor networks. Compared to \cite{karvar13}, here we present a non-convex optimization framework to solve the collaboration problem with nonzero collaboration cost. 
To elaborate, 
we describe collaboration through a collaboration
matrix, in which the nonzero entries characterize the collaboration
topology and the values of these entries characterize
the collaboration weights. We introduce a formulation that
simultaneously optimizes both the collaboration topology and
the collaboration weights. In contrast, %This is in contrast to \cite{karvar13}, where the 
the optimization in \cite{karvar13} was performed in a sequential manner, where a sub-optimal collaboration topology was first obtained, and then the optimal collaboration weights were sought. 
The new formulation leads to a more efficient allocation of energy
resources as evidenced by improved distortion performance in
numerical results.

We study two types of problems while designing optimal collaboration schemes. 
One is the information constrained collaboration problem, where we minimize the energy cost subject to an information constraint. 
The other is the energy constrained collaboration problem, where the
Fisher information is maximized subject to a total energy budget.  
Similar formulations have been considered for the problem of power allocation in parameter estimation \cite{cuixiagolluopoo07} 
and state tracking \cite{leodeyeva11,jiacheswi13}.
Characterization of the collaboration cost in this work in terms of the cardinality function, leads to combinatorial optimization problems. 
For tractability,
we employ the  reweighted $\ell_1$ method \cite{canwakboy08} and alternating directions
method of multipliers (ADMM) \cite{boyparchupeleck11} 
to find a locally optimal solution for the information constrained problem. For the energy constrained problem, we exploit its
relationship with the information constrained problem and
propose a bisection algorithm for its solution. We empirically show that the proposed methods yield near optimal performance. 

In the existing collaborative estimation literature \cite{karvar12isit, karvar12allerton,karvar13} with $N$ sensors, it has been assumed that 
every node can share its information (through the act of collaboration) with other nodes, but only $M$ ($M \leq N$) given nodes have the ability to communicate with the FC.
This is due to the fact that it is usually difficult to coordinate multiple sensors for synchronized transmissions to the FC (as required for an amplify-and-forward scheme) and it may not be possible for all the sensors to participate in that process. In this paper, we formalize this notion by adding a finite cost to selecting each particular sensor for coordinated transmissions to the FC. In this way, the total cost may be reduced if only a small number of sensors are selected to transmit their data. For example, a sensor which is far away from the FC may have a higher sensor selection cost \cite{chechuzha05}. 
This raises some fundamental questions that we try to answer in the second part of this paper: Which $M$ sensors should be selected? And what is the optimal value of $M$? 
%It is worth mentioning that the issue of \textit{sensor selection} is also relevant to other applications such as target tracking, field monitoring, and distributed control \cite{masfarvar12, liumasfarvar14_icassp, Tomlin2012}. 

It is worth mentioning that the problem of \textit{sensor selection} has been widely studied in the context of parameter/state estimation, e.g., \cite{josboy09, jamsimleu14, masfarvar12,liumasfarvar14_icassp,liufarmasvar14,sch13}. In \cite{josboy09}, the sensor selection problem for parameter estimation was elegantly formulated with the help of auxiliary Boolean variables, where each Boolean variable determines whether or not its corresponding sensor is selected. 
%In \cite{moambsin11}, a more general framework that captures the sensor selection problem  was proposed through a reformulation of Kalman filter for linear dynamical systems.
%a distributed algorithm of sensor selection was proposed when the measurement noise 
%is uncorrelated. 
In \cite{jamsimleu14}, a sparsity-aware sensor selection problem was introduced by minimizing the number of selected sensors subject to a certain estimation quality. %And  a distributed algorithm for  sensor selection  was presented for parameter estimation with uncorrelated measurements.
In \cite{masfarvar12,liumasfarvar14_icassp,liufarmasvar14}, the optimal sensor selection schemes were found by promoting the sparsity of estimator gains. In \cite{sch13}, the design of sensor selection scheme was transformed to the recovery of a sparse matrix.  %Although the sensor selection problem was
%studied extensively, our work is different from the aforementioned literature 
Although both our work and the existing literature \cite{josboy09, moambsin11,jamsimleu14, masfarvar12,liumasfarvar14_icassp,liufarmasvar14,sch13}  use the $\ell_0$ norm and $\ell_1$ relaxation in dealing with sensor selection problems, 
our work is significantly different from  \cite{josboy09, moambsin11,jamsimleu14, masfarvar12,liumasfarvar14_icassp,liufarmasvar14,sch13}, since  the sensor activation scheme is \textit{jointly} optimized with the collaboration strategy.
Once the problem of {joint} selection and collaboration is solved, 
we obtain not only the sensor selection scheme but also the collaboration topology and the power allocation scheme for distributed estimation.
We emphasize that the focus of the paper is on sensor collaboration where it is employed in conjunction with sensor selection as well as when no sensor selection is involved.

To determine the optimal sensor selection scheme in a collaborative estimation system,
we associate (a) the cost of sensor selection with the number of nonzero rows of the collaboration matrix (i.e., its row-sparsity), and (b) the cost of sensor collaboration with the number of nonzero entries of the collaboration matrix (i.e., its  overall sparsity).
%we characterize the cost of \textit{sensor selection} through the \textit{row-sparsity} of the collaboration matrix, whose \textit{element-wise} sparsity corresponds to the \textit{sensor collaboration} cost.
Based on these associations, we then present a unified framework that jointly designs the optimal sensor selection and collaboration schemes.  It will be shown that there exists 
a trade-off between %the number of selected sensors and number of collaboration links 
sensor selection and sensor collaboration
for a given estimation performance.

%%%%%%%%%%%%%%%%%%%%%%%%%version 1: preliminary work%%%%%%%%%%%%%%%%%%%%%%%
In a preliminary version of this paper \cite{liukarfarvar14_isit}, we presented
the optimization framework for designing the optimal sensor collaboration strategy with \textit{nonzero} collaboration cost and 
\textit{unknown} collaboration topologies 
in scenarios where the set of sensors that communicate with the FC is given in advance.
In this paper, we have three new contributions.
\begin{itemize}
\item We elaborate on the theoretical foundations of the proposed optimization approaches in \cite{liukarfarvar14_isit}. 
\item We improve the computational efficiency of the approaches in \cite{liukarfarvar14_isit} by proposing a fast algorithm to solve the resulting optimization problem.
\item The issue of sensor selection is taken into account. We present a unified framework
for the joint design of optimal sensor selection and collaboration schemes. 
\end{itemize}

The rest of the paper is organized as follows. In Section\,\ref{sec: sp_sensr_col}, we introduce the collaborative estimation system. 
In Section\,\ref{sec: prob_form}, we formulate the information and energy constrained sensor collaboration problems with non-zero collaboration cost.  
In Section\,\ref{sec: inf_coll}, we develop efficient approaches 
to solve the information constrained collaboration problem.
In Section\,\ref{sec: ene_coll}, the energy constrained collaboration problem is studied. 
In Section\,\ref{sec: sel}, we investigate the issue of sensor selection.
In Section\,\ref{sec: numerical}, we demonstrate the effectiveness of our proposed
framework through numerical examples. Finally, in Section\,\ref{sec: conclusion} we summarize our work and discuss future research
directions.

%%%%%%%%%%%%%%%%%%%%%Section II: Preliminaries: System Model%%%%%%%%%%%%%%%%%%%%%%%%%%%%%%%

\section{Preliminaries: A Model for Sensor Collaboration}
\label{sec: sp_sensr_col}

\begin{figure}[htb]
\begin{center}
 \includegraphics[width=0.99 \columnwidth]{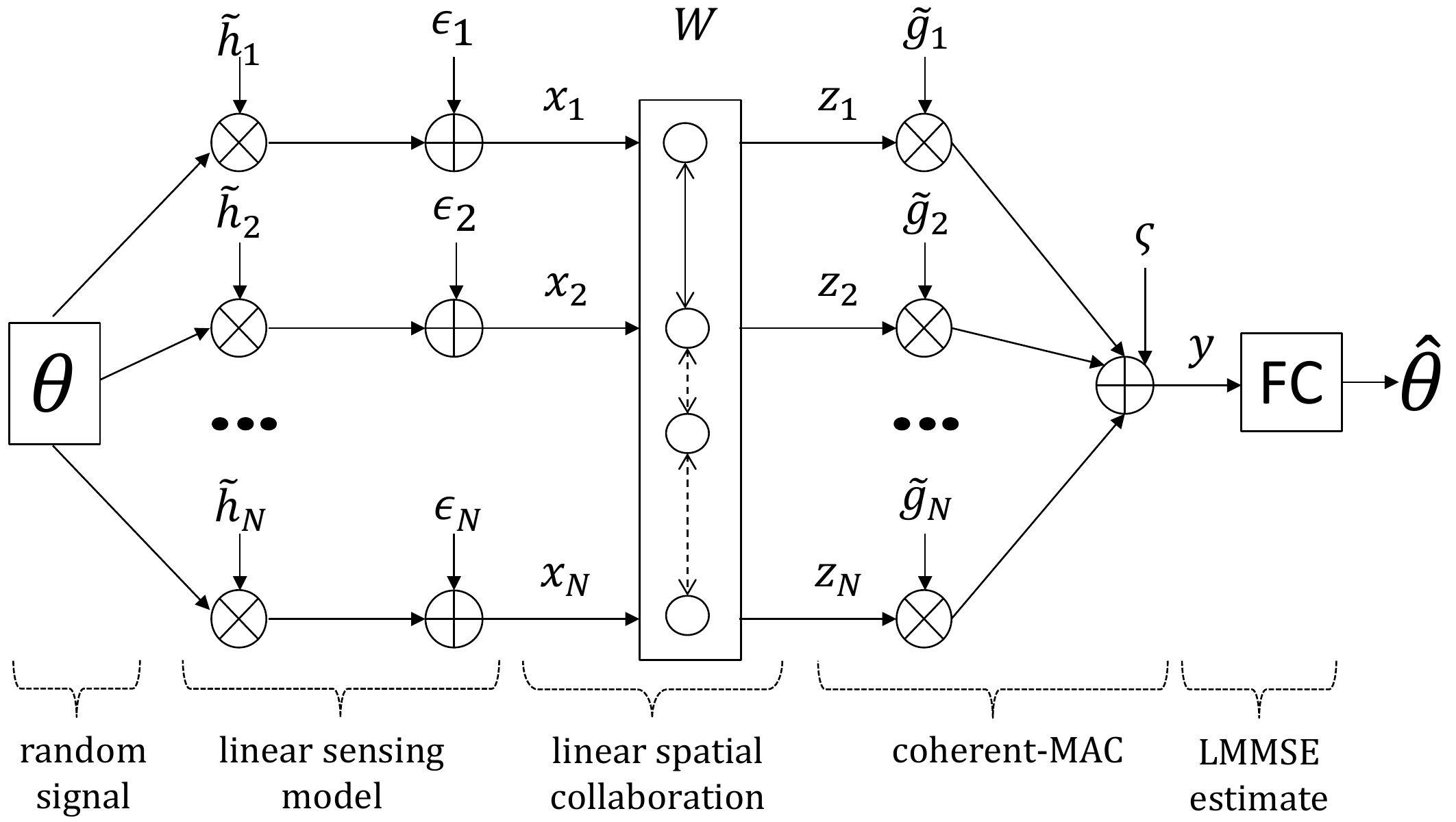}
 \caption{Collaborative estimation architecture showing the sensor measurements, transmitted signals, and the received signal at FC.}
\label{fig:col_sch}
\end{center}
\end{figure}

In this section, we introduce a distributed estimation system that involves inter-sensor collaboration. We assume that 
the task of the sensor network is to estimate a random parameter $\theta$, which 
follows a Gaussian distribution with zero mean and variance $\eta^2$. 
In the estimation system, sensors first report their raw measurements via a linear sensing model. Then, individual sensors can update their
observations through spatial collaboration, which refers to
(linearly) combining observations from other sensors.
The updated measurements are transmitted through a coherent MAC. Finally, the FC determines a global estimate of $\theta$ by using a linear estimator. We show the collaborative estimation system in 
Fig.\,\ref{fig:col_sch}, and in what follows we elaborate on each of its parts.

The linear sensing model is given by 
\begin{align}
\mathbf x = \tilde {\mathbf h} \theta + \boldsymbol \epsilon,
\label{eq: lin_sen}
\end{align}
where $\mathbf x = [x_1,\ldots,x_N]^T$ denotes the vector of measurements from $N$ sensors, $\tilde {\mathbf h}$ is the vector of observation gains with known second order statistics $\mathbb E [ \tilde {\mathbf h} ] = \mathbf h $ and $\mathrm{cov} ( \tilde {\mathbf h}) = \boldsymbol \Sigma_h$,  
and $\boldsymbol \epsilon $ represents the vector of zero-mean Gaussian noise with $\mathrm{cov}(\boldsymbol \epsilon) = \boldsymbol \Sigma_{\epsilon}$.

%{\color{blue}
%We assume that $M$ nodes, out of a total of $N$ sensor nodes ($M \leq N$), communicate with the FC over a coherent MAC, and each sensor is able to pass its observation to the $M$ communicating nodes.  In the existing literature \cite{karvar12isit, karvar12allerton,karvar13}, the $M$ communicating nodes are given in advance. However, in this paper, we  will find the best \textit{subset} of  sensors to communicate with the FC.
%}

%We assume that $M$ nodes, out of a total of $N$ sensor nodes ($M \leq N$), communicate with the FC over a coherent MAC, and each sensor is able to pass its observation to one or more nodes among  the $M$ communicating nodes. In the existing literature \cite{karvar12isit, karvar12allerton,karvar13}, the $M$ communicating nodes are given in advance. But here, we assume $M = N$, and will find the best \textit{subset} of sensors to communicate with the FC.

The \textit{sensor collaboration} process is described by 
\begin{align}
\mathbf z =  \mathbf W \mathbf x,
\label{eq: col_sig}
\end{align}
where $\mathbf z \in \mathbb R^N$ denotes the message after collaboration, and $\mathbf W \in \mathbb R^{N \times N}$ is the collaboration matrix that contains weights used to combine sensor measurements.
In \eqref{eq: col_sig}, we assume that sharing of an observation is realized through a reliable (noiseless) communication link that consumes power $C_{mn}$, regardless of its implementation. And the matrix $\mathbf C \in \mathbb R^{N \times N}$ describing all the collaboration costs among various sensors is assumed to be known in advance. The focus of the present work is on the conceptual aspects of sensor collaboration and not the details of its physical implementation. The proposed ideal and relatively simple collaboration model enables us to readily obtain explicit expressions for the transmission cost and the estimation distortion.

After sensor collaboration, the message $\mathbf z$ is transmitted to the FC through a coherent MAC, so that the received signal is a coherent sum \cite{xiacuiluogol08} 
\begin{align}
y = \tilde{\mathbf g}^T \mathbf z + \varsigma,
\label{eq: channel}
\end{align}
where 
$\tilde{\mathbf g}$ is the vector of
channel gains with known second order statistics $ \mathbb E [\tilde{\mathbf g}] = \mathbf g $
and $\mathrm{cov}(\tilde{\mathbf g}) = \boldsymbol \Sigma_g$, 
and $\varsigma$ is a zero-mean Gaussian noise with variance $\xi^2$. 

%{\color{blue}
The \textit{transmission cost} is given by the energy required for transmitting the message $\mathbf z$ in (\ref{eq: col_sig}), namely, 
\begin{align}
T_{\mathbf W} = &~\mathbb E_{\theta, \tilde {\mathbf h}, \boldsymbol \epsilon } [\mathbf z^T \mathbf z] = \mathbb E [\mathbf x^T \mathbf W^T \mathbf W \mathbf x] 
= \mathbb E[ \tr( \mathbf W \mathbf x \mathbf x^T \mathbf W^T)]. \nonumber
% = \tr [ \mathbf W \mathbf E_{x} \mathbf W^T ], 
%\label{eq: P_W}
\end{align}
From (\ref{eq: lin_sen}), we obtain that
%where from (\ref{eq: lin_sen}) we obtain
\begin{align}
\boldsymbol {\Sigma}_{ x}  \Def \mathbb E_{\theta, \tilde {\mathbf h}, \boldsymbol \epsilon } [\mathbf x \mathbf x^T] = \boldsymbol \Sigma_{\epsilon} + \eta^2(\mathbf h\mathbf h^T + \boldsymbol \Sigma_h).
\label{eq: E_x}
\end{align}
Thus, the transmission cost can be written as
\begin{align}
T_{\mathbf W} = \tr [ \mathbf W \boldsymbol \Sigma_{x} \mathbf W^T ].
\label{eq: P_W}
\end{align} %}

%We assume that the FC knows the \textit{second-order statistics} of system parameters (i.e., observation gain, information gain and additive noises)  and that the corresponding variance and covariance matrices are invertible.
We assume that the FC knows the \textit{second-order statistics} of the observation gain, information gain, and  additive noises, and that the corresponding variance and covariance matrices are invertible.

To estimate the random parameter $\theta$,
we consider  the linear minimum mean square error (LMMSE) estimator  \cite{karbook} %\cite{karvar13}
\begin{align}
\hat \theta = a_{ \text{\tiny {LMMSE}}  } y,
\end{align}
where $a_{ \text{\tiny {LMMSE}}  }$ is determined by
the minimum mean square error criterion.
From the theory of linear Bayesian estimators \cite{karbook}, we can readily obtain $a_{ \text{\tiny {LMMSE}}  } $  and the corresponding estimation distortion
\begin{subequations}
\begin{equation}
 a_{ \text{\tiny {LMMSE}}  } = \argmin_{a} \mathbb E[(\theta - ay)^2] = \frac{\mathbb E[y \theta]}{ \mathbb E [y^2]}, \quad \text{and}
\end{equation}
\begin{equation}
\hspace*{-0.35in} D_{\mathbf W}  = \mathbb E[(\theta - a_{ \text{\tiny {LMMSE}}  } y)^2] = \eta^2 - \frac{(\mathbb{E}[y \theta])^2}{\mathbb{E}[y^2]}.
\label{eq: LMMSEE_DW}
\end{equation}
\label{eq: LMMSEE}
\end{subequations}
In (\ref{eq: LMMSEE}), substituting (\ref{eq: col_sig}) and
(\ref{eq: channel}), we obtain
\begin{align}
\mathbb E[y^2] =& \mathbb E [ \tilde{\mathbf g}^T \mathbf W \mathbf x \mathbf x^T \mathbf W^T \tilde{\mathbf g}] + \xi^2 \nonumber \\
=& \mathbb E [\tr( \tilde{\mathbf g}  \tilde{\mathbf g}^T \mathbf W \mathbf x \mathbf x^T \mathbf W^T )] = \tr [ \boldsymbol \Sigma_{\tilde g} \mathbf W \boldsymbol \Sigma_x \mathbf W^T ],
\label{eq: Ey2}
\end{align}
where $ \boldsymbol \Sigma_{\tilde g} := \mathbb E[\tilde{\mathbf g} \tilde{\mathbf g}^T ]  = \mathbf g \mathbf g^T + \boldsymbol \Sigma_g$, and $\boldsymbol \Sigma_x$ is given by (\ref{eq: E_x}). 
Moreover, it is easy to show that
\begin{align}
\mathbb{E}[y\theta] = \eta^2 \mathbf g^T \mathbf W \mathbf h.
\label{eq: Eytheta}
\end{align}
Now, the coefficient of LMMSE estimator $a_{ \text{\tiny {LMMSE}}  }$ and the corresponding estimation distortion $D_{\mathbf W}$ are determined according to (\ref{eq: LMMSEE}), (\ref{eq: Ey2}) and (\ref{eq: Eytheta}). 

We define an \textit{equivalent} Fisher information $J_{\mathbf W}  $ which is monotonically related to $D_{\mathbf W}$,
\begin{align}
J_{\mathbf W} := & \frac{1}{D_{\mathbf W}} - \frac{1}{\eta^2} \nonumber \\
 = & \frac{(\mathbf g^T \mathbf W \mathbf h)^2}{\tr[\boldsymbol \Sigma_{\tilde g} \mathbf W \boldsymbol \Sigma_x \mathbf W^T] - \eta^2 (\mathbf g^T \mathbf W \mathbf h)^2 + \xi^2},
\label{eq: J_W}
\end{align}
For convenience, we often express the estimation distortion (\ref{eq: LMMSEE_DW}) as a function of the Fisher information
\begin{align}
D_{\mathbf W}  \Def 
\frac{\eta^2}{1+\eta^2 J_{\mathbf W}}.
\label{eq: D_w}
\end{align}

%%%%%%%%%%%%%%%%%%%%%Section III: Problem Formulation%%%%%%%%%%%%%%%%%%%%%%%%%%%%%%%

\section{Optimal Sparse Sensor Collaboration}
\label{sec: prob_form}
In this section, we first make an association between the collaboration topology and the sparsity structure of the collaboration matrix $\mathbf W$. We then define the collaboration cost and sensor selection cost with the help of the cardinality function (also known as the $\ell_0$ norm). For simplicity of presentation, we concatenate the elements of $\mathbf W$ into a vector form, and present two formulations of collaboration problems (without the cost of sensor selection). By further incorporating the cost of sensor selection, we formulate an optimization problem for the joint design of optimal sensor selection and collaboration schemes.

Recalling the collaboration matrix $\mathbf W$ in (\ref{eq: col_sig}), 
we note that the nonzero entries of $\mathbf W$ correspond to the active collaboration links among sensors. 
For example, $W_{mn} = 0$ indicates the absence of a collaboration link from the $n$th sensor to the $m$th sensor, where $ W_{mn}$ is the $(m,n)$th entry of $\mathbf W$. Conversely, 
$W_{mn} \neq 0$ signifies that the $n$th sensor shares its observation with the $m$th sensor. Thus, 
the \textit{sparsity structure} of $\mathbf W$ characterizes the \textit{collaboration topology}.

%where $\mathrm{card}(\cdot)$ is the cardinality function which gives the
%number of nonzero elements of its (in general, matrix) argument, 

%$\mathrm{card}(\cdot)$ denotes the cardinality function which gives the
%number of nonzero elements of its (vector in general) argument. 
%In (\ref{eq: card}), $\mathrm{card} (W_{mn}) = 0$ indicates no collaboration link from the $n$th sensor to the $m$th sensor, and $\mathrm{card} (W_{mn}) = 1$ signifies that the $n$th sensor shares its observation with the $m$th sensor. 
%For instance, a matrix $\mathbf W$ with $\mathrm{card}(W_{mn})=1$ for all $m, n \in \{1,2,\ldots,N\}$ %and $n = 1,\ldots,N$ 
%corresponds to a \textit{fully-connected} network. And a matrix $\mathbf W$ with $\mathrm{card}(W_{mn})=1$  only for $ m = n = \{1,2,\ldots,N\}$  implies a \textit{distributed} network. 

%\begin{remark}
%A matrix $\mathbf W$ with $\mathrm{card}(W_{mn})=1$ for all $m, n \in \{1,2,\ldots,N\}$ %and $n = 1,\ldots,N$ 
%corresponds to a \textit{fully-connected} network. And a matrix $\mathbf W$ with $\mathrm{card}(W_{mn})=1$  only for $ m = n = \{1,2,\ldots,N\}$  implies a \textit{distributed} network. 
%\end{remark}

For a given collaboration topology, the \textit{collaboration cost} is given by
\begin{align}
Q_{\mathbf W} = \sum_{m = 1}^N \sum_{n=1}^N C_{mn} \mathrm{card} (W_{mn}).%A_{mn},
\label{eq: Q_A}
\end{align}
where $C_{mn}$ is the cost of sharing an observation from the $n$th sensor to the $m$th sensor. Note that $C_{nn} = 0$, since each node can collaborate with itself at no cost. To account for an active collaboration link, we use the cardinality function
\begin{align}
\mathrm{card}(W_{mn}) = \left \{ 
\begin{array}{ll}
0 & \quad W_{mn} = 0 \\
1 & \quad W_{mn} \neq 0.
\end{array}
\right.
\label{eq: card}
\end{align}
%which gives the
%number of nonzero elements of its (in general, matrix) argument.

%where $\mathbf C$ is a known cost matrix, $C_{mn}$ corresponds to the cost of sharing an observation from the $n$th sensor to the $m$th sensor. {We assume that $C_{nn} = 0$, since each node can collaborate with itself at no cost}.

Next, we define the sensor selection/activation cost. Partitioning the matrix $\mathbf W$ rowwise, 
the linear spatial collaboration process (\ref{eq: col_sig}) can be written as
\begin{align}
\begin{bmatrix}
z_1 \\
z_2 \\
\vdots \\
z_N
\end{bmatrix} = \mathbf W \mathbf x =   \begin{bmatrix}
\boldsymbol \omega_1^T \\
\boldsymbol \omega_2^T \\
\vdots \\
\boldsymbol \omega_N^T
\end{bmatrix} \mathbf x, 
\label{eq: W_row_sel}
\end{align}
where $\boldsymbol \omega_n^T$ is the $n$th row vector of $\mathbf W$. It is clear from (\ref{eq: W_row_sel}) that the \textit{non-zero rows} of $\mathbf W$ characterize the selected sensors that communicate with the FC.
Suppose, for example, that only the $n$th sensor communicates with the FC. In this case, 
it follows from (\ref{eq: W_row_sel}) that $z_n =  \boldsymbol \omega_n^T \mathbf x$ and $\boldsymbol \omega_m = 0$ for $m \neq n$. The goal of sensor selection is to find the best \textit{subset} of sensors, i.e., $M$ nodes out of a total of $N$ sensor nodes ($M \leq N$), to communicate with the FC. This is in contrast with the existing work \cite{karvar12isit, karvar12allerton,karvar13}, where the $M$ communicating nodes are selected \textit{a priori}.

%{\color{blue}
%We assume that $M$ nodes, out of a total of $N$ sensor nodes ($M \leq N$), communicate with the FC over a coherent MAC, and each sensor is able to pass its observation to the $M$ communicating nodes.  In the existing literature \cite{karvar12isit, karvar12allerton,karvar13}, the $M$ communicating nodes are given in advance. However, in this paper, we  will find the best \textit{subset} of  sensors to communicate with the FC.
%}

The sensor \textit{selection cost} can be defined through the row-wise cardinality of $\mathbf W$
\begin{align}
S_{\mathbf W} = \sum_{n=1}^N d_n \card(\| \boldsymbol \omega_n \|_2),
\label{eq: E_W}
\end{align}
where $\mathbf d = [d_1,d_2,\ldots,d_N]^T$ is a given vector of sensor selection cost, and  $\| \cdot \|_2$ denotes the Euclidean norm (also known as $\ell_2$ norm).
%It has been shown in \cite{chechuzha05} that $d_n$  can be chosen as the value which is proportional to the distance from the FC to the $n$th sensor. Also 
In (\ref{eq: E_W}), the use of $\ell_2$ norm is motivated by the problem of group Lasso \cite{yualin06}, which uses $\ell_2$ norm to promote the group-sparsity of a vector.

Note that both the expressions of transmission cost (\ref{eq: P_W}) and Fisher information (\ref{eq: J_W}) involve a quadratic matrix function\footnote{A quadratic matrix function is a function $f: \mathbb R^{n \times r} \to \mathbb R$ of the form $f(\mathbf X) = \tr(\mathbf X^T \mathbf A \mathbf X) + 2 \tr(\mathbf B^T \mathbf X) + c$, where $\mathbf A \in \mathbb  R^n$ is a symmetric matrix, $\mathbf B \in \mathbb R^{n \times r}$ and $c \in \mathbb R$.} \cite{bec07}. 
For simplicity of presentation, we convert the quadratic matrix function  to a quadratic vector function, where the elements of $\mathbf W$ are concatenated into its row-wise vector $\mathbf w$.
Specifically, the vector $\mathbf w$ is given by
\begin{align}
\mathbf w \hspace*{-0.02in} = \hspace*{-0.02in}[w_1, w_2,\ldots,w_L]^T \hspace*{-0.02in}, ~~ w_l = W_{m_ln_l},
\label{eq: vector_w}
\end{align}
where $L = N^2$,
$m_l = \lceil \frac{l}{N}\rceil$, $n_l = l - (\lceil \frac{l}{N}\rceil - 1)N$ and $\lceil x \rceil$ is the ceiling function that yields the smallest integer not less than $x$. 
%{\color{blue}
% According to \eqref{eq: W_row_sel}, the vector $\mathbf w$ is given by
%\begin{align}
%\mathbf w   =  [\boldsymbol \omega_1^T, \boldsymbol \omega_2^T,\ldots,\boldsymbol \omega_N^T]^T, \nonumber
%%\label{eq: vector_w}
%\end{align}
%where $\boldsymbol \omega_n \hspace*{-0.02in} =  [ W_{n1},  W_{n2}, \ldots,  W_{nN}]^T$.
%We then describe the index mapping of $\mathbf W$ for $\mathbf w$. 
%}

%{\color{blue}

As shown in Appendix\,\ref{appendix: quad_vec},  the expressions of transmission cost (\ref{eq: P_W}), Fisher information (\ref{eq: J_W}), collaboration cost \eqref{eq: Q_A}, and selection cost \eqref{eq: E_W} can be converted into functions of the collaboration vector $\mathbf w$,
\begin{subequations}
\begin{equation}
T(\mathbf w) = \mathbf w^T \boldsymbol \Omega_{\Tsj} \mathbf w, 
\label{eq: Tw_vector}
\end{equation}
\begin{equation}
J(\mathbf w) = \frac{\mathbf w^T \boldsymbol \Omega_{\text{\tiny{JN}}} \mathbf w}{\mathbf w^T \boldsymbol \Omega_{\text{\tiny {JD}}} \mathbf w + \xi^2}, 
\label{eq: Jw_vector}
\end{equation} 
\begin{equation}
Q(\mathbf w) = \sum_{l=1}^L c_l \, \mathrm{card}(w_l), \quad %~\text{and} ~
\label{eq: Qw_vector}
\end{equation}
\begin{equation}
S(\mathbf w) = \sum_{n=1}^N d_n \card(\| \mathbf w_{\Gsj_n} \|_2),
\label{eq: Sw_vector}
\end{equation}
\end{subequations}
where the matrices $\boldsymbol \Omega_{\Tsj}$, $\boldsymbol \Omega_{\text{\tiny{JN}}}$, 
$\boldsymbol \Omega_{\text{\tiny {JD}}}$ are all symmetric positive semidefinite and defined in 
Appendix\,\ref{appendix: quad_vec}, $c_l$ is the $l$th entry of $\mathbf c$ that is the row-wise vector of the known cost matrix $\mathbf C$, and $\text{G}_n$ is an index set such that $
\mathbf w_{\Gsj_n} = [w_{(n-1)N+1}, w_{(n-1)N+2}, \ldots, w_{nN}]^T
$ for $n=1,2,\ldots,N$. It is clear from \eqref{eq: W_row_sel} and \eqref{eq: Sw_vector} that
the row-sparsity of $\mathbf W$ is precisely characterized by the group-sparsity of $\mathbf w$ over the index sets $\{ \mathrm{G}_n \}_{n=1,2,\ldots,N}$.
Based on the transmission cost $T(\mathbf w)$, collaboration cost $Q(\mathbf w)$, and performance measure $J(\mathbf w)$, 
we first pose the sensor collaboration problems by disregarding the cost of sensor selection and assuming that all $N$ sensors are active.
\begin{itemize}
\item \textit{Information constrained} sensor collaboration
%\begin{align}
%\begin{array}{ll}
%\hspace*{-0.4cm} \displaystyle \minimize_{\mathbf w} & \mathbf  w^T \displaystyle \boldsymbol \Omega_{\Tsj} \mathbf w +  \sum_{l=1}^L c_l\, \mathrm{card}(w_l)  \\
%\hspace*{-0.4cm} \st &  
%%\mathbf w^T (J \boldsymbol \Omega_{\JDsj} - \boldsymbol \Omega_{\JNsj})\mathbf w + J \xi^2 \leq 0,
%\displaystyle \frac{\mathbf w^T \boldsymbol \Omega_{\text{\tiny{JN}}} \mathbf w}{\mathbf w^T \boldsymbol \Omega_{\text{\tiny{JD}}} \mathbf w + \xi^2} \geq J,
%\end{array}
%\tag{$\mathrm{P}_1$}
%\label{eq: inf_con_prob}
%\end{align}
\begin{align}
\begin{array}{ll}
\displaystyle \minimize_{\mathbf w} &  P(\mathbf w)  \\
\st &  J(\mathbf w) \geq \check J,
\end{array}
\tag{$\mathrm{P}1$}
\label{eq: inf_con_prob}
\end{align}
\end{itemize}
where 
\[
P(\mathbf w) \Def T(\mathbf w) + Q (\mathbf w),
\] and $\check J > 0 $ is a given information threshold. 
%{\color{blue}where the constraint is equivalent to the information inequality 
%$J_{\mathbf w} \geq J$  (see $J_{\mathbf w}$ in \eqref{eq: Jw_Pw}),} and
%$J > 0 $ is a given information threshold. {\color{blue} Note that \eqref{eq: inf_con_prob} is not a convex problem, since the matrix $J \boldsymbol \Omega_{\JDsj} - \boldsymbol \Omega_{\JNsj}$ is \textit{not} positive semi-definite. It is easy to see that problem (\ref{eq: inf_con_prob}) becomes infeasible if the matrix $J \boldsymbol \Omega_{\JDsj} - \boldsymbol \Omega_{\JNsj}$ is positive semi-definite. }

\begin{itemize}
\item \textit{Energy constrained} sensor collaboration
%\begin{align}
%\begin{array}{ll}
%\hspace*{-0.8cm} \displaystyle \maximize_{\mathbf w}   & \hspace*{-0.07cm} \displaystyle \frac{\mathbf w^T \boldsymbol \Omega_{\text{\tiny{JN}}} \mathbf w}{\mathbf w^T \boldsymbol \Omega_{\text{\tiny{JD}}} \mathbf w + \xi^2}\\
%\hspace*{-0.8cm}  \st   &  \displaystyle \hspace*{-0.07cm}  \mathbf w^T \boldsymbol \Omega_{\Tsj} \mathbf w \hspace*{-0.04cm}  +  \hspace*{-0.04cm}  \sum_{l=1}^L \hspace*{-0.04cm}  c_l \, \mathrm{card}(w_l) \leq P,
%\end{array}
%\tag{$\mathrm{P}_2$}
%\label{eq: ene_con_prob}
%\end{align}
\begin{align}
\begin{array}{ll}
\displaystyle \maximize_{\mathbf w}   & \displaystyle J(\mathbf w)\\
\st   &  P(\mathbf w) \leq \hat P,
\end{array}
\tag{$\mathrm{P}2$}
\label{eq: ene_con_prob}
\end{align}
\end{itemize}
where $\hat P > 0 $ is a given energy budget.

Next, we incorporate the sensor selection cost $S(\mathbf w)$ in \eqref{eq: Sw_vector}, and
pose the optimization problem for the joint design of optimal sensor selection and collaboration schemes.
\begin{itemize}
\item \textit{Joint} sensor selection and collaboration
%\begin{align}
%\hspace*{-0.3in}
%\begin{array}{ll}
%\displaystyle \minimize_{\mathbf w} &  \hspace*{-0.08in} \displaystyle  \mathbf w^T \boldsymbol \Omega_{\Tsj} \mathbf w \hspace*{-0.03in} +  \hspace*{-0.02in} \sum_{l=1}^L \hspace*{-0.01in} c_l \, \mathrm{card}(w_l) %\vspace*{-0.06in} \\
%%& \displaystyle 
%\hspace*{-0.03in} + \hspace*{-0.02in} \sum_{n=1}^N \hspace*{-0.01in} d_n \card(\| \mathbf w_{\Gsj_n} \|_2) \\
%%\vspace*{0.04in} \\
%\st &  \hspace*{-0.08in} \displaystyle \frac{\mathbf w^T \boldsymbol \Omega_{\text{\tiny{JN}}} \mathbf w}{\mathbf w^T \boldsymbol \Omega_{\text{\tiny{JD}}} \mathbf w + \xi^2} \geq J.
%\end{array}
%\tag{$\mathrm{P}_3$}
%\label{eq: inf_sel_ori}
%\end{align}
\begin{align}
\hspace*{-0.3in}
\begin{array}{ll}
\displaystyle \minimize_{\mathbf w} & \displaystyle P(\mathbf w) + S(\mathbf w)  \\
\st &  \displaystyle J(\mathbf w) \geq \check J.
\end{array}
\tag{$\mathrm{P}3$}
\label{eq: inf_sel_ori}
\end{align}
\end{itemize}     
%} 
In (\ref{eq: inf_sel_ori}), we minimize the total energy cost subject to an information constraint. This formulation is motivated by scenarios where saving energy is the major goal 
in the context of sensor selection \cite{liuzhama09}. 
Problem (\ref{eq: inf_sel_ori})
is of a similar form as (\ref{eq: inf_con_prob}) except for the incorporation of sensor selection cost. However, we will show that the presence of the sensor selection cost makes finding the solution of (\ref{eq: inf_sel_ori})  more challenging; see Sec.\,\ref{sec: sel} for details.

In (\ref{eq: inf_con_prob})-(\ref{eq: inf_sel_ori}), the cardinality function, which appears in $Q(\mathbf w)$ and $S(\mathbf w)$, 
promotes the sparsity of $\mathbf w$, and therefore the sparsity of $\mathbf W$.
%determines the sparsity structure of the collaboration matrix.
%Therefore, we call problems (\ref{eq: inf_con_prob})-(\ref{eq: inf_sel_ori}) \textit{sparsity-aware} sensor collaboration problems.
Thus, we refer to (\ref{eq: inf_con_prob})-(\ref{eq: inf_sel_ori}) as sparsity-aware sensor collaboration problems. It is worth mentioning that the proposed sensor collaboration problems are solved in a centralized manner at the FC, whereas the inter-sensor collaboration occurs in a distributed way and among sensors.
We also note that (\ref{eq: inf_con_prob})-(\ref{eq: inf_sel_ori}) are nonconvex optimization problems due to the presence of the cardinality function and the nonconvexity of the expression for the Fisher information (see Remark\,\ref{rk: nonconvexity}). 
In the following sections, we will elaborate on the optimization approaches for solving (\ref{eq: inf_con_prob})-(\ref{eq: inf_sel_ori}).

%%%%%%%%%%%%%%%%%%%%%Section IV: Information constrained problem%%%%%%%%%%%%%%%%%%%%%%%%%%%%%%%

\section{%Problem \eqref{eq: inf_con_prob}: 
Information Constrained Sensor Collaboration}
\label{sec: inf_coll}

In this section, we relax the original information constrained problem (\ref{eq: inf_con_prob}) by using an iterative reweighted $\ell_1$ minimization method. This results in an $\ell_1$ optimization problem, which can be efficiently solved by ADMM.

Due to the presence of the cardinality function, problem (\ref{eq: inf_con_prob}) is combinatorial in nature.
A state-of-the-art method for solving (\ref{eq: inf_con_prob}) is to replace the cardinality function (also referred to as the $\ell_0$ norm) with a weighted $\ell_1$ norm \cite{canwakboy08}. This leads to the following optimization problem
\begin{align}
\begin{array}{ll}
\displaystyle \minimize_{\mathbf w} 
%\displaystyle w^{t+1} = \argmin_{\mathbf w} 
&  \mathbf w^T \boldsymbol \Omega_{\Tsj} \mathbf w + \| \boldsymbol \Omega_{\Csj} \mathbf w  \|_1 %\sum_{l=1}^L C_{m_ln_l} \alpha_l^t |w_l| 
\\
\st 
& \mathbf w^T (\check J \boldsymbol \Omega_{\text{\tiny{JD}}} -  \boldsymbol \Omega_{\text{\tiny{JN}}})\mathbf w + \check J \xi^2 \leq 0,
\end{array}
%\tag{${\mathrm{P}_1^{\ell_1}}$}
\label{eq: inf_con_prob_l1}
\end{align}
where $ \boldsymbol \Omega_{\Csj} = \diag(\alpha_1^t c_1, \alpha_2^t c_2, \ldots, \alpha_L^t c_L)$, and $\{ \alpha_l^t\}_{l=1,2,\ldots,L}$ denote the weights assigned for entries of an $\ell_1$ norm at the $t$th iteration  in Algorithm\,1.
If $\alpha_l  = 1$ for all $l \in \{1,2,\ldots,L \}$, we recover the standard unweighted $\ell_1$ norm. 
Since the $\ell_0$ norm only counts the number of nonzero entries of a vector, 
the use of the $\ell_1$ norm for approximating the $\ell_0$ norm has the disadvantage that   the amplitudes of the nonzero entries come into play. To compensate for the amplitude of nonzero entries, 
we iteratively normalize the entries of the argument of the $\ell_1$ norm, to make this norm a better proxy for the $\ell_0$ norm.
We summarize the reweighted $\ell_1$ method  for solving (\ref{eq: inf_con_prob}) in Algorithm\,1.

%However, the use of the $\ell_1$ norm leads to an undesired
%dependence on the magnitude of elements in a vector. 
%Therefore, the authors in \cite{canwakboy08} proposed an iterative reweighted $\ell_1$ method, which attempts to better approximate the $\ell_0$ norm at the expense of several reweighting iterations. 
%We state the reweighted $\ell_1$ method for solving (\ref{eq: inf_con_prob}) in Algorithm\,1. 

\begin{algorithm}
\caption{Reweighted $\ell_1$ method for solving (\ref{eq: inf_con_prob})}
\begin{algorithmic}[1]
\Require given $\varepsilon > 0$ and $\epsilon_{\mathrm{rw}} > 0$. Set $\alpha_l^0 = 1$ for $l = 1, \ldots, L$ and $ \boldsymbol \Omega_{\Csj} = \diag(\alpha_1^0 c_1, \alpha_2^0 c_2, \ldots, \alpha_L^0 c_L)$.
\For {$t = 0 , 1, \ldots $}
\State solve problem (\ref{eq: inf_con_prob_l1}) to
%the reweighted $\ell_1$-based 
%(\ref{eq: inf_con_prob}) to 
obtain solution
\Statex \hspace*{0.2in}$\mathbf w^t   = [w_1^t, w_2^t, \ldots, w_L^t]^T. $
%\[\mathbf w^t  \hspace*{-0.05in} = \hspace*{-0.02in} [w_1^t, w_2^t, \ldots, w_L^t]^T.\]
\State update the weights $\displaystyle \alpha_l^{t+1} =  \frac{1}{|w_l^t| + \varepsilon}$ and
\Statex \hspace*{0.2in}$ \boldsymbol \Omega_{\Csj} = \diag(\alpha_1^{t+1} c_1, \alpha_2^{t+1}c_2, \ldots, \alpha_L^{t+1} c_L)$.
\State if $\| \mathbf w^{t+1} - \mathbf w^{t}\|_2 < \epsilon_{\mathrm{rw}} $, \textbf{quit}.
\EndFor
\end{algorithmic}
\end{algorithm}

Reference \cite{canwakboy08} shows that much of the benefit of using the reweighted $\ell_1$ method is gained from its first few iterations.
In Step\,3 of Algorithm\,1, the positive scalar $\varepsilon$ is a small number which insures that the denominator is always nonzero, and helps the convergence of Algorithm\,1; for example,
if $w_l^t \to 0$, the weight $\alpha_l^{t+1}$ converges to $\frac{1}{\varepsilon}$.

\begin{remark}
\label{rk: nonconvexity}
In (\ref{eq: inf_con_prob_l1}), the inequality constraint is equivalent to the information inequality in (\ref{eq: inf_con_prob}). According to Lemma\,\ref{lemma2} in Appendix\,\ref{appendix: theorem1}, 
we obtain that the matrix $\check J \boldsymbol \Omega_{\JDsj} - \boldsymbol \Omega_{\JNsj}$ is not positive semidefinite. Indeed, if the matrix   $\check J \boldsymbol \Omega_{\JDsj} - \boldsymbol \Omega_{\JNsj}$ was positive semidefinite, problem (\ref{eq: inf_con_prob_l1}) would have an empty feasible set. 
%Therefore,  
%%the matrix $J \boldsymbol \Omega_{\JDsj} - \boldsymbol \Omega_{\JNsj}$ is \textit{not} positive semi-definite, and 
%the Fisher information constraint is not convex.
\end{remark}

%The reweighted $\ell_1$ minimization method typically takes several iterations to converge \cite{canwakboy08}. In our numerical experiments, only $4$ or $5$ iterations were required for satisfactory accuracy. 

%{\color{blue}where the constraint is equivalent to the information inequality 
%$J_{\mathbf w} \geq J$  (see $J_{\mathbf w}$ in \eqref{eq: Jw_Pw}),} and
%$J > 0 $ is a given information threshold. {\color{blue} Note that \eqref{eq: inf_con_prob} is not a convex problem, since the matrix $J \boldsymbol \Omega_{\JDsj} - \boldsymbol \Omega_{\JNsj}$ is \textit{not} positive semi-definite. It is easy to see that problem (\ref{eq: inf_con_prob}) becomes infeasible if the matrix $J \boldsymbol \Omega_{\JDsj} - \boldsymbol \Omega_{\JNsj}$ is positive semi-definite. }

Given $\{ \alpha_l^t\}_{l=1,\ldots,L}$,
problem (\ref{eq: inf_con_prob_l1}) is a nonconvex optimization problem, %since the matrix $J \boldsymbol \Omega_{\JDsj} - \boldsymbol \Omega_{\JNsj}$ is not positive semidefinite. 
and its objective function is not differentiable. 
In what follows, 
we will employ ADMM to find its locally optimal solutions.
%two efficient optimization methods: ADMM and linearization method.

\subsection*{Alternating Direction Method of Multipliers }
\label{sec: admm}

In our earlier work \cite{liukarfarvar14_isit}, we have applied ADMM to solve problem 
(\ref{eq: inf_con_prob_l1}). 
ADMM is an optimization method well-suited for problems that involve sparsity-inducing regularizers (e.g., cardinality function or $\ell_1$ norm) \cite{boyparchupeleck11,linfarjov13}. The major advantage of  ADMM is that it allows us to split the optimization problem  (\ref{eq: inf_con_prob_l1}) into a nonconvex quadratic program with only one quadratic constraint (QP1QC) and an unconstrained $\ell_1$ norm optimization problem, 
of which the former can be solved efficiently and the latter analytically.
When applied to a non-convex problem such as \eqref{eq: inf_con_prob_l1}, ADMM is not guaranteed to converge and yields locally optimal solutions when it does \cite{boyparchupeleck11}. However, we have found ADMM to both converge and yield satisfactory results for \eqref{eq: inf_con_prob_l1}. Indeed, our numerical observations agree with the literature \cite{boyparchupeleck11,linfarjov13,liufarmasvar14} that demonstrates the power and utility of ADMM in solving nonconvex optimization problems.

We begin by reformulating the optimization problem (\ref{eq: inf_con_prob_l1}) in a way that lends itself to the application of ADMM, 
\begin{align}
\begin{array}{ll}
\displaystyle \minimize_{\mathbf w, \mathbf v} &  \mathbf w^T \boldsymbol \Omega_{\Tsj} \mathbf w +   \|\boldsymbol {\Omega}_{\Csj} \mathbf v  \|_1 %\sum_{l=1}^L C_{m_ln_l} \alpha_l |v_l| 
+ \mathcal I (\mathbf w) 
 \\
\st &  \mathbf w = \mathbf v,
\label{eq: inf_con_prob_wl1_ad}
\end{array}
\end{align}
where we introduce the indicator function $\mathcal I (\mathbf w)$
\begin{align}
\mathcal I (\mathbf w) = \left \{
\begin{array}{ll}
0 & \text{if $\mathbf w^T (\check J \boldsymbol \Omega_{\text{\tiny{JD}}} - \boldsymbol \Omega_{\text{\tiny{JN}}})\mathbf w + J \xi^2 \leq 0$} \\
\infty & \text{otherwise}.
\end{array}
 \right. 
 \label{eq: indicator}
\end{align}

The augmented Lagrangian of (\ref{eq: inf_con_prob_wl1_ad}) is given by
\begin{align}
\mathcal L (\mathbf w, \mathbf v, \boldsymbol \chi) = &~
\mathbf w^T \boldsymbol \Omega_{\text{\tiny{T}}} \mathbf w +   \|\boldsymbol {\Omega}_{\Csj} \mathbf v  \|_1 %\sum_{l=1}^L C_{m_ln_l} \alpha_l |v_l| 
+ \mathcal I (\mathbf w)  \nonumber \\
& ~ +  \boldsymbol \chi^T (\mathbf w - \mathbf v) + \frac{\rho}{2} 
\| \mathbf w - \mathbf v \|_2^2,
\label{eq: ALag}
\end{align}
where the vector $\boldsymbol \chi$ is the Lagrangian multiplier, and the scalar $\rho > 0$ is a penalty weight.
The ADMM algorithm iteratively executes the following three steps \cite{boyparchupeleck11} for  $k=1,2,\ldots$
\begin{align}
\mathbf w^{k+1}
&= \argmin_{\mathbf w} ~ \mathcal{L} (\mathbf w,\mathbf v^k, \boldsymbol \chi^k),
\label{eq: w_step} \\
\mathbf v^{k+1}
&= \argmin_{\mathbf v} ~ \mathcal {L} (\mathbf w^{k+1}, \mathbf v, \boldsymbol \chi^k),
\label{eq: v_step} \\
\boldsymbol \chi^{k+1}
&= \boldsymbol \chi^k + \rho (\mathbf w^{k+1} - \mathbf v^{k+1}), 
\label{eq: dual_step}
\end{align}
%and $\boldsymbol \lambda^{k+1}
%= \boldsymbol \lambda^k + \rho (\mathbf w^{k+1} - \mathbf v^{k+1})$,
until
$\| \mathbf w^{k+1}-\mathbf v^{k+1}\|_2 \leq \epsilon_{\mathrm{ad}}$ and $\| \mathbf v^{k+1} - \mathbf v^{k}\|_2 \leq \epsilon_{\mathrm{ad}}$, where $\epsilon_{\mathrm{ad}}$ is a stopping tolerance. 
%To initialize ADMM, we choose $\mathbf v^0 = \mathbf w^0$ and $\mathbf w^0 = \tilde{\mathbf w}$, where $\tilde{\mathbf w} $ is given by (\ref{eq: eig_P01}); see more details on initialization in Sec.\,\ref{sub: initialization}.
%(\ref{eq: w_fJ}).

It is clear from ADMM steps (\ref{eq: w_step})-(\ref{eq: v_step}) that the original non-differentiable problem can be effectively separated into a `$\mathbf w$-minimization' subproblem (\ref{eq: w_step}) and a `$\mathbf v$-minimization' subproblem (\ref{eq: v_step}), of which the former can be treated as a nonconvex QP1QC and the latter can be solved analytically. 
In the subsections that follow, we will elaborate on  the execution of the minimization problems (\ref{eq: w_step}) and (\ref{eq: v_step}).

\subsubsection{$\mathbf w$-minimization step}
Completing the squares with respect to $\mathbf w$ in (\ref{eq: ALag}), the $\mathbf w$-minimization step (\ref{eq: w_step}) is given by
\begin{align}
\begin{array}{ll}
\displaystyle \minimize_{\mathbf w} &  \mathbf w^T \boldsymbol \Omega_{\text{\tiny{T}}} \mathbf w + \frac{\rho}{2}\| \mathbf w - \mathbf a\|_2^2 \\
\st &  \mathbf w^T (\check J \boldsymbol \Omega_{\text{\tiny{JD}}} - \boldsymbol \Omega_{\text{\tiny{JN}}})\mathbf w + \check J \xi^2 \leq 0 ,
\end{array}
\label{eq: phi_w}
\end{align}
where we have applied the definition of $\mathcal I(\mathbf w)$ in (\ref{eq: indicator}), and $\mathbf a \Def \mathbf v^k - 1/\rho \boldsymbol \chi^k$. 
%For simplicity, we will use $\mathbf a$ instead of $\mathbf a^k$. 
Problem (\ref{eq: phi_w}) is a nonconvex QP1QC.
%%%%%%%%%%%%%%%%%%%%%%%%%%%%%%%%%%%%%%%%%%%%%%%%%%%%%%%%%%%%%%%%%%%%%%
To seek the global minimizer of a nonconvex QP1QC, an approach based on semidefinite program (SDP) relaxation has been used in \cite{liukarfarvar14_isit}. However, computing solutions to SDP problems %, e.g. calling the off-the-shelf solver CVX \cite{cvx}, 
becomes inefficient for problems with hundreds or thousands of variables. Therefore, we develop a  faster approach by exploiting the KKT conditions  of (\ref{eq: phi_w}). This is presented in Prop.\,\ref{prop: sol_phi_w}.

\begin{myprop}
\label{prop: sol_phi_w}
The KKT-based solution of problem (\ref{eq: phi_w}) is given by
\[
\mathbf w^{k+1} = \tilde{\boldsymbol \Omega}_{\Tsj}^{-\frac{1}{2}} \mathbf U \mathbf u ,
\]
where $\tilde{\boldsymbol \Omega}_{\Tsj} \Def \boldsymbol \Omega_{\Tsj} + \frac{\rho}{2} \mathbf I$, $\mathbf U$ is an orthogonal matrix that satisfies the eigenvalue decomposition
\begin{align}
 \frac{1}{\check J \xi^2} \tilde{\boldsymbol \Omega}_{\Tsj}^{-\frac{1}{2}} (\check J \boldsymbol \Omega_{\text{\tiny{JD}}} - \boldsymbol \Omega_{\text{\tiny{JN}}})  \tilde{\boldsymbol \Omega}_{\Tsj}^{-\frac{1}{2}} = \mathbf U \boldsymbol \Lambda \mathbf U^T,
 \label{eq: decom_prop1}
\end{align}
and $\mathbf u$ is given by
\begin{align}
\left \{
\begin{array}{ll}
\mathbf u = -  \mathbf g & \text{if $ \mathbf g^T \boldsymbol \Lambda  \mathbf g+ 1 \leq 0 $} \\
\mathbf u = - (\mathbf I + \mu_0 \boldsymbol \Lambda )^{-1} \mathbf g  &  
\text{otherwise}.
\end{array}
\right.
\label{eq: u_phi_col}
\end{align}
In (\ref{eq: u_phi_col}), $\mathbf g \Def - \frac{\rho}{2}  \mathbf U^T \tilde{\boldsymbol \Omega}_{\Tsj}^{-\frac{1}{2}} \mathbf a$, and
$\mu_0 $ is a \textit{positive} root of the equation in $\mu$
\begin{align}
f(\mu) \Def \sum_{l=1}^L\frac{\lambda_lg_l^2}{( \mu \lambda_l + 1 )^2}  + 1 = 0, 
\label{eq: nl_eig}
\end{align}
where $g_l$ is the $l$th element of $\mathbf g$, and $\lambda_l$ is the $l$th diagonal entry of $\boldsymbol \Lambda $. 
\end{myprop}
\textbf{Proof:} See Appendix\,\ref{appendix: sol_phi_w}, in which
letting $\mathbf A_0 = \tilde{\boldsymbol \Omega}_{\Tsj}$ , $\mathbf b_0 = - \frac{\rho}{2} \mathbf a$, $\mathbf A_1 =\check J \boldsymbol \Omega_{\text{\tiny{JD}}} - \boldsymbol \Omega_{\text{\tiny{JN}}}$, $\mathbf b_1 = \mathbf 0$ and $r_1 = \check J \xi^2$, we can obtain the results given in Prop.\,\ref{prop: sol_phi_w}.
\hfill $\blacksquare$

%\begin{remark}
The rationale behind deriving the eigenvalue decomposition \eqref{eq: decom_prop1} is that by introducing $\mathbf u = \mathbf U^T \tilde{\boldsymbol \Omega}_{\Tsj}^{\frac{1}{2}} \mathbf w$, 
problem (\ref{eq: phi_w}) can be transformed to
\begin{align}
\begin{array}{ll}
\displaystyle \minimize_{\mathbf u} & \quad \mathbf u^T \mathbf u + 2 \mathbf u^T \mathbf g \\
\st & \quad \mathbf u^T \boldsymbol \Lambda  \mathbf u + 1 \leq 0.
\end{array}
\label{eq: prob_simple_prop1}
\end{align}
The benefit of this reformulation is that the KKT conditions of (\ref{eq: prob_simple_prop1}) are more compact and easily solved, since $\boldsymbol \Lambda $ is a diagonal matrix and its inversion is tractable. The KKT conditions of (\ref{eq: prob_simple_prop1}) are precisely depicted by (\ref{eq: u_phi_col}) and (\ref{eq: nl_eig}). We note that solutions of (\ref{eq: nl_eig}) can be found by using the MATLAB function \textsf{fminbnd} or by using Newton's method. 

%obtained  efficiently through the solution of a nonlinear equation with a \textit{scalar} argument
%We note that solutions of (\ref{eq: nl_eig}) can be found by using 
%the MATLAB function \textsf{fminbnd} or by using Newton's method. 

In general, Eq.\,\eqref{eq: nl_eig} is a high-order polynomial function and it is very difficult to  obtain all the positive roots. However, we have observed that numerical searches over small targeted intervals yields satisfactory results. One such interval is given by Lemma\,3, and the other is demonstrated in Remark\,\ref{rk: remark_interval} below.
If we find multiple positive roots, we select the one corresponding  to the lowest objective value of the nonconvex QP1QC \eqref{eq: phi_w}.
%it is intractable to enumerate all the possible positive roots of (\ref{eq: nl_eig}). 
%However,  several insights on the features of (\ref{eq: nl_eig}) can be exploited to
%establish different initialization regions, during which 
%the numerical method (e.g., MATLAB function \textsf{fminbnd} or Newton's method) can be applied to
%find different positive roots. We then 
%choose a \textit{better} positive root which renders a lower objective value of the nonconvex QP1QC \eqref{eq: phi_w}. The particular features of (\ref{eq: nl_eig}) are presented in Lemma\,3. 

\textbf{Lemma\,3:}
\textit{The function  $f(\mu)$  is monotonically decreasing on the interval $(0,- \frac{1}{\lambda_1})$ and the positive root of $f(\mu)=0$ is unique when $f(0) > 0$, where $\lambda_1$ represents the unique  negative eigenvalue in $\{ \lambda_l\}_{l=1,2,\ldots,L}$.
%If $\mu \in (- \frac{1}{\lambda_1}, + \infty )$, the function $f(\mu)$ may not be monotonic, and  the number of positive roots of $f(\mu) = 0$ is uncertain.
}

\textbf{Proof:} See Appendix\,\ref{appendix: lemma3}. \hfill $\blacksquare$

\begin{remark} \label{rk: remark_interval}
Motivated by Lemma\,3, one may inquire about the monotonicity of $f(\mu)$ over the interval
$\mu \in (- \frac{1}{\lambda_1}, \infty)$. In
Appendix\,\ref{appendix: lemma3}, we show that the sign of the first-order derivative of $f(\mu)$ is difficult to determine from \eqref{eq: df_mu} and \eqref{eq: interval_inf}. And our numerical results show that there  may exist other positive roots over the interval $(- \frac{1}{\lambda_1}, \infty)$.
%Therefore, there is no guarantee the number of positive roots of $f(\mu) = 0$ is uncertain.
\end{remark}

In general, we cannot guarantee global optimality for solutions found through KKT, since KKT conditions constitute only  necessary conditions for optimality in nonconvex problems \cite{boyvan04}.
However, our extensive numerical results show that 
 numerical search over several small intervals works effectively for finding the positive roots of Eq.\,\eqref{eq: nl_eig} and the ADMM algorithm always converges to a near-optimal solution of the information constrained   collaboration problem. %As a result, the
%ADMM algorithm converges well in practice. 

\subsubsection{$\mathbf v$-minimization step}
Completing the squares with respect to $\mathbf v$ in (\ref{eq: ALag}), the $\mathbf v$-minimization step (\ref{eq: v_step}) becomes
\begin{align}
\begin{array}{ll}
\displaystyle \minimize_{\mathbf v} &  \|\boldsymbol \Omega_{\Csj} \mathbf v  \|_1 %\sum_{l=1}^L C_{m_ln_l} \alpha_l |v_l| 
+ 
\frac{\rho}{2}\| \mathbf v - \mathbf b\|_2^2,
\end{array}
\label{eq: psi_v}
\end{align}
where $\mathbf b \Def \frac{1}{\rho} \boldsymbol \chi^k + \mathbf w^{k+1}$.  
The solution of (\ref{eq: psi_v}) is given by soft thresholding \cite{linfarjov13}
\begin{align}
v_l = \left\{
\begin{array}{cc}
(1 - \frac{ \alpha_l^t c_l }{\rho |b_l|}) b_l & | b_l| > \frac{\alpha_l^t c_l}{\rho}  \\
0 & | b_l | \leq \frac{\alpha_l^t c_l}{\rho}
\end{array}
\right. 
\label{eq: prox_v}
\end{align}
for $l=1,2,\ldots,L$, where $v_l$ denotes the $l$th element of the vector $\mathbf v$.

\subsubsection{Initialization} 
\label{sub: initialization}
%{\color{blue} 
To initialize ADMM we require a feasible vector.
%For initializing ADMM, we are required to seek a feasible vector $\mathbf w^0$ of problem (\ref{eq: inf_con_prob_l1}). 
%Recalling that  the inequality constraint of (\ref{eq: inf_con_prob_l1}) is equivalent to 
%the information inequality $J_{\mathbf w} \geq J$, where $J_{\mathbf w}$ was defined in \eqref{eq: Jw_Pw}. Thus for a given information threshold $J$, we need to determine $\mathbf w^0$ such that $J_{\mathbf w_0} \geq J$. 
It has been shown in
Theorem\,1 (see Appendix\,\ref{appendix: theorem1}) that the optimal collaboration vector for a {fully-connected} network with an information threshold $\check J$ is a feasible vector
for  (\ref{eq: inf_con_prob_l1}). Thus, we choose $\mathbf v^0 = \mathbf w^0$ and $\mathbf w^0 = \tilde{\mathbf w}$, where $\tilde{\mathbf w}$ is given by \eqref{eq: eig_P01}. %}

%, in terms of the optimal value and solution for problem (\ref{eq: inf_con_prob}) (with a fully-connected topology), are given by
%\begin{equation}
%\left \{ \begin{array}{l}
%\tilde P  =  \lambda_{\mathcal G, \mathrm{min}}^{\mathrm{pos}} \left(  {\boldsymbol \Omega}_{\text{\tiny{P}}},  - { \boldsymbol \Omega}_{\text{\tiny {JD} }} + \frac{{\boldsymbol \Omega}_{\text{\tiny{JN} } }}{J}\right) \xi^2  + \mathbf 1^T \mathbf c \\
%\tilde {\mathbf w} = a  \tilde {\mathbf v},
%\end{array}
%\right.
%\label{eq: eig_P01}
%\end{equation}
%where $\lambda_{\mathcal G, \mathrm{min}}^{\mathrm{pos}} (\mathbf A, \mathbf B)$ denotes the minimum positive eigenvalue of the generalized eigenvalue problem  $\mathbf A \mathbf v = \lambda \mathbf B \mathbf v$, $\tilde {\mathbf v}$ is the corresponding eigenvector, $\mathbf c$ is the known vector of collaboration cost, and
%the scalar $a$ satisfies $\tilde{\mathbf w}^T {\boldsymbol \Omega}_{\Tsj}  \tilde {\mathbf w}   =  \tilde P - \mathbf 1^T \mathbf c$.
%
%Since $J_{\tilde {\mathbf w}} \geq J$, the vector $\tilde {\mathbf w}$ is feasible for the optimization problem (\ref{eq: inf_con_prob_l1}). Thus, we choose
%$\mathbf w^0 = \tilde {\mathbf w}$. } 

\subsubsection{Complexity Analysis}
To solve the information constrained collaboration problem (\ref{eq: inf_con_prob}), the iterative reweighted $\ell_1$ method (Algorithm\,1) is used as the outer loop, and 
the ADMM algorithm constitutes the inner loop. 
%It has been shown in \cite{canwakboy08} that much of the benefit by using reweighted $\ell_1$ method is reaped from
%the first few reweighting iterations. In our numerical examples with $100$ optimization variables, the number of required iterations is less than $5$.  
It is often the case that the iterative reweighted $\ell_1$ method converges  within a few iterations \cite{canwakboy08,liuvemfarmasvar_14,liufarmasvar14,liumasfarvar14_icassp}.
Moreover, it has been shown in \cite{boyparchupeleck11} that the ADMM algorithm typically requires a few tens of iterations for converging with modest accuracy. %\cite{boyparchupeleck11, linfarjov13, masfarvar12}.
At each iteration, the major cost is
associated with solving the KKT conditions in
the w-minimization step. The complexity of
obtaining a KKT-based solution is given by
$O(L^3)$ \cite{panche99}, since the complexity of the  eigenvalue decomposition dominates that of Newton's method. 
%This is in contrast with the SDP-based $\mathbf w$-minimization step in \cite{liukarfarvar14_isit}, where the complexity is approximated by $O(L^{4.5})$.

 %%%%%%%%%%%%%%%%%%%%%Section V: Energy constrained problem%%%%%%%%%%%%%%%%%%%%%%%%%%%%%%%

\section{%Problem (\ref{eq: ene_con_prob}): 
Energy Constrained Sensor Collaboration}
\label{sec: ene_coll}
In this section, we first explore the correspondence between the energy constrained collaboration problem and the information constrained problem. With the help of this correspondence, we propose a bisection algorithm to solve the energy constrained problem. 

According to \eqref{eq: Tw_vector}, \eqref{eq: Jw_vector} and   \eqref{eq: Qw_vector},
the energy constrained sensor collaboration problem (\ref{eq: ene_con_prob}) can be written as
\[
\begin{array}{ll}
\hspace*{-0.8cm} \displaystyle \maximize_{\mathbf w}   & \hspace*{-0.07cm} \displaystyle \frac{\mathbf w^T \boldsymbol \Omega_{\text{\tiny{JN}}} \mathbf w}{\mathbf w^T \boldsymbol \Omega_{\text{\tiny{JD}}} \mathbf w + \xi^2}\\
\hspace*{-0.8cm}  \st   &  \hspace*{-0.07cm}  \mathbf w^T \boldsymbol \Omega_{\Tsj} \mathbf w \hspace*{-0.04cm}  +  \hspace*{-0.04cm}  \sum_{l=1}^L \hspace*{-0.04cm}  c_l\, \mathrm{card}(w_l) \leq \hat P.
\end{array}
\]
Compared to the information constrained problem (\ref{eq: inf_con_prob}), problem (\ref{eq: ene_con_prob}) is more involved  due to the nonconvex objective function and the cardinality function in the inequality constraint. Even if we replace the cardinality function with its $\ell_1$ norm relaxation, the resulting $\ell_1$ optimization problem is still difficult,
since the feasibility of the relaxed constraint does not guarantee the feasibility of the original problem (\ref{eq: ene_con_prob}).

However, if the collaboration topology is {given}, the collaboration cost $\sum_{l=1}^L c_l \, \mathrm{card}(w_l)$ is a constant and the constraint in (\ref{eq: ene_con_prob}) becomes a homogeneous quadratic constraint (i.e., no linear term with respect to $\mathbf w$ is involved). 
In this case, problem (\ref{eq: ene_con_prob}) can be solved by \cite[Theorem\,1]{karvar13}.

%On the other hand, setting $\mathbf w = c \hat{\mathbf w}$ for some fixed vector $\hat{\mathbf w}$, it can be shown that the objective and constraint functions in (\ref{eq: ene_con_prob}) are strictly increasing functions of $c$ when $c > 1$, and strictly decreasing functions of $c$ when $c < 1$.
%{\color{blue}Then, from Prop.\,\ref{prop: converse}, we can establish
%the correspondence between (\ref{eq: inf_con_prob}) and (\ref{eq: ene_con_prob}).}
%we can conclude that (\ref{eq: inf_con_prob}) is a `converse formulation' of (\ref{eq: ene_con_prob}).
%{\color{blue}

%{\color{blue}
%In (\ref{eq: ene_con_prob}), we note that by setting $\mathbf w = c \hat{\mathbf w}$ for some fixed vector $\hat{\mathbf w}$, the objective and constraint functions are strictly increasing functions of $c$ when $c > 1$, and strictly decreasing functions of $c$ when $c < 1$. Then, 
%}
In Prop\,\ref{prop: converse}, we present the relationship between  the energy constrained problem (\ref{eq: ene_con_prob}) and the information constrained problem (\ref{eq: inf_con_prob}).
Motivated by this relationship, we then take advantage of the solution of (\ref{eq: inf_con_prob}) to obtain the collaboration topology for (\ref{eq: ene_con_prob}).  This idea will be elaborated on later.

\begin{myprop}
\label{prop: converse}
Consider the two problems (\ref{eq: inf_con_prob}) and (\ref{eq: ene_con_prob})
\begin{align}
\begin{array}{ll}
\displaystyle \minimize_{\mathbf w} &  P(\mathbf w)  \\
\st &  J(\mathbf w) \geq \check J
\end{array}
~\text{and}~
\begin{array}{ll}
\displaystyle \maximize_{\mathbf w} &  J(\mathbf w)  \\
\st &  P(\mathbf w) \leq \hat P,
\end{array}
\nonumber
\end{align}
where the optimal solutions are denoted by $\mathbf w_1$ and $\mathbf w_2$, respectively.
If $\check J = J(\mathbf w_2)$, then $\mathbf w_1 = \mathbf w_2$;
If $\hat P = P(\mathbf w_1)$, then $\mathbf w_2 = \mathbf w_1$.
%the two problems are converses of each other in the sense that 
%if $J = J_{\mathrm{opt}}(P)$, 
%the optimal solution of (\ref{eq: inf_con_prob}) is equal to the solution of (\ref{eq: ene_con_prob});
%if  $P = P_{\mathrm{opt}}(J)$, the optimal solution of (\ref{eq: ene_con_prob}) is equal to the solution of (\ref{eq: inf_con_prob}).
\end{myprop}
\textbf{Proof:}
%The proof is  straightforward by using the method of contradiction. 
%Details of the proof are reported in \cite{liuswafarvar14_TechRep} and omitted here for brevity.
See Appendix\,\ref{appendix: converse}. 
%Similar conclusions have been shown in \cite{jiacheswi13,karvar13}. Details of the proof are reported in [Arxiv] and omitted here for brevity.
\hfill $\blacksquare$
%}

Prop.\,\ref{prop: converse} implies that the solution of the  energy constrained problem (\ref{eq: ene_con_prob}) can be obtained by seeking the global minimizer of the information constrained problem (\ref{eq: inf_con_prob}), if  the information threshold in (\ref{eq: inf_con_prob}) is set by using the optimal value of (\ref{eq: ene_con_prob}). However, this methodology is intractable in practice since the optimal value of (\ref{eq: ene_con_prob}) is unknown in advance, and the globally optimal solution of problem (\ref{eq: inf_con_prob}) may not be found using reweighted $\ell_1$-based methods.

Instead of deriving the solution of  (\ref{eq: ene_con_prob}) from  (\ref{eq: inf_con_prob}), we can infer
the collaboration topology of the energy constrained problem (\ref{eq: ene_con_prob}) from the sparsity structure of the solution to the information constrained problem (\ref{eq: inf_con_prob}) using a bisection algorithm. 
%\textcolor{blue}{We remark that the
%objective function of (\ref{eq: ene_con_prob}) (in terms of Fisher information) is upper bounded
%by the value of Fisher information (denoted by $J_0$) that is achieved when the network is \textit{fully connected} and the energy budget $P$ goes to \textit{infinity}. Thus, the objective value of (\ref{eq: ene_con_prob})  belongs to 
%an interval $[0, J_0]$, where according to \cite[Theorem\,1]{karvar13},
%\begin{align}
%J_0 = \lambda_{\mathcal G, \mathrm{max}} ( { \boldsymbol \Omega}_{\JNsj},   {\boldsymbol \Omega}_{\JDsj} ).
%\label{eq: form_J0}
%\end{align} } 
According to Lemma\,\ref{lemma1} in Appendix\,\ref{appendix: theorem1}, 
the objective function of (\ref{eq: ene_con_prob}) (in terms of Fisher information) is bounded over an interval $[0, J_0)$. 
%And there is a one-to-one correspondence between the Fisher information and energy budget $\hat P$ in (\ref{eq: ene_con_prob}). 
%{\color{blue} And the Fisher information is a monotonic function of the energy budget.}
And there is a one-to-one correspondence between the value of Fisher information evaluated at the optimal solution of (\ref{eq: ene_con_prob}) and energy budget $\hat P$.
Therefore we perform a bisection algorithm on the interval, and then solve
the information constrained problem to obtain the resulting energy cost and collaboration topology. The procedure terminates if the resulting energy cost is  close to the energy budget $\hat P$.
We summarize the bisection algorithm in Algorithm\,2.

%Also, there exists a one-to-one correspondence between Fisher information and energy budget in (\ref{eq: ene_con_prob}). 
%Therefore, a bisection procedure can be performed on the interval $[0, J_0]$, and then we solve
%the information constrained problem to obtain the resulting energy cost and collaboration topology. The procedure terminates if the resulting energy cost is reasonably close to the energy budget $P$.
%We summarize the bisection algorithm in Algorithm\,4.
\begin{algorithm}
\caption{Bisection algorithm for seeking the optimal collaboration topology of (\ref{eq: ene_con_prob})}
\begin{algorithmic}[1]
\Require given $\epsilon_{\mathrm{bi}}>0$,  $\underline J = 0$ and $\overline J = J_0$.
\Repeat $~ \check J = \frac{\underline J+ \overline J}{2}$
\State for a given $\check J$, solve (\ref{eq: inf_con_prob}) using Algorithm\,1 to obtain 
\hspace*{0.16in} the collaboration topology (in terms of the sparsity 
\hspace*{0.16in} structure of $\mathbf w$) and 
 the resulting energy cost $P$.
%}
\If{$ P < \hat P$} $\underline J = \check J$ \Else $~\overline J = \check J$ \EndIf
\Until {$\overline J - \underline J < \epsilon_{\mathrm{bi}}$ or $| \hat P - P | < \epsilon_{\mathrm{bi}}$}
\end{algorithmic}
\end{algorithm}

We remark that the collaboration topology obtained in Step\,2 of Algorithm\,2 is not globally optimal. However, we have observed that  for the information constrained sensor collaboration (\ref{eq: inf_con_prob}),
%the Fisher information is a monotonic function of the energy budget;
the value of energy cost $P$ is monotonically related to the value of desired estimation distortion; see Fig.\,\ref{fig: Inf_prob_MComp} for an example. 
%As a result, a larger (or smaller) $\check J$ leads to a larger (or smaller) $P$ in Step\,2 of Algorithm\,2.
Therefore, the proposed bisection algorithm converges in practice and at most requires $\lceil \mathrm{log}_2(\overline J/\epsilon_{\mathrm{bi}})\rceil$ iterations. Once the bisection procedure terminates, we obtain a locally optimal collaboration topology for (\ref{eq: ene_con_prob}). Given this topology, 
the energy constrained problem (\ref{eq: ene_con_prob}) becomes a problem with a quadratic constraint and an objective that is a ratio of homogeneous quadratic functions, whose analytical solution is given by \cite[Theorem\,1]{karvar13}.
Through the aforementioned procedure, we obtain a locally optimal solution to problem (\ref{eq: ene_con_prob}).

%similarly defined by \eqref{eq: eig_P02}. We refer readers to
%\cite[Sec.\,III-B]{karvar13} for more details.

%%%%%%%%%%%%%%%%%%%%%Section VI: Sensor Selection%%%%%%%%%%%%%%%%%%%%%%%%%%%%%%%

\section{%Problem (\ref{eq: inf_sel_ori}): 
Joint Sensor Selection and Collaboration}
\label{sec: sel}
In this section, we study the problem of the joint design of optimal sensor selection and collaboration schemes, which we formulated in 
(\ref{eq: inf_sel_ori}).
Similar to solving the information constrained collaboration problem (\ref{eq: inf_con_prob}), we first relax the original problem to a nonconvex $\ell_1$  optimization problem. 
However, in contrast to Section\,\ref{sec: inf_coll}, we observe that ADMM fails to converge
%However, unlike in Section \ref{sec: inf_coll}, the application of ADMM directly was found to exhibit convergence issues 
(a possible reason is explored later). To circumvent this, we adopt an iterative method %(called \emph{linearization}) 
to solve the nonconvex $\ell_1$ optimization problem. 

Using the reweighted $\ell_1$ minimization method, we replace the cardinality function with the weighted $\ell_1$ norm, which yields the following $\ell_1$ optimization problem
at each reweighting iteration %Then, problem (\ref{eq: inf_sel_ori}) can be relaxed to
\begin{align}
\hspace*{-0.15in}
\begin{array}{ll}
\displaystyle \minimize_{\mathbf w} &   \displaystyle \mathbf w^T \boldsymbol \Omega_{\Tsj} \mathbf w + 
\| \tilde{\boldsymbol \Omega}_{\Csj}  \mathbf w\|_1 + %\gamma 
\sum_{n=1}^N \tilde d_n \| \mathbf w_{\Gsj_n} \|_2\\ 
\st &  \mathbf w^T (\check J \boldsymbol \Omega_{\JDsj} - \boldsymbol \Omega_{\JNsj})\mathbf w + \check J \xi^2 \leq 0,
\end{array}
\label{eq: inf_sel_ori_l1}
\end{align}
where $\tilde{\boldsymbol \Omega}_{\Csj} \Def \diag(\tau_1^t c_1 , \tau_2^t c_2 , \ldots,  \tau_L^tc_L )$,  
$ \tilde d_n \Def \delta_n^{t} d_n$,
$\tau_l^t  $ and $\delta_n^{t}$ are the positive weights with respect to 
$w_l$ and $\mathbf w_{\Gsj_n}$ at the reweighting iteration $t$, respectively.
Let the solution of (\ref{eq: inf_sel_ori_l1}) be $\mathbf w^t$, then the weights 
$\tau_l^{t+1}  $ and $\delta_n^{t+1}$ for the next reweighting iteration are updated as
\[
\tau_l^{t+1} = \frac{1}{ | w_l^{t}| + \varepsilon},~~\delta_n^{t+1} = \frac{1}{ \| \mathbf w_{\Gsj_n}^{t}\|_2 + \varepsilon},
\]
where we recall that $\varepsilon$ is a small positive number that insures a nonzero denominator.
%In (\ref{eq: inf_sel_ori_l1}), the $\ell_1$ norms $\{ |w_l| \}_{l=1,2,\ldots,L}$ and $\ell_2$ norms $\{ \| \mathbf w_{\Gsj_n} \|_2 \}_{n=1,2,\ldots,N}$  determine the sparsity of $\mathbf w$ at the single- and group-coefficient levels, respectively. 

\subsection{Convex restriction}
\label{sec: sel_restrict}
Problem (\ref{eq: inf_sel_ori_l1}) is a nonconvex optimization problem. 
Similar to our approach in Section\,\ref{sec: admm} to find solutions of \eqref{eq: inf_con_prob_l1},
one can use ADMM to 
split (\ref{eq: inf_sel_ori_l1}) into a nonconvex QP1QC ($\mathbf w$-minimization step) and an unconstrained optimization problem with an objective function composed of $\ell_1$ and $\ell_2$ norms ($\mathbf v$-minimization step), where the latter can be solved analytically. However, our numerical examples show that the resulting ADMM algorithm fails to converge. 
Note that in the $\mathbf w$-minimization step, the sensor selection cost is excluded and 
each sensor collaborates with itself at no cost. 
Therefore, to achieve an information threshold, there exist scenarios in which 
the collaboration matrix yields nonzero 
diagonal entries. This implies that the $\mathbf w$-minimization step does not produce group-sparse solutions (i.e., row-wise sparse collaboration matrices).
%These facts lead to nonzero 
%diagonal entries of the obtained collaboration matrix, which implies that
%the $\mathbf w$-minimization step does not produce group-sparse solutions (i.e., rowwise sparse collaboration matrices).
%Since the row-sparsity of $\mathbf W$ is precisely the group-sparsity of $\mathbf w$, the $\mathbf w$-minimization step does not produce group-sparse solutions. 
However, the $\mathbf v$-minimization step always leads to group-sparse solutions. The mismatched sparsity structures of solutions in the subproblems of ADMM cause
the issue of nonconvergence, which we circumvent by using a linearization method to convexify the optimization problem.
%This is possibly because the $\mathbf w$-minimization step does not produce group-sparse solutions, but the incorporation of $\ell_2$ norm in  the $\mathbf v$-minimization step always leads to the group-sparse solutions. 
%To cope with the issue of nonconvergence, 
%we use a linearization method to convexify the optimization problem (\ref{eq: inf_sel_ori_l1}). 

A linearization method is introduced in \cite{aspboy03report}
for solving the nonconvex quadratically constrained quadratic program (QCQP) by 
\textit{linearizing}  the nonconvex parts of quadratic constraints, thus rendering a convex QCQP.
In (\ref{eq: inf_sel_ori_l1}), the nonconvex constraint is given by
\begin{align}
\mathbf w^T \check J \boldsymbol \Omega_{\JDsj} \mathbf w + \check J \xi^2 \leq  \mathbf w^T \boldsymbol \Omega_{\JNsj} \mathbf w,
\label{eq: non_ineq}
\end{align}
where $\boldsymbol  \Omega_{\JDsj} $ and $\boldsymbol \Omega_{\JNsj}$ are positive semidefinite; see (\ref{eq: O_P})-(\ref{eq: O_JD}). 
We linearize the right hand side of (\ref{eq: non_ineq}) around a feasible point $\boldsymbol \beta$%$\bar{\mathbf w}$
%\begin{align}
%\mathbf w^T J \boldsymbol \Omega_{\JDsj} \mathbf w + J \xi^2 \leq \bar{\mathbf w}^T \boldsymbol \Omega_{\JNsj} \bar{\mathbf w} + 2 \bar{\mathbf w}^T \boldsymbol \Omega_{\JNsj} (\mathbf w- \bar{\mathbf w}).
%\label{eq: linear_cons}
%\end{align}
\begin{align}
\mathbf w^T \check J \boldsymbol \Omega_{\JDsj} \mathbf w + \check J \xi^2 \leq {\boldsymbol \beta}^T \boldsymbol \Omega_{\JNsj} \boldsymbol \beta + 2 {\boldsymbol \beta}^T \boldsymbol \Omega_{\JNsj} (\mathbf w- \boldsymbol \beta).
\label{eq: linear_cons}
\end{align}
Note that the right hand side of (\ref{eq: linear_cons}) is an affine \textit{lower bound} on the convex function $\mathbf w^T \boldsymbol {\Omega}_{\JNsj} \mathbf w$. 
This implies that the set of $\mathbf w$ that satisfy (\ref{eq: linear_cons}) is a strict subset of the set of $\mathbf w$ that satisfy (\ref{eq: non_ineq}).

%This implies that the feasible set (\ref{eq: linear_cons}) can be regarded as a convex subset of the original constraint set (\ref{eq: non_ineq}). 

By replacing (\ref{eq: non_ineq}) with (\ref{eq: linear_cons}), we obtain a `restricted' convex version of problem (\ref{eq: inf_sel_ori_l1})
\begin{align}
%\hspace*{-0.3in}
\begin{array}{ll}
\displaystyle \minimize_{\mathbf w} &  \hspace*{-0.05in} \displaystyle   \varphi (\mathbf w) \Def \mathbf w^T \boldsymbol \Omega_{\Tsj} \mathbf w +  \hspace*{-0.03in}
\| \tilde {\boldsymbol \Omega}_{\Csj} \mathbf w\|_1 + \hspace*{-0.03in} %\gamma 
\sum_{n=1}^N \tilde d_n \| \mathbf w_{\Gsj_n} \|_2\\ 
\st &  \hspace*{-0.05in} \mathbf w^T   \tilde{\boldsymbol \Omega}_{\JDsj} \mathbf w \hspace*{-0.02in} - 2   \tilde{\boldsymbol \beta}^T 
\mathbf w  \hspace*{-0.02in} + \tilde \gamma  \leq \hspace*{-0.02in} 0,
\end{array}
%\hspace*{-0.3in}
\label{eq: inf_sel_ori_l1_convex}
\end{align}
where $ \tilde{\boldsymbol \Omega}_{\JDsj} \Def \check J \boldsymbol \Omega_{\JDsj} $, $ \tilde{\boldsymbol \beta}\Def \boldsymbol \Omega_{\JNsj} \boldsymbol \beta $, and $ \tilde \gamma \Def  {\boldsymbol \beta}^T \boldsymbol \Omega_{\JNsj} \boldsymbol \beta + \check J \xi^2 $. 
Different from \eqref{eq: non_ineq}, the inequality constraint in \eqref{eq: inf_sel_ori_l1_convex} no longer represents the information inequality but becomes a convex quadratic constraint. Since 
problem (\ref{eq: inf_sel_ori_l1_convex}) is convex, the convergence of ADMM %to solve 
is now guaranteed \cite{boyparchupeleck11}. And the optimal value of (\ref{eq: inf_sel_ori_l1_convex}) yields an upper bound on that of \eqref{eq: inf_sel_ori_l1}.

We summarize the linearization method in Algorithm\,3. In the following subsection, we
will elaborate on the implementation of ADMM in Step\,3 of Algorithm\,3. We remark that 
the convergence of Algorithm\,3 is guaranteed \cite{aspboy03report}, 
since Algorithm\,3 starts from a feasible point $\mathbf w^0$ that satisfies \eqref{eq: non_ineq} and at each iteration, we solve a linearized convex problem  with a smaller feasible set which contains the linearization point (i.e., the solution at the previous iteration).
In other words, for a given linearization point $\boldsymbol \beta  = \mathbf w^{s-1}$, we always obtain a new  feasible point $\mathbf w^{s}$ with a lower or equal objective value at each iteration.
%the feasible set at the new iteration contains the solution at the previous iteration, and consequently, for a given linearizing point $\boldsymbol \beta  = \mathbf w^{s-1}$, we always obtain a new  feasible point $\mathbf w^{s}$ with a lower objective value at each iteration.
Finally, we note that the iterative algorithm can be initialized at $\tilde{\mathbf w}$, which is given by (\ref{eq: eig_P01}).

\begin{algorithm}
\caption{Linearization method on solving (\ref{eq: inf_sel_ori_l1})}
\begin{algorithmic}[1]
\Require given $\epsilon_{\mathrm{li}} > 0$ and $\mathbf w^0  = \tilde{\mathbf w} $.
\For{iteration $s=1,2,\ldots$}
\State set $\boldsymbol \beta = \mathbf w^{s-1} $.
\State solve (\ref{eq: inf_sel_ori_l1_convex})  for the  solution $\mathbf w^s$ by using Algorithm\,4.
%\State Update the linearizing point $\bar{\mathbf w} = \mathbf w^k$.
\State \textbf{until} $\| \varphi (\mathbf w^{s}) - \varphi(\mathbf w^{s-1}) \| < \epsilon_{\mathrm{li}}$.
\EndFor
\end{algorithmic}
\end{algorithm}

\subsection{Solution via ADMM}
%We follow the same procedure of implementing ADMM as in Sec.\,\ref{sec: admm}. 
%%But compared to Sec.\,\ref{sec: admm}, here ADMM yields different subproblems.
%
%First, we reformulate problem (\ref{eq: inf_sel_ori_l1_convex}) as follows
%\begin{align}
%\hspace*{-0.1in}
%\begin{array}{ll}
%\displaystyle \minimize_{\mathbf w, \mathbf v} &   \hspace*{-0.05in} \displaystyle \mathbf w^T \boldsymbol \Omega_{\Tsj} \mathbf w \hspace*{-0.03in} +  \hspace*{-0.03in} \mathcal I^{\prime}(\mathbf w) \hspace*{-0.03in} + \hspace*{-0.03in}
%\| \tilde {\boldsymbol \Omega}_{\Csj} \mathbf v\|_1 \hspace*{-0.03in} + \hspace*{-0.03in}
%\sum_{n=1}^N \tilde d_n \| \mathbf v_{\Gsj_n} \|_2\\ 
%\st &   \hspace*{-0.05in} \mathbf w = \mathbf v,
%\end{array}
%\hspace*{-0.1in} \label{eq: inf_sel_ori_l1_convex_ad}
%\end{align}
%where $\mathcal I^{\prime} (\mathbf w)$ is an indicator function 
%\[
%\mathcal I^{\prime} (\mathbf w) \Def \left \{
%\begin{array}{ll}
%0 &  \mathbf w^T  \tilde{\boldsymbol \Omega}_{\JDsj} \mathbf w \hspace*{-0.02in} - \hspace*{-0.02in}  2  \tilde{\boldsymbol \beta}^T 
%\mathbf w  \hspace*{-0.02in} + \tilde \gamma \leq \hspace*{-0.02in} 0 \\
%\infty & \text{otherwise}.
%\end{array}
% \right. 
%% \label{eq: indicator_sel}
%\]

Similar to \eqref{eq: inf_con_prob_wl1_ad} in Sec.\,\ref{sec: admm}, 
we introduce the auxiliary variable $\mathbf v$ to replace $\mathbf w$ in the $\ell_1$ and $\ell_2$ norms in \eqref{eq: inf_sel_ori_l1_convex} while adding the constraint $\mathbf w = \mathbf v$, and split problem \eqref{eq: inf_sel_ori_l1_convex} into 
a sequence of subproblems as in (\ref{eq: w_step})-(\ref{eq: dual_step}). However, compared to Sec.\,\ref{sec: admm}, the current ADMM algorithm yields different subproblems due to the presence of the $\ell_2$ norm in the objective function and the convexification in the  constraint.

\subsubsection{$\mathbf w$-minimization step}

According to (\ref{eq: w_step}), the $\mathbf w$-minimization step is given by
\begin{align}
\hspace*{-0.1in}
\begin{array}{ll}
\displaystyle \minimize_{\mathbf w} &   \mathbf w^T \tilde{\boldsymbol \Omega}_{\Tsj}  \mathbf w - \rho \mathbf a^T \mathbf w \\
%\hspace*{-0.03in} \frac{\rho}{2}\| \mathbf w - \mathbf a\|_2^2 \\
\st &   \mathbf w^T \tilde{\boldsymbol \Omega}_{\JDsj} \mathbf w \hspace*{-0.02in} - \hspace*{-0.02in}  2  \tilde{\boldsymbol \beta}^T 
\mathbf w  \hspace*{-0.02in} + \tilde \gamma \leq \hspace*{-0.03in} 0,
\end{array}
\hspace*{-0.1in} \label{eq: phi_w_sel}
\end{align}
where $\tilde{\boldsymbol \Omega}_{\Tsj} = \boldsymbol \Omega_{\Tsj} + \frac{\rho}{2} \mathbf I$, $\mathbf a = \mathbf v^{k} - 1/\rho \boldsymbol \chi^{k}$, $\mathbf v^k$ and $\boldsymbol \chi^k$ denote the value of $\mathbf v$ and $\boldsymbol \chi$ at the $k$th iteration of ADMM, and $\boldsymbol \chi$ is the dual variable. Note that different from problem (\ref{eq: phi_w}), problem (\ref{eq: phi_w_sel}) is a convex QCQP, which can be efficiently solved.
%{\color{blue}the inequality constraint in \eqref{eq: phi_w_sel} no longer represents the information inequality, however, problem (\ref{eq: phi_w_sel}) is a convex QCQP which can be efficiently solved.}
 %by the current optimization solvers, such as CVX \cite{cvx}.

%\begin{remark}
%\label{rk: QCQP_efficient}
To solve the convex QCQP (\ref{eq: phi_w_sel}), the complexity of using interior-point method in standard solvers is roughly $O(L^{3.5})$ \cite{nem12}. 
To reduce the computational complexity, we can derive the KKT-based solution of  (\ref{eq: phi_w_sel}). Since problem (\ref{eq: phi_w_sel}) is convex, KKT conditions are both necessary and sufficient for optimality. Similar to Prop.\,\ref{prop: sol_phi_w}, we can apply the eigenvalue decomposition technique to simplify (\ref{eq: phi_w_sel}) and find its solution.
This is summarized in Prop.\,\ref{prop: sol_phi_u_sel}.

\begin{myprop}
\label{prop: sol_phi_u_sel}
The optimal solution of  problem (\ref{eq: phi_w_sel}) is given by 
\[
\mathbf w^{k+1} = {\tilde{\boldsymbol \Omega}_{\Tsj}}^{-\frac{1}{2}} {\mathbf U} {\mathbf u} , 
\]
where
${\mathbf U}$ is given by the following  eigenvalue decomposition
\[
\frac{1}{\tilde \gamma} {\tilde{\boldsymbol \Omega}_{\Tsj}}^{-\frac{1}{2}} \tilde{\boldsymbol \Omega}_{\JDsj} \tilde{\boldsymbol \Omega}_{\Tsj}^{-\frac{1}{2}}  = {\mathbf U} {\boldsymbol \Lambda } {\mathbf U}^T,
\] and
${\mathbf u}$ is given by
\begin{align}
\left \{
\begin{array}{ll}
 {\mathbf u} = - { \mathbf g} & \text{if $ { \mathbf g}^T {\boldsymbol \Lambda } { \mathbf g} + 2 { \mathbf g}^T {\mathbf e} +1 \leq 0 $} \\
 {\mathbf u} = - (\mathbf I +  \mu_1 {\boldsymbol \Lambda })^{-1} ( {\mathbf g} +  \mu_1 {\mathbf e}) &  
\text{otherwise}.
\end{array}
\right.
\label{eq: u_sel_sol_KKT}
\end{align}
In (\ref{eq: u_sel_sol_KKT}),  $ {\mathbf g} = - \rho {\mathbf U}^T \tilde{\boldsymbol \Omega}_{\Tsj}^{-\frac{1}{2}} \mathbf a / 2$,
${\mathbf e} = -  {\mathbf U}^T \tilde{\boldsymbol \Omega}_{\Tsj}^{-\frac{1}{2}} \tilde {\boldsymbol \beta} / {\tilde \gamma}  $, 
$ \mu_1 $ is \textit{the} positive root of the equation in $\mu$
%\[ \displaystyle
%\sum_{l=1}^L \left( \frac{ \lambda_l( \mu  e_l +  g_l)^2}{( \mu  \lambda_l + 1 )^2} - \frac{2  e_l (\mu  e_l +  g_l)}{ \mu  \lambda_l + 1} \right) + 1 = 0,
%\]
\[
\sum_{l=1}^L \frac{(\lambda_l g_l -e_l)^2}{\lambda_l(1+\mu \lambda_l)^2} - \sum_{l=1}^L\frac{e_l^2}{\lambda_l} +1 = 0,
\]
$e_l$ and $g_l$ are the $l$th elements of $\mathbf e$ and $\mathbf g$, respectively, and $\lambda_l$ is the $l$th diagonal entry of $\boldsymbol \Lambda $. 
\end{myprop}
\textbf{Proof:} See Appendix\,\ref{appendix: sol_phi_w}, in which letting
$\mathbf A_0 = \tilde{\boldsymbol \Omega}_{\Tsj}$,$\mathbf b_0 = -\frac{\rho}{2}\mathbf a$, $\mathbf A_1 = \tilde{\boldsymbol \Omega}_{\JDsj}$, $\mathbf b_1 = - \tilde{\boldsymbol \beta}$ and $r_1 = \tilde{\gamma}$, we obtain the result in Prop.\,\ref{prop: sol_phi_u_sel}.
 \hfill $\blacksquare$

We remark that 
%the solution of problem (\ref{eq: phi_w_sel}) given by Prop.\,\ref{prop: sol_phi_u_sel} yields a  similar form as that  of problem \eqref{eq: phi_w} in Prop.\,\ref{prop: sol_phi_w}. 
Prop.\,\ref{prop: sol_phi_u_sel} is similar to Prop.\,\ref{prop: sol_phi_w} except for
the presence of $\mathbf e$ and positive eigenvalues $\{ \lambda_l \}_{l=1,2,\ldots,L}$.
%However, a vector $\mathbf e$ is involved in Prop.\,\ref{prop: sol_phi_u_sel}. This is different from Prop.\,\ref{prop: sol_phi_w}, where $\mathbf e$ is assumed to be a zero vector. 
These difference are caused by the convex restriction of \eqref{eq: inf_sel_ori_l1}. 
%Indeed, if $\tilde{\boldsymbol \beta} = 0$ then expression (\ref{eq: u_sel_sol_KKT}) reduces to (\ref{eq: u_phi_col}).

\subsubsection{$\mathbf v$-minimization step}
According to (\ref{eq: v_step}), the $\mathbf v$-minimization step is given by
\begin{align}
\hspace*{-0.15in}
\begin{array}{ll}
\displaystyle \minimize_{\mathbf v} &  \displaystyle \| \tilde{\boldsymbol \Omega}_{\Csj} \mathbf v\|_1 + %\gamma  
\sum_{n=1}^N \tilde d_n\| \mathbf v_{\Gsj_n} \|_2 + \frac{\rho}{2}\|\mathbf v - \mathbf b \|_2^2,
\end{array}
\hspace*{-0.15in} \label{eq: phi_v_sel}
\end{align}
where  $\mathbf b = \mathbf w^{k+1}+1/\rho \boldsymbol \chi^{k}$.

We recall that $\tilde {\boldsymbol \Omega}_{\Csj}$ defined in (\ref{eq: inf_sel_ori_l1}) is a diagonal matrix. 
Let the vector $\mathbf f$ be composed of the diagonal entries of $\tilde {\boldsymbol \Omega}_{\Csj}$. We then
define a sequence of diagonal matrices $\mathbf F_n:=\diag(\mathbf f_{\Gsj_n})$ for $n = 1,2,\ldots, N$, where $\mathbf f_{\Gsj_n}$ is a vector composed of those entries of $\mathbf f$ whose indices belong to the set $ \mathrm{G}_n $.
Since the index sets $\{ \mathrm{G}_n \}_{j=1,2,\ldots, N}$ are disjoint,  problem (\ref{eq: phi_v_sel}) can be decomposed into a sequence of subproblems for $n=1,2,\ldots,N$,
\begin{align}
\hspace*{-0.1in} 
\begin{array}{ll}
\displaystyle \minimize_{\mathbf v_{\Gsj_n}} &  \hspace*{-0.03in}  \displaystyle \| \mathbf F_n \mathbf v_{\Gsj_n}\|_1 \hspace*{-0.02in}  + \hspace*{-0.02in}  %\gamma  
\tilde d_n \| \mathbf v_{\Gsj_n} \|_2 \hspace*{-0.02in}  + \hspace*{-0.02in}  \frac{\rho}{2}\|\mathbf v_{\Gsj_n} \hspace*{-0.03in} - \mathbf b_{\Gsj_n} \|_2^2.
\end{array}
\label{eq: phi_v_sel_sub}
\end{align}
Problem \eqref{eq: phi_v_sel_sub} can be solved analytically via the following proposition.
\begin{myprop}
\label{prop: v_sel}
The minimizer of (\ref{eq: phi_v_sel_sub}) is given by
\begin{align}
\mathbf v_{\Gsj_n} =  \left \{ 
\begin{array}{cl}
(1 - \frac{ %\gamma  
\tilde d_n}{\rho \| \boldsymbol \nu \|_2}) \boldsymbol \nu & \quad \| \boldsymbol \nu \|_2 \geq \frac{%\gamma  
\tilde d_n}{\rho} \vspace*{0.05in}\\
0 & \quad \| \boldsymbol \nu \|_2 < \frac{%\gamma  
\tilde d_n}{\rho},
\end{array}
\right.
\end{align}
where
$
\boldsymbol \nu = \mathrm{sgn}(\mathbf b_{\Gsj_n}) \odot \mathrm{max}(|\mathbf b_{\Gsj_n}| - \frac{1}{\rho} \mathbf f_{\Gsj_n}, 0),
%\label{eq: u}
$
the operator $\mathrm{sgn}(\cdot)$ is defined in a componentwise fashion as
\[
\mathrm{sgn}(x) = \left \{ 
\begin{array}{l l}
     1 &   x > 0 \\
    0 &  x = 0 \\
     -1  &  x < 0,
  \end{array} 
\right.
\]
$\odot$ denotes the point-wise product,
and the operator $\mathrm{max}(\mathbf x,\mathbf y)$ returns a vector whose entries are the pointwise maximum of the entries of $\mathbf x$ and $\mathbf y$.
\end{myprop}
\textbf{Proof:}  
The main idea of the proof,  motivated by \cite[Theorem\,1]{yualiuye13}, is to study the subgradient of a nonsmooth objective function. However, for problem  \eqref{eq: phi_v_sel_sub}, we require to exploit  its particular features:
weighted $\ell_1$ norm $ \| \tilde{\boldsymbol \Omega}_{\Csj} \mathbf v\|_1 $, and disjoint sub-vectors $\{ \mathbf v_{\Gsj_n} \}_{n=1,2,\ldots,N}$. 
See Appendix\,\ref{appendix: v_sel} for the complete proof. 
%Details of the proof are reported in \cite{liuswafarvar14_TechRep}.
\hfill $\blacksquare$

%We remark that the objective function in (\ref{eq: phi_v_sel_sub}) is strongly convex, and thus, the minimizer of (\ref{eq: phi_v_sel_sub}) is unique \cite{boyvan04}. 

%\begin{remark}
%A relevant problem to (\ref{eq: phi_v_sel_sub}) is called `overlapping group Lasso' in \cite{yualiuye13}, where the `overlapping' refers to the scenarios in which 
%the index sets $\{ \mathrm{G}_n \}_{n=1,2,\ldots, N}$ may overlap.
%It was shown in \cite{yualiuye13} that the problem of overlapping group Lasso cannot be solved analytically. However, in problem (\ref{eq: phi_v_sel_sub}),  the index sets $\{ \mathrm{G}_n \}_{n=1,2,\ldots, N}$ disjoint. %This property  makes the analytical solution of problem (\ref{eq: phi_v_sel_sub}) tractable. 
%Further different from  \cite{yualiuye13}, a linear composite $\ell_1$ norm, $\| \mathbf F_n \mathbf v_{\Gsj_n}\|_1$,
%is involved in (\ref{eq: phi_v_sel_sub}). It was shown in \cite{parboy13} that a nonsmooth function with a linear composite argument in an optimization problem is much more difficult to handle even though this nonsmooth function itself is easy to evaluate. However, the linear mapping matrix  $\mathbf F_n$ considered here is a diagonal matrix.
%The facts of disjoint groups and diagonal linear mapping matrix make
%the analytical solution of (\ref{eq: phi_v_sel_sub}) tractable.
%\end{remark}

In Algorithm\,4, we present our proposed ADMM algorithm for solving
(\ref{eq: inf_sel_ori_l1_convex}).
 
\begin{algorithm}
\caption{Solving problem (\ref{eq: inf_sel_ori_l1_convex}) via ADMM}
\begin{algorithmic}[1]
\Require given $\rho$, $\epsilon_{\mathrm{ad}}$, $\boldsymbol \chi^0 = \mathbf 0$ and $\mathbf w^0 = \mathbf v^0 = \tilde {\mathbf w} $.
\For{$k=0,1,\ldots$}
\State obtain $\mathbf w^{k+1}$ from a standard QCQP solver or Prop.\,\ref{prop: sol_phi_u_sel}.
\State obtain $\mathbf v^{k+1} \hspace*{-0.03in}= \hspace*{-0.03in} [ (\mathbf v_{\Gsj_1}^{k+1})^T,  \hspace*{-0.03in} \ldots, \hspace*{-0.03in} (\mathbf v_{\Gsj_N}^{k+1})^T]^T$ from Prop.\,\ref{prop: v_sel}.
\State update dual variable $\boldsymbol \chi^{k+1} = \boldsymbol \chi^k + \rho (\mathbf w^{k+1} - \mathbf v^{k+1}).$
\State \textbf{until} $
\|\mathbf w^{k+1}-\mathbf v^{k+1}\|_2 \leq \epsilon_{\mathrm{ad}},
~
\|\mathbf v^{k+1} - \mathbf v^{k}\|_2 \leq \epsilon_{\mathrm{ad}}
$.
\EndFor
\end{algorithmic}
\end{algorithm}

%\subsection{Complexity analysis}
%For solving the \textit{original} problem (\ref{eq: inf_sel_ori}), 
%we first replace the cardinality function with weighted $\ell_1$ norm, which yields the nonconvex problem (\ref{eq: inf_sel_ori_l1}). Then, we use the method of linearization to convexify (\ref{eq: inf_sel_ori_l1}). This leads to a convex problem (\ref{eq: inf_sel_ori_l1_convex}), which is solved by ADMM. 
%During the aforementioned procedure, the iterative reweighted $\ell_1$ 
%method is used as the outer loop, the method of linearization forms the middle loop, and  
%the ADMM algorithm constitutes the inner loop. 
%Compared to the complexity of solving the information constrained problem \eqref{eq: inf_con_prob}, 
%an extra computation cost is taken due to the presence of middle loop. However, the method of linearization formed as the middle loop only requires a few iterations for convergence. In our experiments, the number of iterations is less than $10$.

To summarize, for solving the original problem (\ref{eq: inf_sel_ori}) we first
replace the cardinality function with the weighted $\ell_1$ norm, which
yields the nonconvex problem (\ref{eq: inf_sel_ori_l1}). We then use the linearization
method to convexify (\ref{eq: inf_sel_ori_l1}). The resulting convex problem
(\ref{eq: inf_sel_ori_l1_convex}) is  solved by ADMM as outlined in Algorithm\,4.
 
%\begin{algorithm}
%\caption{Algorithm for the joint design of sensor selection and collaboration schemes}
%\begin{algorithmic}[1]
%\Require Given $\kappa > 0$, $\epsilon_{\mathrm{rw}} > 0$ and $\epsilon_{\mathrm{li}} > 0$. Set $\tau_l^0 = 1$  and $\delta_j^0 = 1$ for $l = 1, \ldots, L$ and $j=1,2,\ldots,N$. Let $\bar{\mathbf w} = \mathbf w_{\mathrm{full},J}^* $, where $\mathbf w_{\mathrm{full},J}^* $ is given in (\ref{eq: w_fJ}). 
%\For {$t = 0 , 1, \ldots $}
%\For {$i = 0, 1, \ldots $}
%\State Given $\tau_l^t $, $\delta_j^t $ and $\bar{\mathbf w}$, apply Algorithm\,7 to solve problem (\ref{eq: inf_sel_ori_l1_convex}) for the optimal solution $\mathbf w^{(i,t)}$.
%\State $\bar{\mathbf w} =\mathbf w^{(i,t)} $.
%\State \textbf{until} $\| \phi (\mathbf w^{(i+1,t)}) - \phi(\mathbf w^{(i,t)}) \| < \epsilon_{\mathrm{li}}$, where
%$\phi (\mathbf w)$ denotes the objective function of (\ref{eq: inf_sel_ori_l1_convex}).
%\EndFor
%\State Update the weights $\tau_l^{t+1} = \frac{1}{ |w_l^{(i,t)}| + \kappa}$ and $\delta_j^{t+1} = \frac{1}{ \| \mathbf w_{\Gsj_j}^{(i,t)}\|_2 + \kappa}$.
%\State if $\| \boldsymbol \tau^{t+1} - \boldsymbol \tau^{t}\|_2 < \epsilon_{\mathrm{rw}} $ and $\| \boldsymbol \delta^{t+1} - \boldsymbol \delta^{t}\|_2 < \epsilon_{\mathrm{rw}} $, \textbf{quit}.
%\EndFor
%\end{algorithmic}
%\end{algorithm}

%%%%%%%%%%%%%%%%%%%%%Section VII: Numerical Example%%%%%%%%%%%%%%%%%%%%%%%%%%%%%%%

\section{Numerical Results}
\label{sec: numerical}
%To demonstrate the efficacy of sparsity-aware sensor collaboration, 
In this section, we will illustrate the performance of our proposed sparsity-aware sensor collaboration methods through numerical examples. The estimation system considered here is shown in Fig.\ref{fig:col_sch}, where for simplicity, we assume that the channel gain and uncertainties are such that the network is homogeneous and equicorrelated. 
%and the system parameters are given by
%\cite[Example\,3]{karvar13},
As in \cite[Example\,3]{karvar13}, we denote the expected observation and channel gains by $h_0$ and $g_0$, the observation and channel gain uncertainties by $\alpha_h$ and $\alpha_g$, 
the measurement noise variance and correlation by $\zeta^2 $ and $\rho_{\mathrm{corr}}$, 
and thereby assume
\begin{align}
\left \{ 
\begin{array}{l}
 \mathbf h = h_0 \sqrt{\alpha_h} \mathbf 1,~ \boldsymbol \Sigma_h = h_0^2 (1-\alpha_h) \mathbf I,  \\
\boldsymbol \Sigma_{\epsilon} = \zeta^2 [(1-\rho_{\mathrm{corr}})\mathbf I + \rho_{\mathrm{corr}} \mathbf 1 \mathbf 1^T], \\
\mathbf g = g_0 \sqrt{\alpha_g} \mathbf 1,~ \boldsymbol \Sigma_g = g_0^2 (1-\alpha_g) \mathbf I.
\end{array}
\right.
\label{eq: sys_para}
\end{align}
Note that channel gains can also be calculated based on path loss models \cite{gold_bk} but we chose the homogeneous model for the sake of simplicity.
%In \eqref{eq: sys_para}, 
%unless specified otherwise, we shall assume that $h_0 = g_0  = 1$, $\alpha_h  = \alpha_g = 0.7$, $\rho_{\mathrm{corr}}=0.5 $, and $\zeta^2 = 1$.
The collaboration cost matrix $\mathbf C$ is given by
\begin{align}
C_{mn} = \alpha_{\mathrm{c}} \| \mathbf s_m - \mathbf s_n \|_2
\label{eq: coll_cost}
\end{align}
for $m, n = 1,2,\ldots,N$, where  $\alpha_{\mathrm{c}}$ is a positive parameter and  $\mathbf s_n$ denotes the location of sensor $n$. 
The vector of sensor selection cost $\mathbf d $ is give by
\begin{align}
d_n = \alpha_{\mathrm{s}}  \| \mathbf s_n - \mathbf s_{\mathrm{fc}}\|_2
\label{eq: sel_cost}
\end{align}
for $n=1,2,\ldots,N$, where $\alpha_{\mathrm{s}}$ is a positive parameter and $\mathbf s_{\mathrm{fc}}$ denotes the location of the FC.

In our experiments, 
unless specified otherwise, we shall assume that $h_0 = g_0  = 1$, $\alpha_h  = \alpha_g = 0.7$, $\rho_{\mathrm{corr}}=0.5 $, $ \xi^2 = \zeta^2 = 1$, $\eta^2 = 0.1$, and $\alpha_{\mathrm{c}} = \alpha_{\mathrm{s}} = 0.01$. The FC and $N$ sensors are randomly deployed on a $10 \times 10$ grid, where the value of $N$ will be specified in different examples. 
While employing the proposed optimization methods, we select $\rho \geq 20$ in ADMM,
$\varepsilon = 10^{-3}$ in the reweighted $\ell_1$ method
 and $ \epsilon_{\mathrm{rw}} = \epsilon_{\mathrm{bi}} =  \epsilon_{\mathrm{li}} = \epsilon_{\mathrm{ad}} = 10^{-3}$ for the stopping tolerance. In our numerical examples, the reweighted $\ell_1$ method (Algorithm\,1), the bisection algorithm (Algorithm\,2) and the linearization method (Algorithm\,3) converge within $10$ iterations. For ADMM, the required number of iterations is less than $100$.

For a better depiction of estimation performance, since $ D_0 < D_{\mathbf w} \le \eta^2$ (see Lemma\,\ref{lemma1} in Appendix\,\ref{appendix: theorem1}), we display the \emph{normalized distortion} 
\begin{align}
D_{\mathrm{norm}}:=\frac{D(\mathbf w) -  D_0}{ \eta^2 -  D_0} \in (0,1], \label{eq: D_norm}
\end{align}
%{\color{blue} where $D_{\mathbf w}$
%defined in (\ref{eq: D_w}) is monotonically related to the value of Fisher information $J_{\mathbf w}$, $D_0$ is the minimum estimation distortion which corresponds to $J_0$ in \eqref{eq: form_J0}, and $\eta^2$ signifies the maximum distortion (i.e., $J_{\mathbf w} = 0$) which is determined by the prior information of $\theta$. }
where $D(\mathbf w)$ defined in (\ref{eq: D_w}) is monotonically related to the value
of Fisher information, and $D_0$ is the minimum estimation distortion given by Lemma\,\ref{lemma1}.
Further to characterize the number of established collaboration links, we define the percentage of collaboration links
\begin{align}
\mathrm{Per}_{\mathbf w} := \frac{\sum_{l=1}^L\mathrm{card}(w_l)- N }{L - N} \times 100 ~~ (\%),
\end{align}
where $L  = N^2$ is the dimension of $\mathbf w$, and  
$ \mathrm{Per}_{\mathbf w}$ belongs to  $[0, 100\%]$. When $ \mathrm{Per}_{\mathbf w} = 0$, the network operates in a distributed manner (i.e., only the diagonal entries of $\mathbf W$ are nonzero). When $ \mathrm{Per}_{\mathbf w} = 100\%$, the network is fully-connected (i.e., $\mathbf W$ has no zero entries).

\begin{figure}[htb]
\centering
\includegraphics[width=0.5\textwidth,height=!]{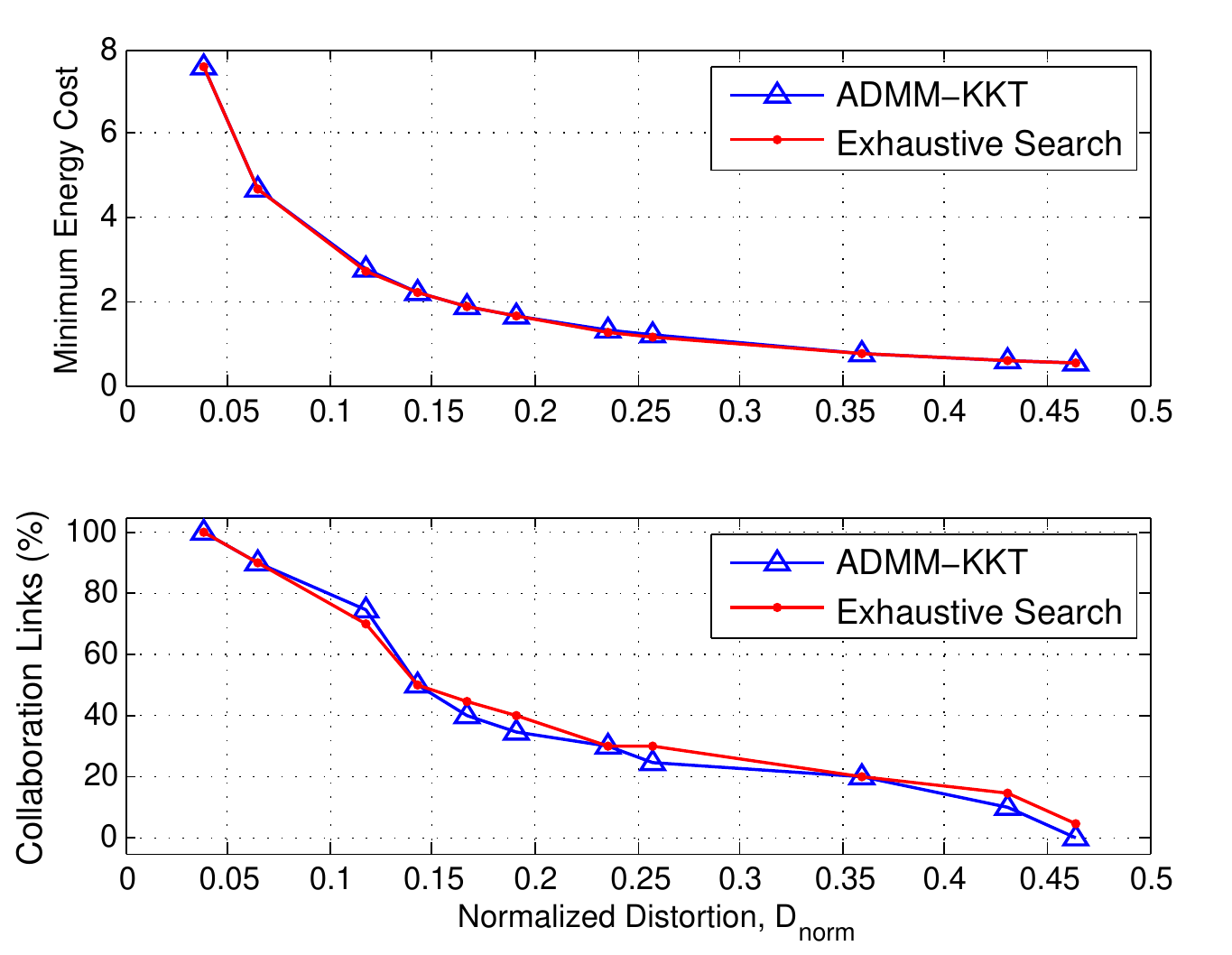} %0.46
\caption{\footnotesize{Performance evaluation for information constrained sensor collaboration.}}
\label{fig: Inf_prob_MComp}
\end{figure}

\begin{figure*}[htb]
\centerline{ \begin{tabular}{ccc}
\includegraphics[width=.32\textwidth,height=!]{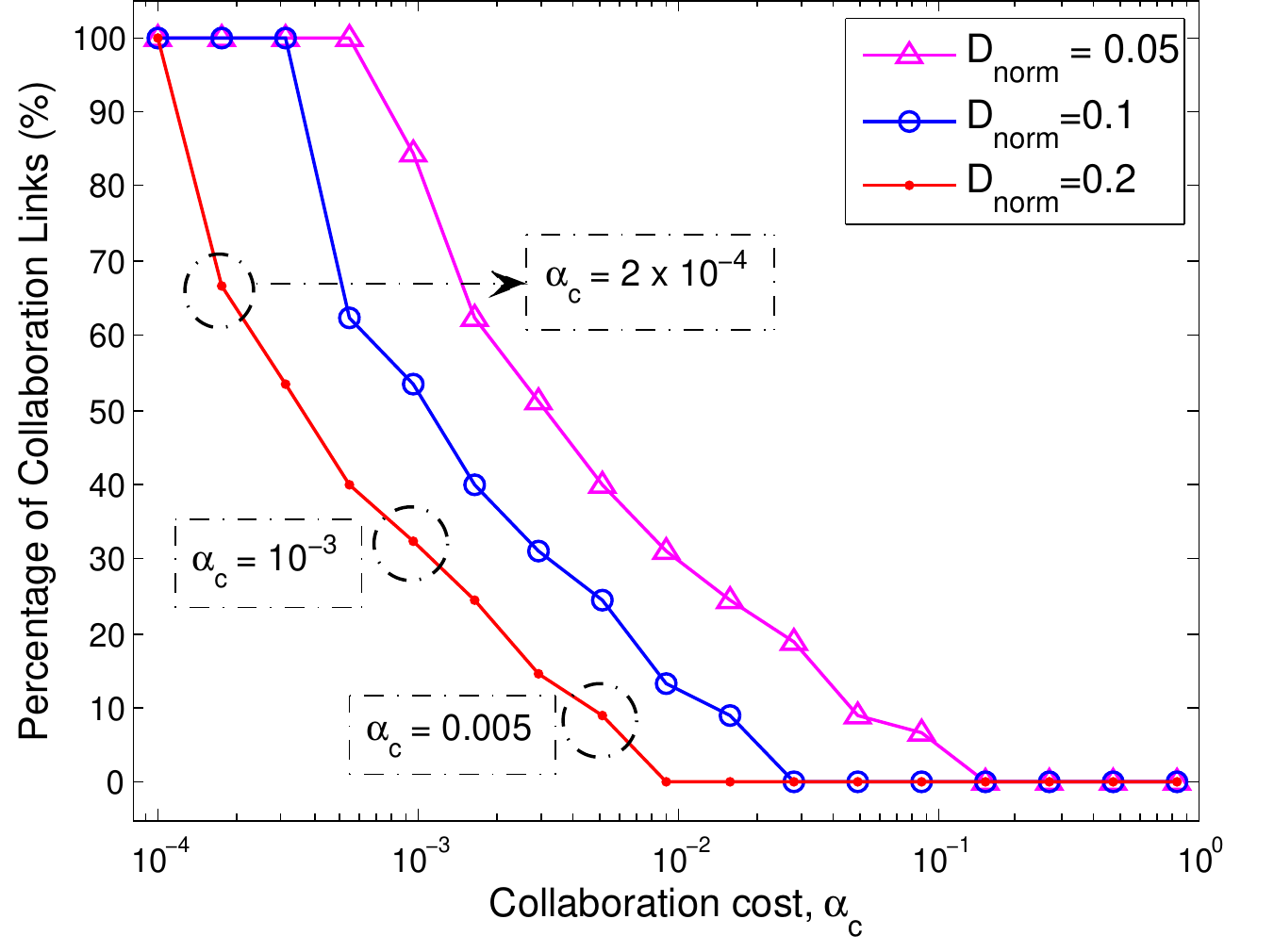}
& 
\includegraphics[width=.32\textwidth,height=!]{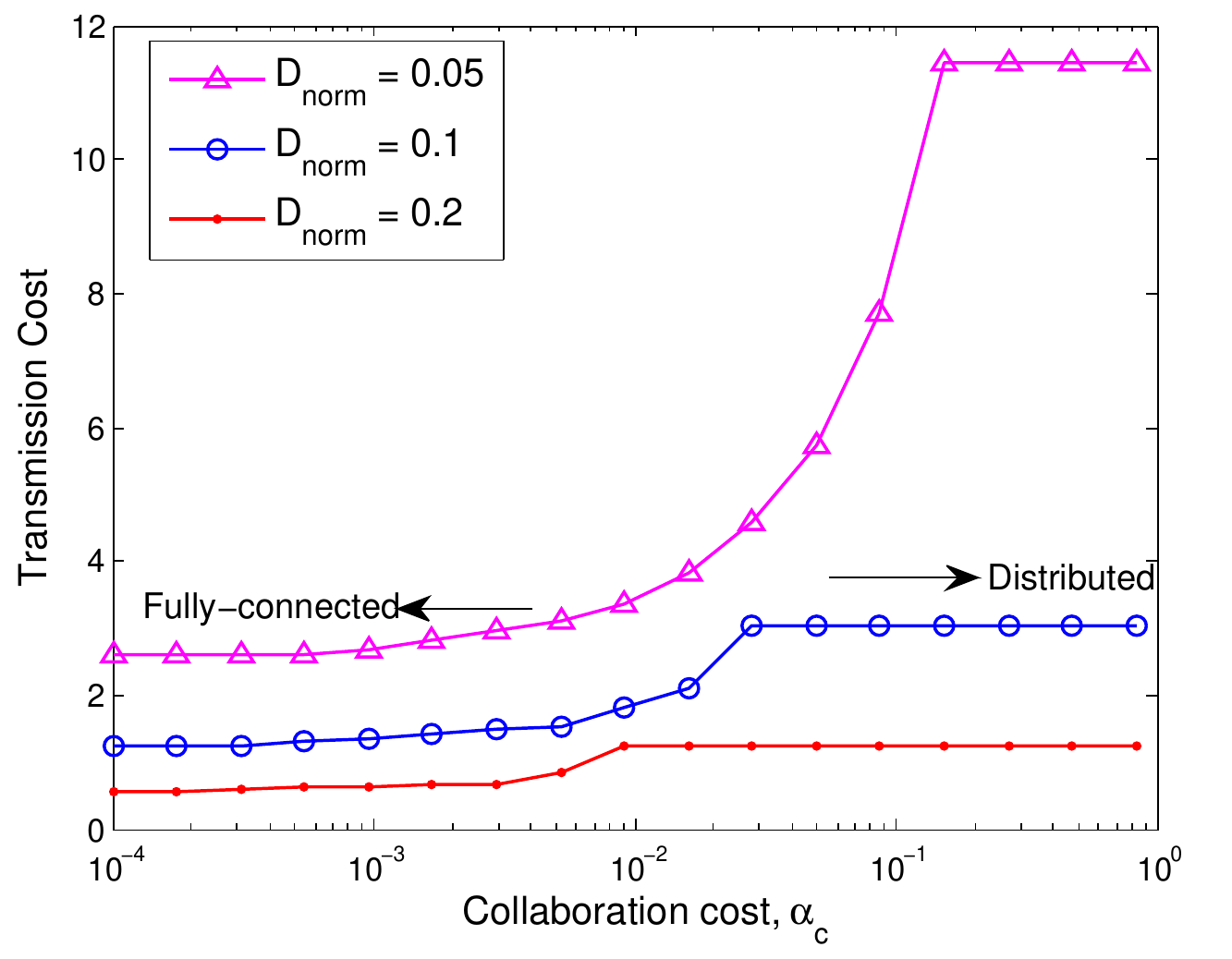}
&
\includegraphics[width=.32\textwidth,height=!]{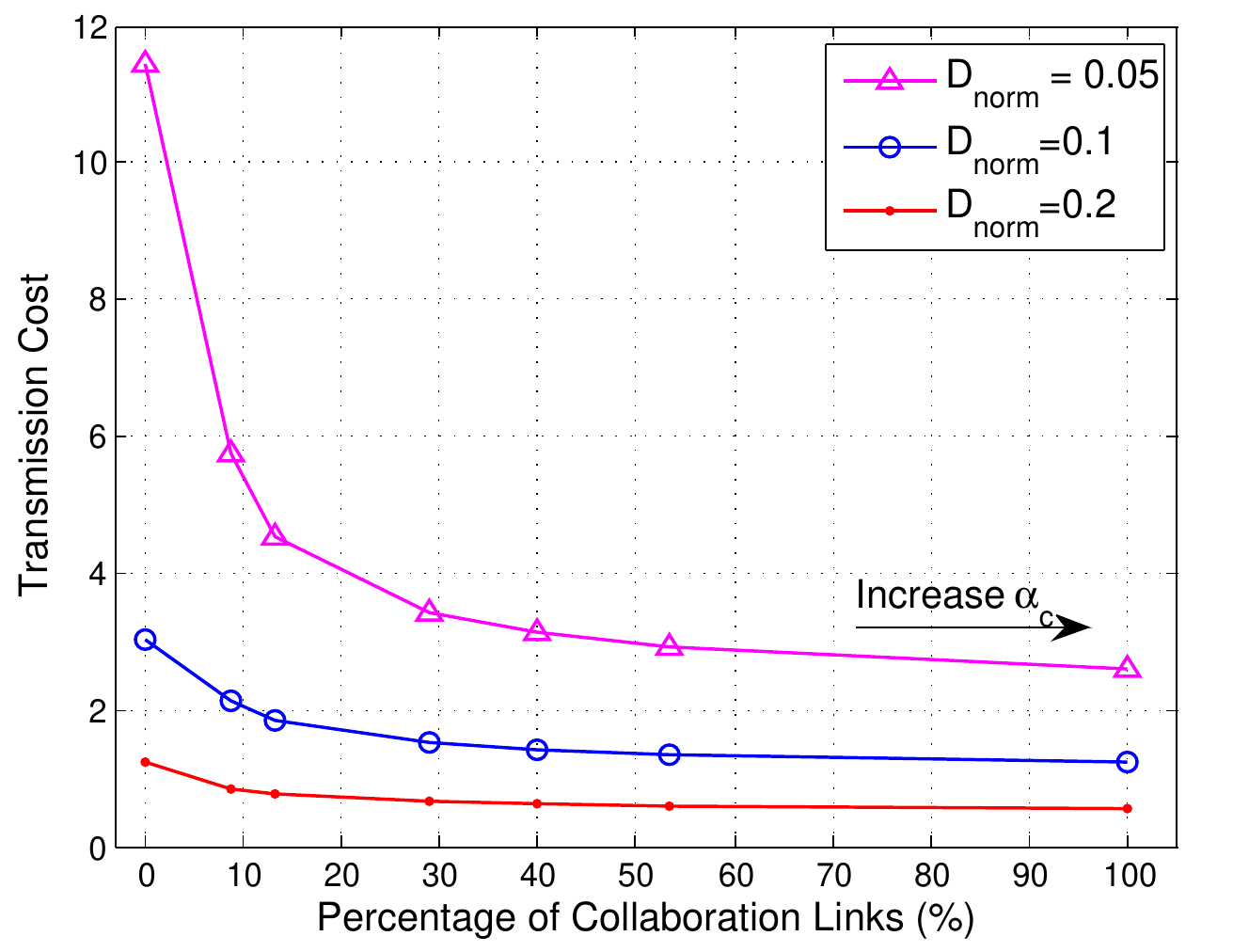}
\\
(a) &
(b)&
(c)
\end{tabular}}
\caption{\footnotesize{ %Performance of solving (\ref{eq: inf_con_prob}) 
Information constrained collaboration for different values of
collaboration cost parameter $\alpha_{\mathrm{c}}$ as $D_{\mathrm{norm}} \in \{ 0.05,0.1, 0.2 \}$: (a) percentage of collaboration links, (b) transmission cost, (c) trade-off between collaboration links and transmission cost.
}}
  \label{fig: Inf_prob_alphaC}
\end{figure*}

\begin{figure*}[htb]
\centerline{ \begin{tabular}{ccc}
\includegraphics[width=.31\textwidth,height=!]{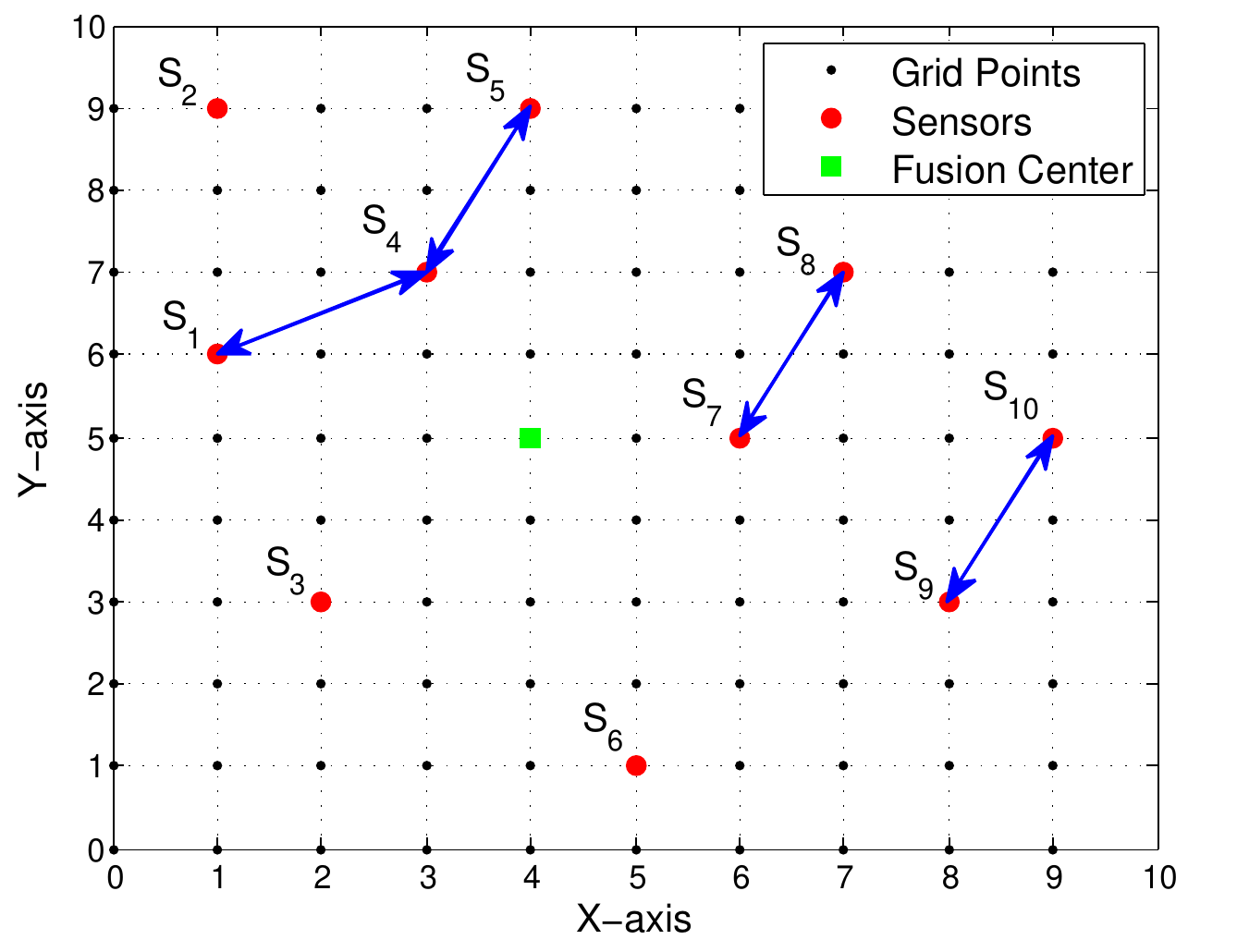}
& \includegraphics[width=.31\textwidth,height=!]{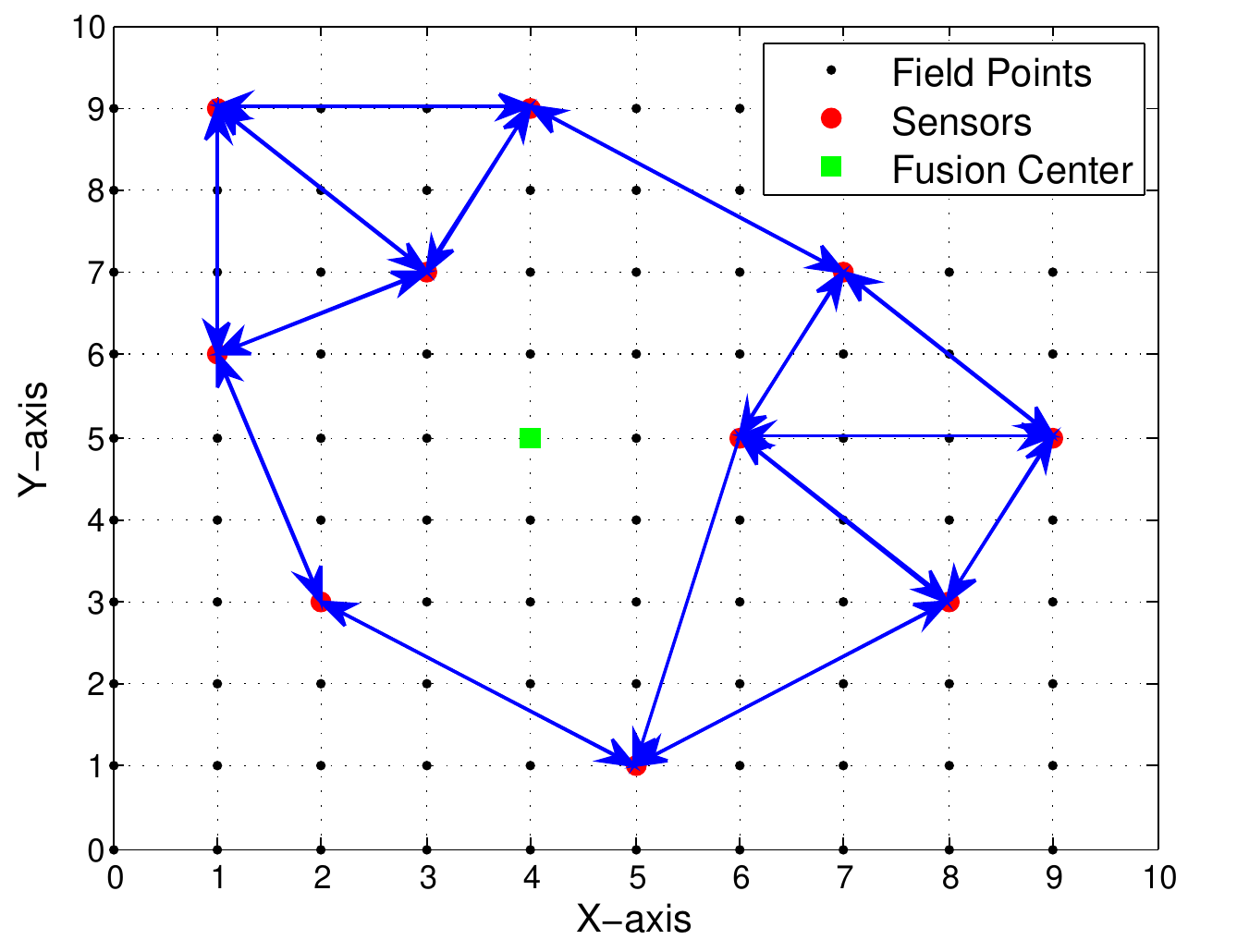}
& \includegraphics[width=.31\textwidth,height=!]{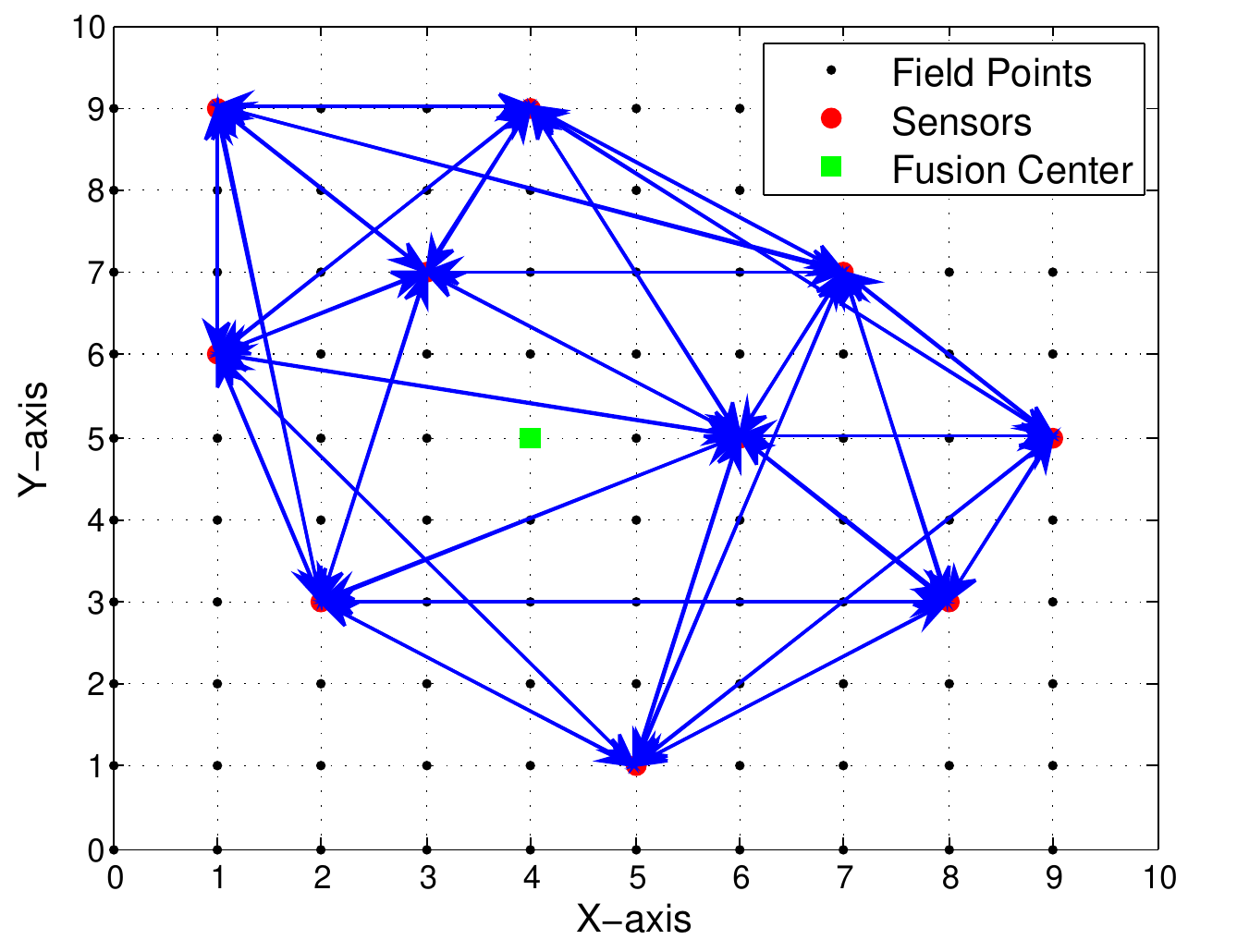}
\\
(a) & (b) & (c) %\\
%\includegraphics[width=.3\textwidth,height=!]{Figs/Top_38_Jr.pdf}
%& \includegraphics[width=.3\textwidth,height=!]{Figs/Top_70_Jr.pdf}
%& \includegraphics[width=.3\textwidth,height=!]{Figs/Top_full_100_Jr.pdf}
%\\
%(d) & (e) & (f)
\end{tabular}}
\caption{\footnotesize{
Collaboration topologies: %(a) $\alpha_{\mathrm{c}} = 0.1$, $\mathrm{card}(\mathbf w) = 10$; 
(a) $\alpha_{\mathrm{c}} = 5 \times 10^{-3}$, $\mathrm{card}(\mathbf w) = 18$; 
%(c) $\alpha_{\mathrm{c}} = 5 \times 10^{-3}$, $\mathrm{card}(\mathbf w) =22$; 
(b) $\alpha_{\mathrm{c}} = 10^{-3}$, $\mathrm{card}(\mathbf w) = 39$; 
(c) $\alpha_{\mathrm{c}} = 2 \times 10^{-4}$, $\mathrm{card}(\mathbf w) = 70$. 
%(f) $\alpha_{\mathrm{c}} = 1 \times 10^{-4}$, $\mathrm{card}(\mathbf w) = 100$.
}}
  \label{fig: topology}
\end{figure*}

In Fig.\,\ref{fig: Inf_prob_MComp}, we present results when we apply the reweighted $\ell_1$-based ADMM algorithm to solve the information constrained problem (\ref{eq: inf_con_prob}).
%, where the $\mathbf w$-minimization step of ADMM is performed through SDP and KKT conditions.
%These results are labeled as `ADMM-SDP' and `ADMM-KKT' in this figure. 
For comparison, we also show the results of using an exhaustive search that enumerates all possible sensor collaboration schemes, where for the tractability of an exhaustive search, we consider a small sized sensor network with $N = 5$. 
In the top subplot of Fig.\,\ref{fig: Inf_prob_MComp}, we present the minimum energy cost  as a function of $D_{\mathrm{norm}}$. 
We can see that the energy cost and estimation distortion is monotonically related, and the proposed approach assures near optimal performance compared to the results of exhaustive search. 
%The `ADMM-KKT' method yields almost the same performance as
%`ADMM-SDP'.
In the bottom subplot, we show  the number of active collaboration links as a function of normalized distortion. Note that a larger estimation distortion corresponds to fewer collaboration links. %In the extreme case, the network performs in a distributed manner when $D_{\mathrm{norm}} > 0.08$.

In Fig.\,\ref{fig: Inf_prob_alphaC}, we solve the information constrained problem
for a relatively large network with $N  =  10$ nodes, and  present the number of collaboration links and the required transmission cost as a function of the collaboration cost parameter $\alpha_{\mathrm{c}}$ for different values of estimation distortion $D_{\mathrm{norm}} \in \{ 0.05,0.1, 0.2 \}$. 
Fig.\,\ref{fig: Inf_prob_alphaC}-(a) shows that the number of collaboration links increases as 
$\alpha_{\mathrm{c}}$ decreases. This is expected, since a smaller value of $\alpha_{\mathrm{c}}$ corresponds to a smaller cost of sensor collaboration, and thus encourages a larger number of collaboration links to be established.  
If we fix the value of $\alpha_{\mathrm{c}}$, we also observe that
the number of collaboration links increases 
as $D_{\mathrm{norm}}$ decreases. 
This is consistent with the results in the bottom subplot of Fig\,\ref{fig: Inf_prob_MComp}.
We show the specific collaboration topologies that correspond to the marked values of $\alpha_c$ in Fig.\,\ref{fig: topology}. These will be discussed in detail later.

Fig.\,\ref{fig: Inf_prob_alphaC}-(b) 
shows that the transmission cost increases as $\alpha_{\mathrm{c}}$ increases for a given estimation distortion. Note that a larger value of $\alpha_{\mathrm{c}}$ indicates a higher cost of sensor collaboration. 
Therefore, to achieve a certain estimation performance, more transmission cost would be consumed instead of sensor collaboration. 
This implies that the transmission cost and collaboration cost are two conflicting 
terms. As we continue to increase $\alpha_{\mathrm{c}}$, the transmission cost converges to a fixed value for a given $D_{\mathrm{norm}}$. This is because the network topology cannot be changed any further (converges to the distributed network), where the transmission cost is deterministic for the given topology and distortion. 

Fig.\,\ref{fig: Inf_prob_alphaC}-(c) shows  the trade-off between 
 the number of collaboration links and the consumed transmission cost by varying
 the parameter $\alpha_{\mathrm{c}}$. One interesting observation is that the transmission cost ceases to decrease significantly when over $50\%$ collaboration links are established. 
The reason is that the transmission cost is characterized 
by the magnitude of nonzero entries in $\mathbf w$, which has very small increment as the number of active links  is relatively large. %Therefore, the transmission cost has a relatively small increase beyond  $60\%$  active links.

In Fig.\,\ref{fig: topology}, 
we present the collaboration topologies obtained from solutions of the information constrained problem (with $D_{\mathrm{norm}} = 0.2$) by varying the parameter of collaboration cost $\alpha_{\mathrm{c}}$; see the labeled points in Fig.\,\ref{fig: Inf_prob_alphaC}-(a). 
In each subplot,  the solid lines with arrows represent the collaboration links among local sensors.
For example in Fig.\,\ref{fig: topology}-(a), the line from sensor $1$ to sensor $4$ indicates that sensor $1$ shares its observation with sensor $4$.
%Fig.\,\ref{fig: topology}-(a) shows a distributed network  for very large value of $\alpha_{\mathrm{c}} $. 
Fig.\,\ref{fig: topology}-(a) shows that the nearest neighboring sensors collaborate initially because of the lower collaboration cost. 
We continue to decrease $ \alpha_{\mathrm{c}} $, 
Fig.\,\ref{fig: topology}-(c) and (d) show that more collaboration links are established, and  sensors tends to collaborate over the entire spatial field rather than aggregating in a small neighbourhood. 
%Finally,
%Figs.\,\ref{fig: topology}-(e) and (f) show that 
%the network tends to operate in a fully-connected manner for very small values of $ \alpha_{\mathrm{c}} $.

\begin{figure}[htb]
\centering
\includegraphics[width=0.5\textwidth,height=!]{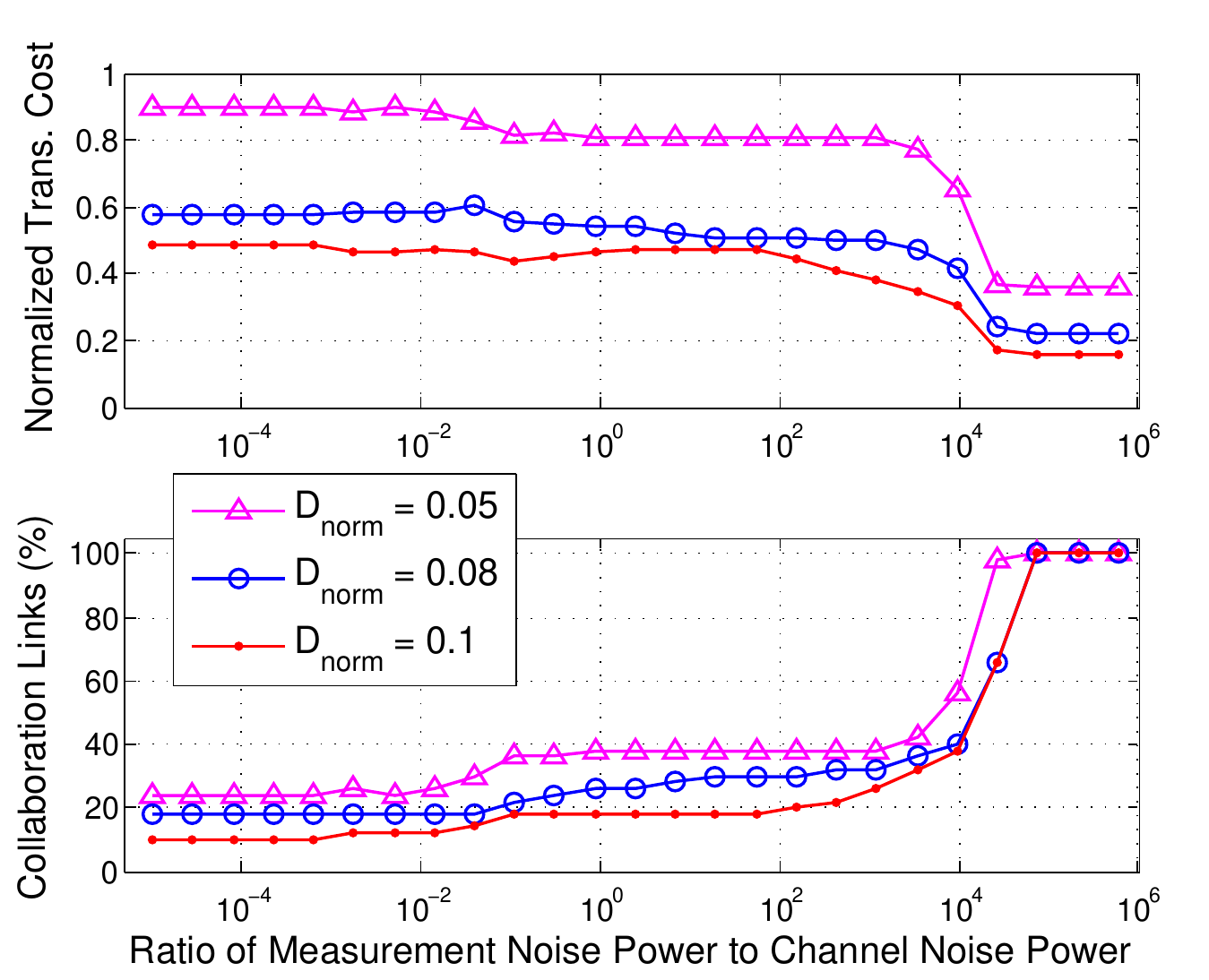} %0.46
\caption{\footnotesize{Transmission cost and collaboration link percentage versus the ratio of measurement noise power and channel noise power.}}
\label{fig: Noise_power}
\end{figure} 

In Fig.\,\ref{fig: Noise_power}, we solve the information constrained problem with $N = 10$ and $D_{\text{norm}} \in \{ 0.05, 0.08, 0.1 \}$ to demonstrate the power allocation schemes when the ratio $\frac{\zeta^2}{\xi^2}$ of measurement noise power to channel noise power varies.
In the top plots of Fig.\,\ref{fig: Noise_power}, we present the transmission cost  which is normalized over the obtained total energy cost.
Given a value of $D_{\text{norm}}$, 
we observe that when the measurement noise power is much less than the channel noise power, the transmission cost dominates the total energy cost since
transmission over noisy channels requires more transmission power. Conversely, when the measurement noise power is much larger than the channel noise power, most of the energy is allocated for sensor collaboration. This is because the act of sensor collaboration can be regarded as a kind of local averaging that effectively reduces the measurement uncertainty at each node. If we fix the ratio of noise power $\frac{\zeta^2}{\xi^2}$, we note that 
to improve the estimation performance, more transmission and collaboration power are consumed.

\begin{figure}[htb]
\centering
\includegraphics[width=0.5\textwidth,height=!]{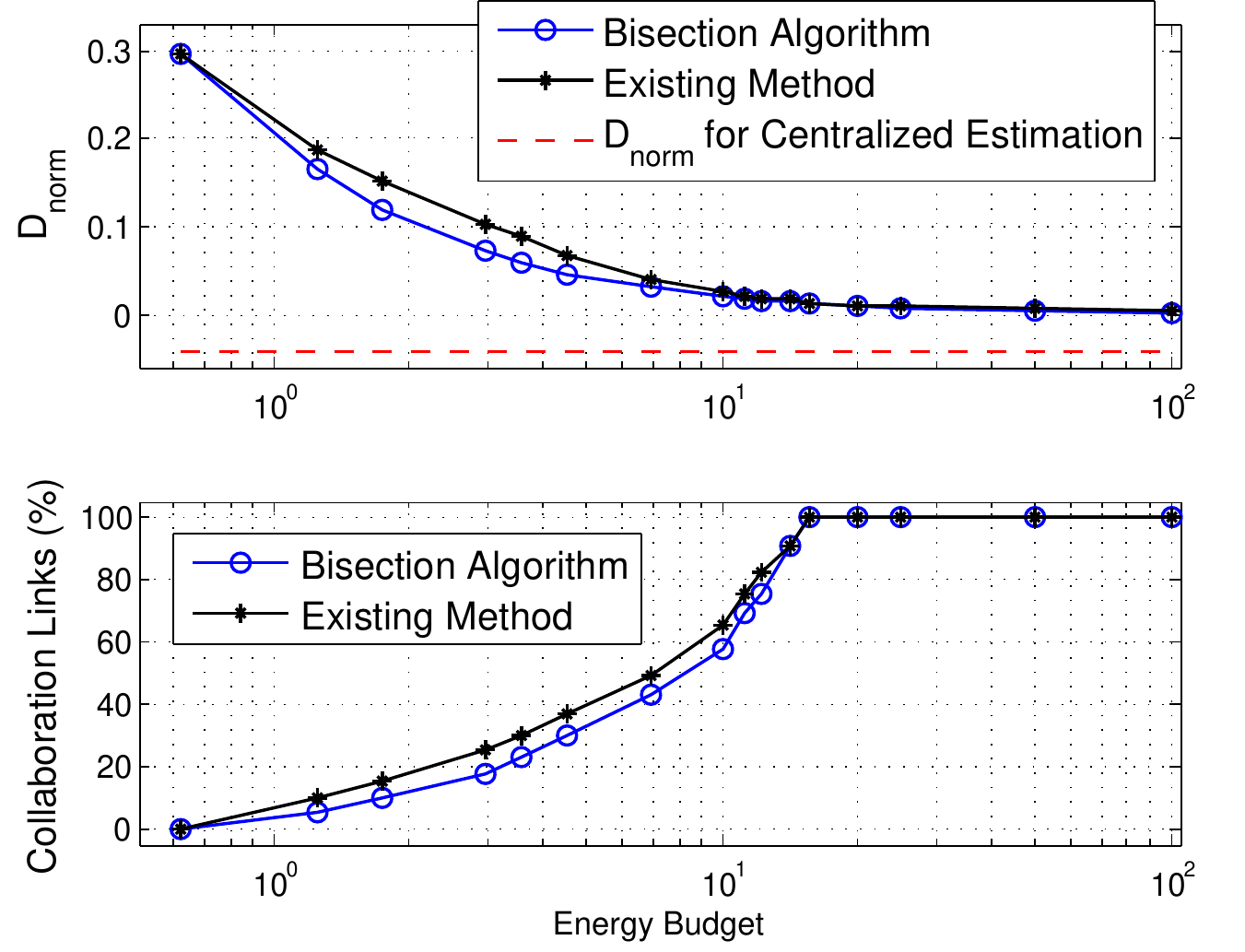} %0.46
\caption{\footnotesize{Performance evaluation for energy constrained sensor collaboration.}}
\label{fig: Ene_prob_MComp}
\end{figure} 

\begin{figure}[htb]
\centering
\includegraphics[width=0.5\textwidth,height=!]{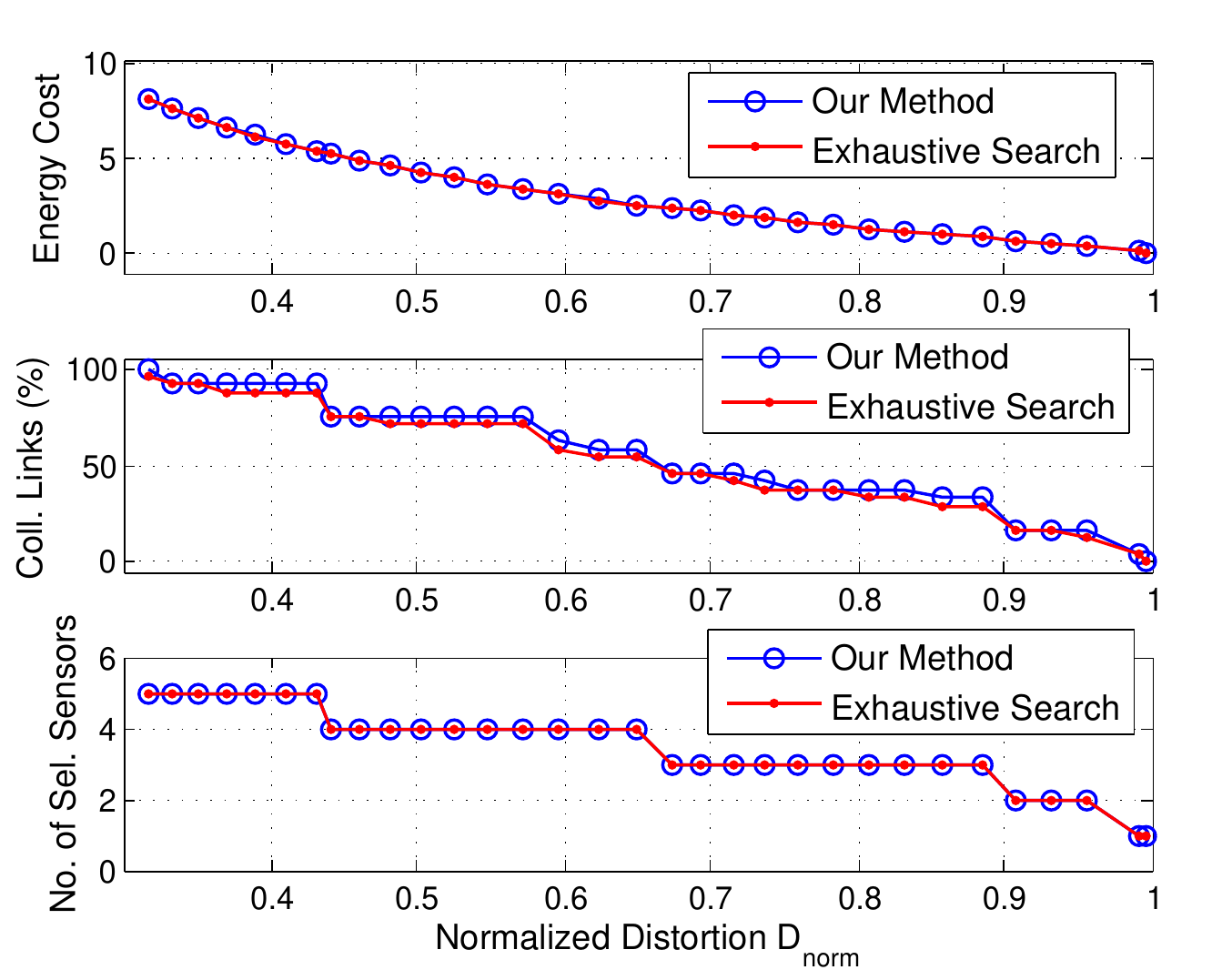}
\caption{\footnotesize{Performance evaluation of sensor selection and collaboration.}}
\label{fig: perf_Sel_J}
\end{figure}

In Fig.\,\ref{fig: Ene_prob_MComp}, we employ the proposed bisection algorithm to solve the energy constrained problem (\ref{eq: ene_con_prob}). We present
the obtained estimation distortion and number of  collaboration links as functions of the energy budget with $N  =  10$. 
For comparison, we also show
the results of using the greedy method in \cite{karvar13} and the normalized  distortion of using centralized estimation, in which  the sensor measurements   are received at the FC in a lossless manner by disregarding channel fading and noise.
As we can see, the method in \cite{karvar13} yields worse estimation performance than our approach, even though its resulting number of collaboration links is larger.  
This indicates that
compared to the number of collaboration links, the optimality of collaboration topology has a more significant impact on the estimation performance. Moreover when the energy budget increases, the estimation distortion converges to zero and the network tends to perform in a fully-connected manner. We further note that  $D_{\text{norm}}$ for centralized estimation is negative, which implies that  the centralized estimation scheme outperforms the proposed distributed estimation scheme. This is because  the former excludes channel impairments,  such as fading and noise.
%the transmission noise is excluded and all the sensor measurements are available at the FC.
%This implies that there exists a performance gap between the smallest estimation distortion of distributed estimation with sensor collaboration and that of centralized estimation, 

\begin{figure}[htb]
\centering
\includegraphics[width=0.47\textwidth,height=!]{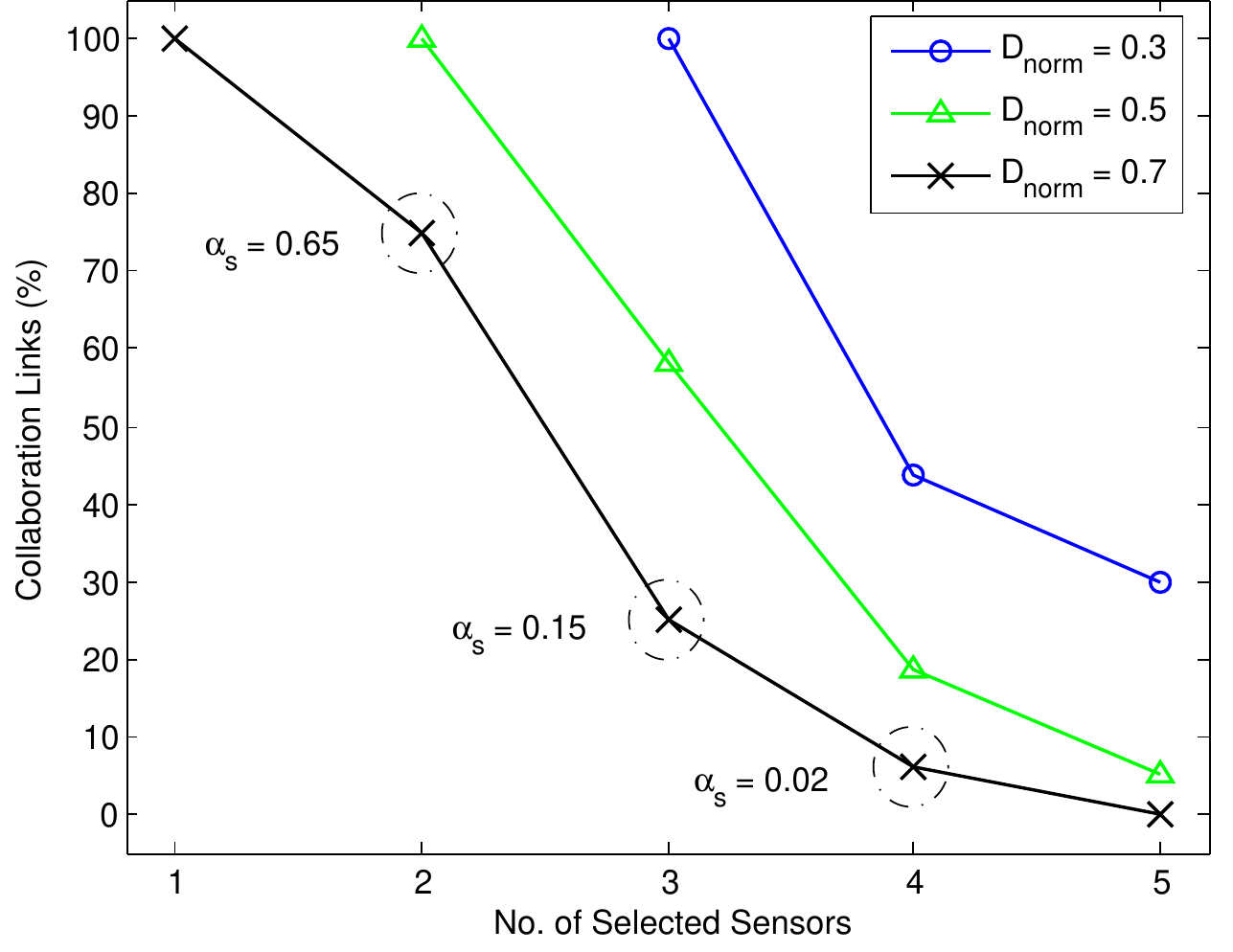}
\caption{\footnotesize{Trade-off between collaboration links and selected sensors.}}
\label{fig: trf_Sel_Col}
\end{figure} 

In Fig.\,\ref{fig: perf_Sel_J}, we employ the convex restriction based ADMM method to solve  
 problem (\ref{eq: inf_sel_ori}) for joint sensor selection and collaboration. We show the resulting energy cost, number of  collaboration links and selected sensors as functions of estimation distortion $D_\mathrm{norm}$.
For comparison, we also present the optimal results obtained from an exhaustive search, where
$N = 5$ sensors are assumed in this example.
We observe that the proposed approach assures near optimal performance for all values of
$D_\mathrm{norm}$.
Moreover, the energy cost, number of collaboration links and selected sensors increases as $D_\mathrm{norm}$ decreases, since a smaller estimation distortion enforces more collaboration links and activated sensors.

\begin{figure*}[htb]
\centerline{ \begin{tabular}{ccc}
\includegraphics[width=.3\textwidth,height=!]{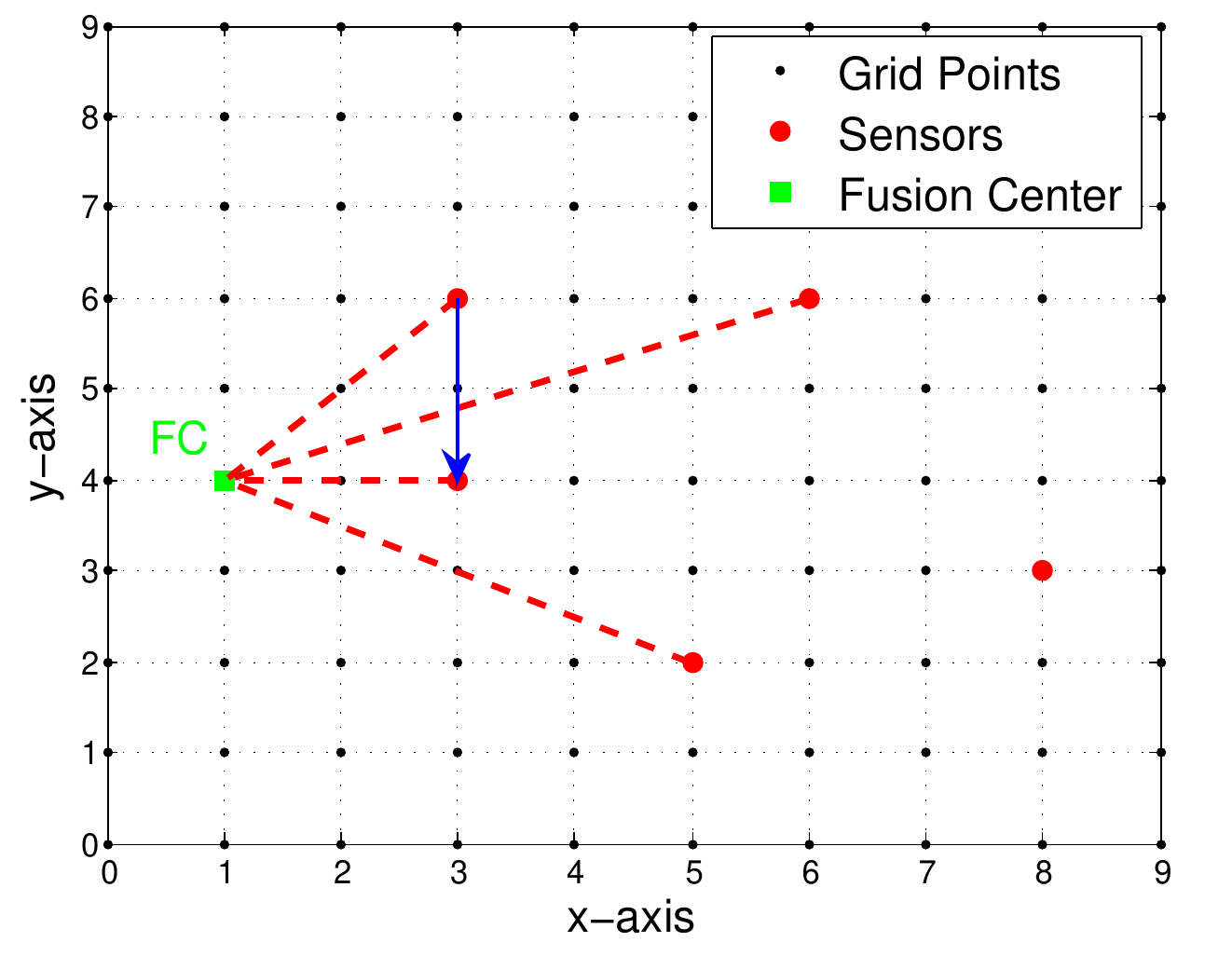}  & 
\includegraphics[width=.3\textwidth,height=!]{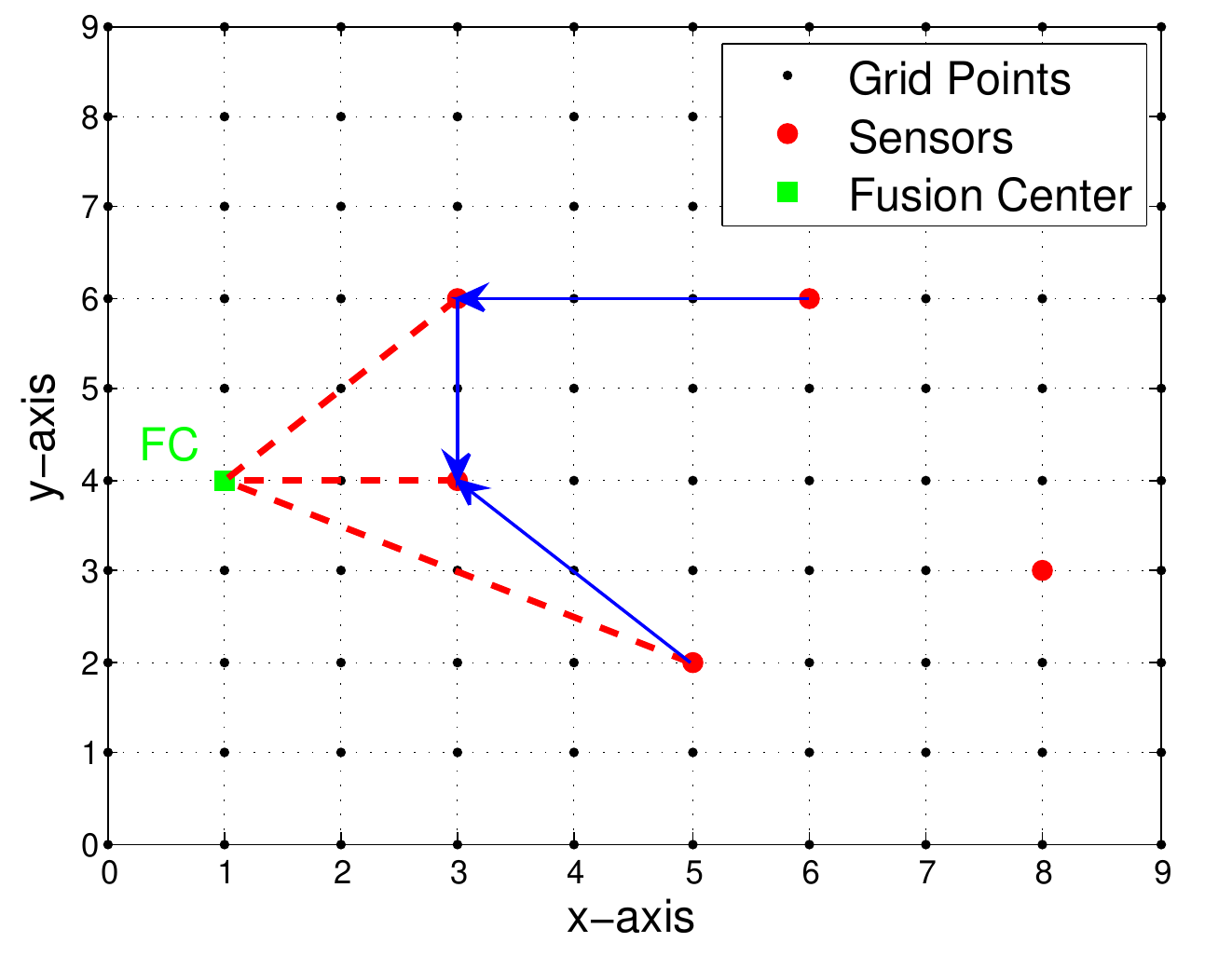} & 
\includegraphics[width=.3\textwidth,height=!]{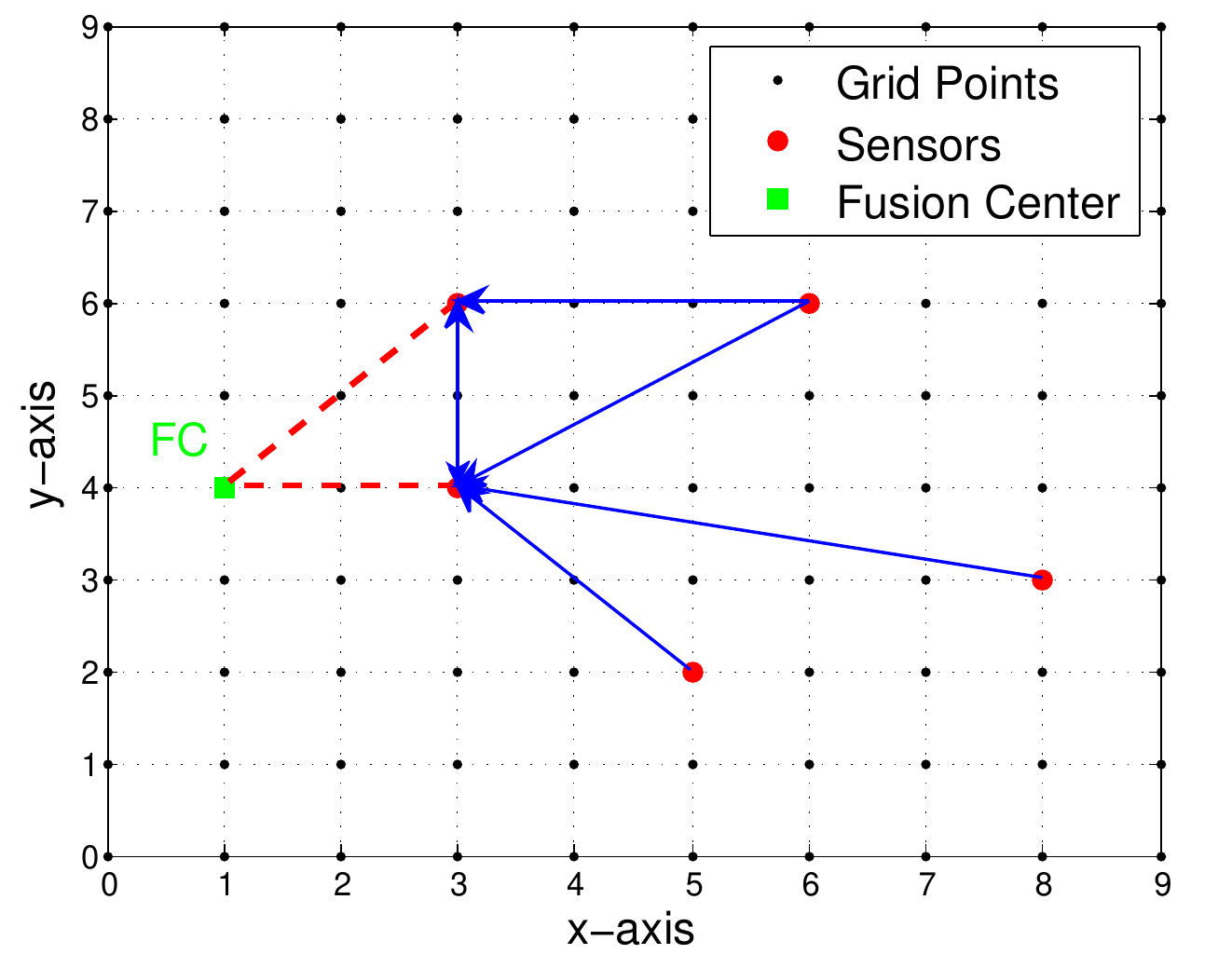}
\\
(a) &
(b) &
(c) 
\end{tabular}}
\caption{\footnotesize{
Network topologies for $D_{\mathrm{norm}} = 0.7$ and $\alpha_{\mathrm{s}} = 0.02 $, $0.15$ and $0.65$.
}}
  \label{fig: Top_SelCol}
\end{figure*}

In Fig.\,\ref{fig: trf_Sel_Col}, we present the trade-offs between the established collaboration links and selected sensors that communicate with the FC. These trade-offs are achieved by fixing $\alpha_c = 0.1$ and varying the  parameter of sensor selection cost
$\alpha_{\mathrm{s}}$ for $D_{\mathrm{norm}} = 0.3$, $0.5$ and $0.7$. 
We fix $D_{\mathrm{norm}}$ and decrease $\alpha_{\mathrm{s}}$, which leads to an increase in the number of selected sensors, meanwhile, the number of collaboration links decreases. That is because to achieve a given estimation distortion, less collaboration links are required if more sensors are selected to communicate with the FC. 
If we fix the number of collaboration links, the number of selected sensors increases as $D_{\mathrm{norm}}$ decreases, since a smaller $D_{\mathrm{norm}}$ enforces more activated sensors. For the marked points as $D_{\mathrm{norm}} = 0.7$, we show the specific sensor collaboration and selection schemes in Fig.\,\ref{fig: Top_SelCol}, where
 the solid line with an arrow represents the collaboration link between two sensors, and the dashed line from one sensor to the FC signifies that this sensor is selected to communicate with the FC.
%
%
%
%In Fig.\,\ref{fig: Top_SelCol}, we show the proposed sensor collaboration and selection schemes when $D_{\mathrm{norm}} = 0.3$ and $\alpha_{\mathrm{s}} \in \{ 0.5, 1, 5 \}$.
%In each subplot, the solid line with an arrow represents the collaboration link between two sensors. And the dashed line from one sensor to the FC signifies that this sensor is selected to communicate with the FC.
Clearly, more collaboration links are established as fewer sensors are selected to communicate with the FC.

\section{Conclusion}
\label{sec: conclusion}

In this paper, we studied the problem of sensor
collaboration with {nonzero} collaboration cost for distributed estimation over a coherent MAC. By making a one-to-one correspondence between
the collaboration topology and the sparsity structure of
collaboration matrix, the sensor collaboration
problems can be interpreted as sparsity-aware optimization problems.
In particular, we studied two types of sensor collaboration problems:
information constrained problem and energy constrained
problem, where we employed the reweighted $\ell_1$-based
ADMM and the bisection algorithm to find their locally optimal
solutions. Further, we investigated the issue of sensor selection in  the proposed collaborative estimation system, where the optimal sensor collaboration and selection schemes can be designed jointly through the entry- and group-level sparsity of the collaboration vector. We  empirically showed that there exists a trade-off between sensor collaboration and sensor selection.

In this paper, we assumed that a random parameter was estimated at one snapshot. 
In future work, we will explore the problem of state tracking for a dynamical system. Also, we will generalize the procedure of sensor collaboration by incorporating the additive noise, and develop a distributed algorithm for sensor collaboration. Lastly, since sensors may have individual power budgets, we will consider the collaboration problem with individual power constraints.

\appendices

\section{Quadratic Functions Transformation}
\label{appendix: quad_vec}
According to \cite[Sec.\,III]{karvar13}, it is straightforward to derive the quadratic vector functions 
\eqref{eq: Tw_vector} and \eqref{eq: Jw_vector} and the corresponding coefficient matrices are given by 
\begin{align}
& \hspace*{0.2cm} \boldsymbol  \Omega_{\Tsj} = \mathbf I_{\text{\tiny{N}}} \otimes \boldsymbol \Sigma_x, ~
\text{$\mathbf I_{\text{\tiny{N}}}$ is the $N \times N$ identity matrix},
\label{eq: O_P} \\
& \begin{array}{ll}
\boldsymbol \Omega_{\text{\tiny {JN}}} = \mathbf G \mathbf h \mathbf h^T \mathbf G^T, & 
\left[ \mathbf G \right]_{l,n} = \left\{ \begin{array}{cl}
g_{m_l}              & n=n_l, \\
0                       & \mbox{otherwise},
\end{array} \right. 
\end{array} \label{eq: O_JN}\\
& \hspace*{0.2cm}  \boldsymbol \Omega_{\text{\tiny {JD}}} =  \mathbf G (\boldsymbol \Sigma_{\epsilon} + \eta^2 \boldsymbol \Sigma_h) \mathbf G^T \hspace*{-0.035in} + \hspace*{-0.015in} \eta^2 \mathbf H \boldsymbol \Sigma_g \mathbf H^T \hspace*{-0.035in}
+ \hspace*{-0.015in} \eta^2 \boldsymbol \Sigma_g \otimes \boldsymbol \Sigma_h   \nonumber \\
& \hspace*{1.15cm} + \hspace*{-0.015in} \boldsymbol \Sigma_g \otimes \boldsymbol \Sigma_{\epsilon}, \quad \mathbf H = \mathbf I_N \otimes \mathbf h,
\label{eq: O_JD}
\end{align}
where $\otimes $ denotes the 
Kronecker product,
$\boldsymbol \Sigma_x$ is defined in (\ref{eq: E_x}), and
$(m_l,n_l)$ is given by \eqref{eq: vector_w}. 
%$\boldsymbol \Sigma_{\tilde g}$ is defined in (\ref{eq: Ey2}). 
It is clear from (\ref{eq: O_P})-(\ref{eq: O_JD})  that 
$\boldsymbol \Omega_{\Tsj}$, $\boldsymbol \Omega_{\JNsj}$, and $\boldsymbol \Omega_{\JDsj}$ are all symmetric positive semidefinite matrices, and $\boldsymbol \Omega_{\JNsj}$ is of rank one.
%$\boldsymbol \Omega_{\Tsj}  \succeq 0$, $\boldsymbol \Omega_{\JNsj} \succeq 0$, 
%%$\mathrm{rank}(\boldsymbol \Omega_{\JNsj}) = 1$  
%and $\boldsymbol \Omega_{\JDsj}  \succeq 0$, where $\mathbf A \succeq 0$ denotes a  positive semidefinite matrix $\mathbf A$. 

%{\color{blue}
\section{Review of \cite[Theorem\,1]{karvar13} for Sensor Collaboration with Given Topologies}
\label{appendix: theorem1}
It has been shown in \cite{karvar13} that the sensor collaboration problems with \textit{given} collaboration topologies can be solved analytically, since the collaboration
cost $Q(\mathbf w) = \sum_{l=1}^L c_l \mathrm{card}(w_l)$ is a constant, and
(\ref{eq: inf_con_prob}) and (\ref{eq: ene_con_prob}) become problems with {homogeneous} quadratic functions,  in which no linear term with respect to $\mathbf w$ is involved. The solutions of (\ref{eq: inf_con_prob}) and (\ref{eq: ene_con_prob}) for a fully-connected network are shown in Theorem\,1.

\textbf{Theorem\,1 \cite[Theorem\,1]{karvar13}:}
For a fully-connected network,  
the optimal values ($\tilde P$ and $J^*$) and solutions ($\tilde {\mathbf w} $ and ${\mathbf w}^{*}$) of (\ref{eq: inf_con_prob}) and (\ref{eq: ene_con_prob})
are given by
\begin{subequations}
\begin{equation}
\left \{ \begin{array}{l}
\displaystyle \tilde P  =  \lambda_{\mathrm{min}}^{\mathrm{pos}} \left(  {\boldsymbol \Omega}_{\Tsj},  - { \boldsymbol \Omega}_{\text{\tiny {JD} }} + \frac{{\boldsymbol \Omega}_{\text{\tiny{JN} } }}{\check J}\right) \xi^2  + \mathbf 1^T \mathbf c \\
\displaystyle \tilde {\mathbf w} = \sqrt{ \frac{\tilde P - \mathbf 1^T \mathbf c}{\tilde {\mathbf v}^T {\boldsymbol \Omega}_{\Tsj} \tilde {\mathbf v}} } ~ \tilde {\mathbf v}, 
\end{array}
\right.
\label{eq: eig_P01}
\end{equation}
\begin{equation}
\text{and} \quad \left \{ \begin{array}{l}
\displaystyle J^* =  \lambda_{ \mathrm{max}} \left( { \boldsymbol \Omega}_{\JNsj},   {\boldsymbol \Omega}_{\JDsj} +  \frac{ \xi^2   {\boldsymbol \Omega}_{\Tsj}  }{\hat P - \mathbf 1^T \mathbf c }\right)\\
\displaystyle {\mathbf w}^*=  \sqrt{\frac{\hat P - \mathbf 1^T \mathbf c}{ ({\mathbf v^*})^T {\boldsymbol \Omega}_{\Tsj}  {\mathbf v^*}} }~ {\mathbf v}^*,
\end{array}
\right.
\label{eq: eig_P02}
\end{equation}
\end{subequations}
where $\lambda_{\mathrm{min}}^{\mathrm{pos}} (\mathbf A, \mathbf B)$ 
and $ \lambda_{\mathrm{max}} (\mathbf A, \mathbf B)$ denote the minimum positive eigenvalue and
the maximum eigenvalue of the generalized eigenvalue problem  $\mathbf A \mathbf v = \lambda \mathbf B \mathbf v$, respectively, and $ \tilde {\mathbf v}$ and $ {\mathbf v}^*$ are the corresponding eigenvectors. \hfill $\blacksquare$

It is clear from (\ref{eq: eig_P02}) that
the optimal Fisher information is upper bounded by $J_0 \Def \lambda_{\mathrm{max}} ( { \boldsymbol \Omega}_{\JNsj},   {\boldsymbol \Omega}_{\JDsj} )$ as $P \to +\infty$.
This implies that to guarantee the \textit{feasibility} of (\ref{eq: inf_con_prob}), the information threshold $\check J$ must lie in the interval $[0, J_0)$. Accordingly, the estimation distortion  in (\ref{eq: D_w}) belongs to $(D_0,\eta^2]$, where  $D_0 = {\eta^2}/{(1+\eta^2 J_0)}$ denotes the minimum distortion, and $\eta^2$ signifies the maximum distortion which is determined by the prior information of $\theta$.
We summarize the boundedness of Fisher information and estimation distortion in Lemma\,\ref{lemma1}.

\begin{lemma}
\label{lemma1}
For problems (\ref{eq: inf_con_prob}) and (\ref{eq: ene_con_prob}), the values of Fisher information and estimation distortion are bounded as $J(\mathbf w) \in [0, J_0)$ and $ D(\mathbf w) \in (D_0,\eta^2]$, respectively, where  $J_0 = \lambda_{\mathrm{max}} ( { \boldsymbol \Omega}_{\JNsj},   {\boldsymbol \Omega}_{\JDsj} )$, $D_0 = {\eta^2}/{(1+\eta^2 J_0)}$, and $\eta^2$ is the variance of the random parameter to be estimated.
\end{lemma}

%\textbf{Lemma\,1}: \textit{For problems (\ref{eq: inf_con_prob}) and (\ref{eq: ene_con_prob}), the values of Fisher information and estimation distortion are bounded as $J(\mathbf w) \in [0, J_0)$ and $ D(\mathbf w) \in (D_0,\eta^2]$, respectively, where  $J_0 = \lambda_{\mathrm{max}} ( { \boldsymbol \Omega}_{\JNsj},   {\boldsymbol \Omega}_{\JDsj} )$, $D_0 = {\eta^2}/{(1+\eta^2 J_0)}$, and $\eta^2$ is the variance of the random parameter to be estimated.}

%}

For a given $\check J \in [0, J_0)$, we demonstrate in Lemma\,\ref{lemma2} that the matrix $\check J \boldsymbol \Omega_{\JDsj}  - \boldsymbol \Omega_{\JNsj}$ is not positive semidefinite.

\begin{lemma}
\label{lemma2}
Given $\check J \in [0, J_0)$, the matrix $\check J \boldsymbol \Omega_{\JDsj}  - \boldsymbol \Omega_{\JNsj}$ is not positive semidefinite.
\end{lemma}

\textbf{Proof:} 
Since $J_0 = \lambda_{\mathrm{max}} ( { \boldsymbol \Omega}_{\JNsj},   {\boldsymbol \Omega}_{\JDsj} )$, there exists an eigenvector $\mathbf v_0$ such that
$
{ \boldsymbol \Omega}_{\JNsj} \mathbf v_0 = J_0 {\boldsymbol \Omega}_{\JDsj} \mathbf v_0
$,
which yields
$
\mathbf v_0^T { \boldsymbol \Omega}_{\JNsj} \mathbf v_0 = \mathbf v_0^T J_0 {\boldsymbol \Omega}_{\JDsj} \mathbf v_0 
$.
Since $\check J \in [0, J_0)$, we obtain
$
\mathbf v_0^T { \boldsymbol \Omega}_{\JNsj} \mathbf v_0 > \mathbf v_0^T \check J {\boldsymbol \Omega}_{\JDsj} \mathbf v_0 
$, namely,
$
\mathbf v_0^T (\check J {\boldsymbol \Omega}_{\JDsj} - { \boldsymbol \Omega}_{\JNsj}) \mathbf v_0 < 0
$.
Therefore, we find a vector $\mathbf v_0$ such that $\mathbf v_0^T (\check J {\boldsymbol \Omega}_{\JDsj} - { \boldsymbol \Omega}_{\JNsj}) \mathbf v_0 < 0$. Namely,  $\check J \boldsymbol \Omega_{\JDsj}- \boldsymbol \Omega_{\JNsj}$ is not positive semidefinite.
\hfill $\blacksquare$

\section{The KKT-Based Solution for A QP1QC}
\label{appendix: sol_phi_w}
To prove the Prop.\,\ref{prop: sol_phi_w} or \ref{prop: sol_phi_u_sel}, we consider a more general case of QP1QC
\begin{align}
\begin{array}{ll}
\displaystyle \minimize_{\mathbf w} & ~ \mathbf w^T \mathbf A_0 \mathbf w + 2 \mathbf b_0^T \mathbf w\\
\st &~ \mathbf w^T \mathbf A_1 \mathbf w + 2 \mathbf b_1^T \mathbf w + r_1 \leq 0,
\end{array}
\label{eq: QP1QC_general}
\end{align}
where $\mathbf A_0$ is a symmetric positive definite matrix, and 
$\mathbf A_1$ is a symmetric matrix.

Upon defining $\mathbf P \Def \frac{1}{r_1} \mathbf A_0^{-\frac{1}{2}} \mathbf A_1 \mathbf A_0^{-\frac{1}{2}} $, we obtain the eigenvalue decomposition of $\mathbf P $
\[
\mathbf P = \mathbf U \boldsymbol \Lambda  \mathbf U^T,
\]
where $\mathbf U$ is an orthogonal matrix that includes the eigenvectors of $\mathbf P$, and $\boldsymbol \Lambda $ is a diagonal matrix that includes the eigenvalues of $\mathbf P$.

Let $\mathbf u \Def \mathbf U^T \mathbf A_0^{\frac{1}{2}} \mathbf w$, $\mathbf g \Def \mathbf U^T  \mathbf A_0^{-\frac{1}{2}} \mathbf b_0$ and $\mathbf e \Def \mathbf U^T \mathbf A_0^{-\frac{1}{2}} \frac{\mathbf b_1}{r_1}$, then problem  (\ref{eq: QP1QC_general}) can be written as
\begin{align}
\begin{array}{ll}
\displaystyle \minimize_{\mathbf u} & ~ \mathbf u^T \mathbf u + 2 \mathbf u^T \mathbf g\\
\st &~ \mathbf u^T \boldsymbol \Lambda  \mathbf u + 2 \mathbf u^T \mathbf e + 1 \leq 0.
\end{array}
\label{eq: QP1QC_general_simple}
\end{align}
The rationale behind using the eigenvalue decomposition technique to reformulate (\ref{eq: QP1QC_general})  is that the KKT conditions of (\ref{eq: QP1QC_general_simple}) are more compact and easily solved since $\boldsymbol \Lambda $ is a diagonal matrix. 

We demonstrate the KKT conditions for  problem (\ref{eq: QP1QC_general_simple}).

\textit{Primal feasibility}: ~$\mathbf u^T \boldsymbol \Lambda \mathbf u + 2 \mathbf u^T \mathbf e + 1 \leq 0$. 

\textit{Dual feasibility}: ~~ $\mu \geq 0$, where $\mu$ is the dual variable.

\textit{Complementary slackness}:
$
\mu ( \mathbf u^T \boldsymbol \Lambda  \mathbf u+ 2 \mathbf u^T \mathbf e +1 ) = 0. %\label{eq: comp_phi_u}
$

\textit{Stationary of the Lagrangian}:
$
%\mathbf u + \mathbf g + \mu (\boldsymbol \Lambda  \mathbf u + \mathbf e) = 0.
\mathbf u = - (\mathbf I +  \mu \boldsymbol \Lambda )^{-1} (\mathbf g + \mu \mathbf e).
%\label{eq: lag_phi_u}
$

If $\mu = 0$, we have
\begin{align}
\mathbf u = -  \mathbf g,
\label{eq: u_opt_case1}
\end{align}
where $\mathbf u^T \boldsymbol \Lambda  \mathbf u+ 2 \mathbf u^T \mathbf e +1 \leq 0$.

If $\mu > 0$, eliminating $\mathbf u$ by substituting  stationary condition into complementary slackness, we have
\begin{align*}
&(\mathbf g + \mu \mathbf e )^T (\mathbf I + \mu \boldsymbol \Lambda )^{-1} \boldsymbol \Lambda 
  (\mathbf I + \mu \boldsymbol \Lambda )^{-1} (\mathbf g + \mu \mathbf e) \\
&  -  2 (\mathbf g + \mu \mathbf e )^T (\mathbf I +\mu \boldsymbol \Lambda )^{-1} \mathbf e + 1 = 0.
\end{align*}
Since $\boldsymbol \Lambda $ is a diagonal matrix, we finally obtain that
\begin{align}
\sum_{l=1}^L \left( \frac{\lambda_l(\mu e_l + g_l)^2}{( \mu \lambda_l + 1 )^2} - \frac{2 e_l ( \mu e_l + g_l)}{\mu \lambda_l + 1} \right) + 1 = 0,
\label{eq: u_opt_case2}
\end{align}
where $\lambda_l$ is the $l$th diagonal element of $\boldsymbol \Lambda $.

If $\lambda_l >0$ for $l = 1,2,\ldots,L$ and thus $\mathbf A_1$ is positive definite, Eq.\,\eqref{eq: u_opt_case2} can be simplified as
\begin{align}
&\sum_{l=1}^L \left \{ \left [ \frac{\sqrt{\lambda_l} (\mu e_l + g_l)}{1+\mu \lambda_l} - \frac{e_l}{\sqrt{\lambda_l}} \right ]^2 - \frac{e_l^2}{\lambda_l}\right \} + 1 \nonumber \\
= &\sum_{l=1}^L \frac{(\lambda_l g_l -e_l)^2}{\lambda_l(1+\mu \lambda_l)^2} - \sum_{l=1}^L\frac{e_l^2}{\lambda_l} +1 = 0.
\label{eq: u_opt_case2_simple}
\end{align}
In \eqref{eq: u_opt_case2_simple}, the function $\frac{(\lambda_l g_l -e_l)^2}{\lambda_l(1+\mu \lambda_l)^2}$ is monotonically decreasing with respect to $\mu$ when $\mu > 0$.
This implies that there exists only one positive root for $f(\mu) = 0$
if $\mu > 0$  satisfies KKT condition. The proof is now complete.
\hfill $\blacksquare$

\section{Proof of Lemma\,3}
\label{appendix: lemma3}
We recall that the eigenvalues $\{ \lambda_l \}_{l = 1,2,\ldots,L}$ in \eqref{eq: decom_prop1}  are obtained from the eigenvalue decomposition of the positive definite matrix  $\boldsymbol \Omega_{\JDsj}$ modified by the rank one matrix  $\boldsymbol \Omega_{\JNsj}$. Then it  can be concluded that there exists only one negative eigenvalue in $\{ \lambda_l \}_{l = 1,2,\ldots,L}$ \cite[Sec.\,5]{gol73}.

Without loss of generality, we assume that $\lambda_1 < 0 < \lambda_2 \leq \lambda_3 < \ldots < \lambda_N$, where the case of $\lambda_l = 0$ ($l>1$) is excluded since it is trivial to obtain that
$\frac{\lambda_lg_l^2}{( \mu \lambda_l + 1 )^2} = 0$ in \eqref{eq: nl_eig}.
Given $ - \frac{1}{\lambda_1} > 0 > - \frac{1}{\lambda_L} \geq \ldots \geq - \frac{1}{\lambda_3} \geq - \frac{1}{\lambda_2}$, we have
$f( - \frac{1}{\lambda_1}) \to -\infty$ and $f( - \frac{1}{\lambda_l}) \to \infty$ for $l = 2,3,\ldots,L$.

Next, we take the first-order derivative of $f(\mu)$, which yields
\begin{align}
\frac{d f(\mu)}{d \mu} = \sum_{l=1}^L [-2 \lambda_l^2 g_l^2 (1 + \mu \lambda_l)^{-3}]. \label{eq: df_mu}
\end{align}
When $\mu \in (0, - \frac{1}{\lambda_1})$, we have
\[
1+ \mu \lambda_1 > 0, ~ \text{and}~ 1+\mu \lambda_l > 0 ~~ \text{for}~ i =2,3,\ldots,L.
\]
From \eqref{eq: df_mu} we obtain $\frac{d f(\mu)}{d \mu} \leq 0 $. Therefore, 
$f(\mu)$ is monotonically decreasing as $\mu \in (0,  - \frac{1}{\lambda_1})$. 
Together with $f( - \frac{1}{\lambda_1}) \to -\infty$ and $f( - \frac{1}{\lambda_L}) \to \infty$, 
we  conclude that
there exists only one positive root of $f(\mu) = 0$ if  $f(0) > 0$. %Now, the proof of Lemma\,3 is complete.

When $\mu \in (- \frac{1}{\lambda_1}, \infty)$, we have
\begin{align}
1+ \mu \lambda_1 < 0, ~ \text{and}~ 1+\mu \lambda_l > 0 ~~ \text{for}~ i =2,3,\ldots,L. \label{eq: interval_inf}
\end{align}
It is clear from \eqref{eq: df_mu} that the sign of $\frac{d f(\mu)}{d \mu}$ is difficult to determine since
$-2 \lambda_1^2 g_1^2 (1 + \mu \lambda_1)^{-3} > 0$ and 
$
-2 \lambda_l^2 g_l^2 (1 + \mu \lambda_l)^{-3} < 0
$ for $l = 2,3,\ldots,L$. Therefore, 
the function $f(\mu)$ may not be monotonic, and  the number of positive roots of $f(\mu) = 0$ is uncertain. The proof  is now complete. \hfill $\blacksquare$

\section{Proof of Proposition\,\ref{prop: converse}}
\label{appendix: converse}
%{\color{blue}
From (\ref{eq: Jw_vector}), \eqref{eq: Tw_vector} and \eqref{eq: Qw_vector}, we have
\[
J(\mathbf w) = \frac{\mathbf w^T \boldsymbol \Omega_{\JNsj} \mathbf w}{\mathbf w^T \boldsymbol \Omega_{\JDsj} \mathbf w + \xi^2},
\]
and
\[
P(\mathbf w) = \mathbf w^T \boldsymbol \Omega_{\Tsj} \mathbf w + \sum_{l=1}^L c_l \card(w_l).
\]

Setting $\mathbf w = c \hat{\mathbf w}$ for some fixed vector $\hat{\mathbf w}$, $J(\mathbf w)$ and $P(\mathbf w)$ are strictly increasing functions of $c$ when $c > 1$, and strictly decreasing functions of $c$ when $c < 1$.
%Since the functions $J (\mathbf w)$ and $P(\mathbf w)$  are strictly increasing (or decreasing) as multiplying $\mathbf w$ by a scalar $c > 1$ (or $c <1$), 
Thus, the optimality is achieved for  (\ref{eq: inf_con_prob}) or  (\ref{eq: ene_con_prob}) when the inequality constraints are satisfied with equality.

Given the energy budget $\hat P$,  we have 
$
P(\mathbf w_2) = \hat P, 
$
where  $\mathbf w_2$ is the optimal solution of the energy constrained problem (\ref{eq: ene_con_prob}).
Our goal is to show $\mathbf w_2$ is also a solution of the information constrained problem (\ref{eq: inf_con_prob}) when $\check J = J(\mathbf w_2)$.

If $\mathbf w_2$ is not the solution of (\ref{eq: inf_con_prob}), we assume a better solution $\mathbf w_2^{\prime}$ such that $P(\mathbf w_2^{\prime}) < P(\mathbf w_2)$. 
Since $P(\cdot)$ strictly increases as multiplying the optimization variables by a scalar $c > 1$,
there exists a scalar $c>1$  such that 
\begin{align}
P(\mathbf w_2^{\prime}) < P(c \mathbf w_2^{\prime}) \leq P(\mathbf w_2).
\label{eq: inq1}
\end{align}

On the other hand, since $J(\cdot)$ strictly increases as multiplying the optimization variables by a scalar $c > 1$, we have $ J(c \mathbf w_2^{\prime}) >  J(\mathbf w_2^{\prime})$.
Further, because $\mathbf w_2^{\prime}$ is a feasible vector for (\ref{eq: inf_con_prob}), we have 
$J(\mathbf  w_2^{\prime} ) \geq \check J $, where recalling that $\check J = J(\mathbf w_2)$. We can then conclude that
\begin{align}
J(c \mathbf w_2^{\prime}) >  J(\mathbf w_2^{\prime}) \geq J (\mathbf w_2).
\label{eq: inq2}
\end{align}

From (\ref{eq: inq1}) and $P(\mathbf w_2) = \hat P$, we obtain that $P(c \mathbf w_2^{\prime}) \leq P$, which implies $c \mathbf w_2^{\prime}$ is a feasible point for (\ref{eq: ene_con_prob}).
From (\ref{eq: inq2}), we have $J(c \mathbf w_2^{\prime}) > J (\mathbf w_2)$, which implies  
$c \mathbf w_2^{\prime}$ yields a higher objective value of  (\ref{eq: ene_con_prob}) than 
$\mathbf w_2$. 
This  \textit{contradicts} to the fact that $\mathbf w_2$ is the optimal solution of (\ref{eq: ene_con_prob}). 
Therefore, we can conclude that $\mathbf w_2$ is the solution of (\ref{eq: inf_con_prob}).

On the other hand, if $\mathbf w_1$ is the solution of (\ref{eq: inf_con_prob}),  it is similar to prove that $\mathbf w_1$ is the solution of (\ref{eq: ene_con_prob}) when $\hat P = P(\mathbf w_1)$. The proof is now complete.
\hfill $\blacksquare$
%}

\section{Proof of Proposition\,\ref{prop: v_sel}}
\label{appendix: v_sel}

For notational convenience, we define $\kappa_1 : = \frac{1}{\rho}$, $\kappa_2 : = \frac{ %\gamma 
\tilde d_n}{\rho}$ and 
$
h_{\kappa_2}^{\kappa_1} (\mathbf v_{\Gsj_n}) := \kappa_1 \| \mathbf F_n \mathbf v_{\Gsj_n}\|_1 + \kappa_2 \| \mathbf v_{\Gsj_n} \|_2.
$
Then problem (\ref{eq: phi_v_sel_sub}) can be written as
\begin{align}
\hspace*{-0.1in}
\begin{array}{ll}
\displaystyle \minimize_{\mathbf v_{\Gsj_n}} &  \phi_{\kappa_2}^{\kappa_1} (\mathbf v_{\Gsj_n}) := h_{\kappa_2}^{\kappa_1}  (\mathbf v_{\Gsj_n}) + \frac{1}{2}\|\mathbf v_{\Gsj_n} - \mathbf b_{\Gsj_n} \|_2^2.
\end{array}
\label{eq: phi_v_sel_sub_v1}
\end{align}
Let $\mathbf v_{\Gsj_n}^* $ be the unique minimizer of the following problem
\begin{align}
\begin{array}{ll}
\displaystyle \minimize_{\mathbf v_{\Gsj_n}} & \quad %\psi_{c_2} (\mathbf v_{\Gsj_j}) :=
h_{\kappa_2}^{0} (\mathbf v_{\Gsj_n}) + \frac{1}{2}\|\mathbf v_{\Gsj_n} - \boldsymbol \nu \|_2^2,
\end{array}
\label{eq: phi_c2}
\end{align}
where 
$
\boldsymbol \nu  = \mathrm{sgn}(\mathbf b_{\Gsj_n}) \odot \mathrm{max}(|\mathbf b_{\Gsj_n}| - \kappa_1\mathbf f_{\Gsj_n}, 0).
$ 

We aim to show $\mathbf v_{\Gsj_n}^* $ is also the minimizer of problem (\ref{eq: phi_v_sel_sub_v1}).
The optimality of $\mathbf v_{\Gsj_n}^* $ for problem (\ref{eq: phi_c2}) yields
\begin{align}
\mathbf 0 \in \mathbf v_{\Gsj_n}^* - \boldsymbol \nu + \partial h_{\kappa_2}^{0} (\mathbf v_{\Gsj_n}^*),
\label{eq: grad_phi_c2}
\end{align}
where $\partial h_{\kappa_2}^{0} (\cdot )$ denotes the subgradient of $ h_{\kappa_2}^{0}(\cdot)$.
We then derive the subgradient of $\phi_{\kappa_2}^{\kappa_1} (\mathbf v_{\Gsj_n})$ at $\mathbf v_{\Gsj_n}^* $ 
\begin{align}
\hspace*{-0.1in}
\partial \phi_{\kappa_2}^{\kappa_1} (\mathbf v_{\Gsj_n}^*) \hspace*{-0.02in} = \hspace*{-0.02in}  \mathbf v_{\Gsj_n}^* \hspace*{-0.02in}  - \hspace*{-0.02in}  \mathbf b_{\Gsj_n} 
\hspace*{-0.02in}  + \hspace*{-0.02in}  \kappa_1 \mathbf F_n  \mathrm{SGN} (\mathbf v_{\Gsj_n}^*) \hspace*{-0.02in}  + \hspace*{-0.02in}  \partial h_{\kappa_2}^{0} (\mathbf v_{\Gsj_n}^*),
\label{eq: grad_phi_c2_c1}
\end{align}
where 
$\mathrm{SGN}(\cdot)$ is defined in a component-wise fashion
$
\mathrm{SGN}(x) = \left \{ 
\begin{array}{l l}
    \{ 1\} & \quad  x > 0 \\
    \left [ -1, 1\right ] & \quad x = 0 \\
    \{ -1 \} & \quad x < 0
  \end{array} 
\right.
$ for $\forall x \in \mathbb R$.
%we use the fact that $\mathrm{SGN} (\mathbf F_j \mathbf v_{\Gsj_j}^*) = \mathrm{SGN} (\mathbf v_{\Gsj_j}^*)$.

The definition of $\boldsymbol \nu = \mathrm{sgn}(\mathbf b_{\Gsj_n}) \odot \mathrm{max}(|\mathbf b_{\Gsj_n}| - \kappa_1 \mathbf f_{\Gsj_n}, 0)$ implies
\begin{align}
\nu_i = \left \{ 
\begin{array}{cl}
\vspace*{0.02in}
 \left [ \mathbf b_{\Gsj_n} \right ]_i - \kappa_1 \left [\mathbf f_{\Gsj_n} \right ]_i & \quad \left [\mathbf b_{\Gsj_n}  \right ]_i > \kappa_1 [\mathbf f_{\Gsj_n}]_i \\
  \vspace*{0.02in}
0 & \quad  \left | \left [\mathbf b_{\Gsj_n} \right ]_i \right | \leq \kappa_1  \left[\mathbf f_{\Gsj_n} \right ]_i  \\
 \left [\mathbf b_{\Gsj_n} \right ]_i + \kappa_1  \left [\mathbf f_{\Gsj_n} \right ]_i &  \quad  \left [\mathbf b_{\Gsj_n} \right]_i < - \kappa_1 \left [\mathbf f_{\Gsj_n} \right ]_i,
\end{array}
\right.
\label{eq: u_def_expand}
\end{align}
where $[\mathbf x]_i$ denotes the $i$th entry of a vector $\mathbf x$. %and $\mathbf f_{\Gsj_j}$ is the vector consisting of the diagonal elements of $\mathbf F_j$.

From (\ref{eq: u_def_expand}), we have 
$
\boldsymbol \nu \in \mathbf b_{\Gsj_n} - \kappa_1 \mathbf F_n \mathrm{SGN}(\boldsymbol \nu)
$.
Since $\mathbf v_{\Gsj_n}^* $ is the minimizer of problem (\ref{eq: phi_c2}), 
according to \cite[Lemma\,1]{yualiuye13}, we can obtain that $\mathrm{SGN}(\boldsymbol \nu ) \subseteq \mathrm{SGN}(\mathbf v_{\Gsj_n}^*)$. Thus,
\begin{align}
\boldsymbol \nu \in \mathbf b_{\Gsj_n} - \kappa_1 \mathbf F_n \mathrm{SGN}(\mathbf v_{\Gsj_n}^*).
\label{eq: u_relation}
\end{align}
Combining (\ref{eq: grad_phi_c2}) and (\ref{eq: u_relation}), we obtain that
\[
\mathbf 0 \in \mathbf v_{\Gsj_n}^* - \mathbf b_{\Gsj_n} + \kappa_1 \mathbf F_n \mathrm{SGN}(\mathbf v_{\Gsj_n}^*) + \partial h_{\kappa_2}^{0} (\mathbf v_{\Gsj_n}^*),
\]
which implies that $\mathbf 0 \in \partial \phi_{\kappa_2}^{\kappa_1} (\mathbf v_{\Gsj_n}^*)$ from (\ref{eq: grad_phi_c2_c1}).
Thus,  $\mathbf v_{\Gsj_n}^* $ is  the minimizer of problem (\ref{eq: phi_v_sel_sub_v1}).

Finally, the closed form of $\mathbf v_{\Gsj_n}^* $ in problem (\ref{eq: phi_c2}) is given by a block soft thresholding operator \cite{parboy13}
\[
\mathbf v_{\Gsj_n}^* = \left \{ 
\begin{array}{ll}
(1 - \frac{\kappa_2}{\| \boldsymbol \nu \|_2})\boldsymbol \nu &  \| \boldsymbol \nu\|_2 \geq \kappa_2 \\
0 &  \| \boldsymbol \nu\|_2 < \kappa_2.
\end{array}
\right.
\]
Now, the proof is complete. \hfill $\blacksquare$

\ifCLASSOPTIONcaptionsoff
  \newpage
\fi



\bibliographystyle{IEEEbib}
\bibliography{journal}  

\begin{thebibliography}{10}

\bibitem{olirod11}
L.~Oliveira and J.~Rodrigues,
\newblock ``Wireless sensor networks: a survey on environmental monitoring,''
\newblock {\em Journal of Communications}, vol. 6, no. 2, 2011.

\bibitem{hevicyan06}
T.~He, P.~Vicaire, T.~Yan, L.~Luo, L.~Gu, G.~Zhou, S.~Stoleru, Q.~Cao, J.~A.
  Stankovic, and T.~Abdelzaher,
\newblock ``Achieving real-time target tracking using wireless sensor
  networks,''
\newblock in {\em Proceedings of IEEE Real Time Technology and Applications
  Symposium}, 2006, pp. 37--48.

\bibitem{caochegao09}
X.~Cao, J.~Chen, C.~Gao, and Y.~Sun,
\newblock ``An optimal control method for applications using wireless
  sensor/actuator networks,''
\newblock {\em Computers \& Electrical Engineering}, vol. 35, no. 5, pp.
  748--756, 2009.

\bibitem{fanli09}
J.~Fang and H.~Li,
\newblock ``Power constrained distributed estimation with cluster-based sensor
  collaboration,''
\newblock {\em IEEE Transactions on Wireless Communications}, vol. 8, no. 7,
  pp. 3822--3832, 2009.

\bibitem{cuixiagolluopoo07}
S.~Cui, J.-J. Xiao, A.~J. Goldsmith, Z.-Q. Luo, and H.~V. Poor,
\newblock ``Estimation diversity and energy efficiency in distributed
  sensing,''
\newblock {\em IEEE Transactions on Signal Processing}, vol. 55, no. 9, pp.
  4683--4695, 2007.

\bibitem{MAH11}
J.~Matamoros and C.~Anton-Haro,
\newblock ``Scaling law of an opportunistic power allocation scheme for
  amplify-and-forward wireless sensor networks,''
\newblock {\em IEEE Communications Letters}, vol. 15, no. 2, pp. 169--171,
  February 2011.

\bibitem{jiacheswi13}
F.~Jiang, J.~Chen, and A.~L. Swindlehurst,
\newblock ``Optimal power allocation for parameter tracking in a distributed
  amplify-and-forward sensor network,''
\newblock {\em IEEE Transactions on Signal Processing}, vol. 62, no. 9, pp.
  2200--2211, May 2014.

\bibitem{gasrimvet03}
M.~Gastpar, B.~Rimoldi, and M.~Vetterli,
\newblock ``To code, or not to code: lossy source-channel communication
  revisited,''
\newblock {\em IEEE Transactions on Information Theory}, vol. 49, no. 5, pp.
  1147--1158, May 2003.

\bibitem{ribgia06}
A.~Ribeiro and G.~B. Giannakis,
\newblock ``Bandwidth-constrained distributed estimation for wireless sensor
  networks{-}part i: Gaussian case,''
\newblock {\em IEEE Transactions on Signal Processing}, vol. 54, no. 3, pp.
  1131--1143, 2006.

\bibitem{xiacuiluogol08}
J.-J. Xiao, S.~Cui, Z.-Q. Luo, and A.~J. Goldsmith,
\newblock ``Linear coherent decentralized estimation,''
\newblock {\em IEEE Transactions on Signal Processing}, vol. 56, no. 2, pp.
  757--770, 2008.

\bibitem{luogiazha05}
Z.-Q. Luo, G.~B. Giannakis, and S.~Zhang,
\newblock ``Optimal linear decentralized estimation in a bandwidth constrained
  sensor network,''
\newblock in {\em Proceedings of IEEE International Symposium on Information
  Theor (ISIT)}, Sept 2005, pp. 1441--1445.

\bibitem{leodeyeva11}
A.~S. Leong, S.~Dey, and J.~S. Evans,
\newblock ``Asymptotics and power allocation for state estimation over fading
  channels,''
\newblock {\em IEEE Transactions on Aerospace and Electronic Systems}, vol. 47,
  no. 1, pp. 611--633, January 2011.

\bibitem{thamit08}
G.~Thatte and U.~Mitra,
\newblock ``Sensor selection and power allocation for distributed estimation in
  sensor networks: Beyond the star topology,''
\newblock {\em IEEE Transactions on Signal Processing}, vol. 56, no. 7, pp.
  2649--2661, July 2008.

\bibitem{thamit06_asilomar}
G.~Thatte and U.~Mitra,
\newblock ``Power allocation in linear and tree wsn topologies,''
\newblock in {\em Proceedings of Asilomar Conference on Signals, Systems and
  Computers}, Oct 2006, pp. 1342--1346.

\bibitem{karvar12isit}
S.~Kar and P.~K. Varshney,
\newblock ``On linear coherent estimation with spatial collaboration,''
\newblock in {\em Proceedings of the 2012 IEEE International Symposium on
  Information Theory Proceedings (ISIT)}, 2012, pp. 1448--1452.

\bibitem{karvar12allerton}
S.~Kar and P.K. Varshney,
\newblock ``Controlled collaboration for linear coherent estimation in wireless
  sensor networks,''
\newblock in {\em Proceedings of the 50th Annual Allerton Conference on
  Communication, Control, and Computing (Allerton)}, 2012, pp. 334--341.

\bibitem{fanvaljamsch_14}
M.~Fanaei, M.~C. Valenti, A.~Jamalipour, and N.~A. Schmid,
\newblock ``Optimal power allocation for distributed blue estimation with
  linear spatial collaboration,''
\newblock in {\em Proceedings of IEEE International Conference on Acoustics,
  Speech and Signal Processing (ICASSP)}, May 2014, pp. 5452--5456.

\bibitem{karvar13}
S.~Kar and P.~Varshney,
\newblock ``Linear coherent estimation with spatial collaboration,''
\newblock {\em IEEE Transactions on Information Theory}, vol. 59, no. 6, pp.
  3532--3553, 2013.

\bibitem{canwakboy08}
E.~Candes, M.~Wakin, and S.~Boyd,
\newblock ``Enhancing sparsity by reweighted $\ell_1$ minimization,''
\newblock {\em Journal of Fourier Analysis and Applications}, vol. 14, pp.
  877--905, 2008.

\bibitem{boyparchupeleck11}
S.~Boyd, N.~Parikh, E.~Chu, B.~Peleato, and J.~Eckstein,
\newblock ``Distributed optimization and statistical learning via the
  alternating direction method of multipliers,''
\newblock {\em Foundations and Trends in Machine Learning}, vol. 3, no. 1, pp.
  1--122, 2011.

\bibitem{chechuzha05}
Y.~Chen, C.~Chuah, and Q.~Zhao,
\newblock ``Sensor placement for maximizing lifetime per unit cost in wireless
  sensor networks,''
\newblock in {\em Proc. IEEE Military Communications Conference}, Oct 2005, pp.
  1097--1102.

\bibitem{josboy09}
S.~Joshi and S.~Boyd,
\newblock ``Sensor selection via convex optimization,''
\newblock {\em IEEE Transactions on Signal Processing}, vol. 57, no. 2, pp.
  451--462, Feb. 2009.

\bibitem{jamsimleu14}
H.~Jamali-Rad, A.~Simonetto, and G.~Leus,
\newblock ``Sparsity{-}aware sensor selection: Centralized and distributed
  algorithms,''
\newblock {\em IEEE Signal Processing Letters}, vol. 21, no. 2, pp. 217--220,
  Feb 2014.

\bibitem{masfarvar12}
E.~Masazade, M.~Fardad, and P.~K. Varshney,
\newblock ``Sparsity-promoting extended {K}alman filtering for target tracking
  in wireless sensor networks,''
\newblock {\em IEEE Signal Processing Letters}, vol. 19, no. 12, pp. 845--848,
  Dec. 2012.

\bibitem{liumasfarvar14_icassp}
S.~Liu, E.~Masazade, M.~Fardad, and P.~K. Varshney,
\newblock ``Sparsity-aware field estimation via ordinary kriging,''
\newblock in {\em Proceedings of IEEE International Conference on Acoustics,
  Speech, and Signal Processing (ICASSP)}, May 2014, pp. 3948--3952.

\bibitem{liufarmasvar14}
S.~Liu, M.~Fardad, E.~Masazade, and P.~K. Varshney,
\newblock ``Optimal periodic sensor scheduling in networks of dynamical
  systems,''
\newblock {\em IEEE Trans. Signal Process.}, vol. 62, no. 12, pp. 3055--3068,
  June 2014.

\bibitem{sch13}
I.~D. Schizas,
\newblock ``Distributed informative-sensor identification via sparsity-aware
  matrix decomposition,''
\newblock {\em IEEE Transactions on Signal Processing}, vol. 61, no. 18, pp.
  4610--4624, Sept 2013.

\bibitem{moambsin11}
Y.~Mo, R.~Ambrosino, and B.~Sinopoli,
\newblock ``Sensor selection strategies for state estimation in energy
  constrained wireless sensor networks,''
\newblock {\em Automatica}, vol. 47, no. 7, pp. 1330--1338, 2011.

\bibitem{liukarfarvar14_isit}
S.~Liu, S.~Kar, M.~Fardad, and P.~K. Varshney,
\newblock ``On optimal sensor collaboration topologies for linear coherent
  estimation,''
\newblock in {\em Proceedings of IEEE International Symposium on Information
  Theory (ISIT)}, 2014, pp. 2624--2628.

\bibitem{karbook}
S.~M. Kay,
\newblock {\em Fundamentals of Statistical Signal Processing: Estimation
  Theory},
\newblock Prentice Hall, Englewood Cliffs, NJ, 1993.

\bibitem{yualin06}
M.~Yuan and Y.~Lin,
\newblock ``Model selection and estimation in regression with grouped
  variables,''
\newblock {\em Journal of the Royal Statistical Society: Series B (Statistical
  Methodology)}, vol. 68, no. 1, pp. 49--67, 2006.

\bibitem{bec07}
A.~Beck,
\newblock ``Quadratic matrix programming,''
\newblock {\em SIAM Journal on Optimization}, vol. 17, no. 4, pp. 1224--1238,
  2007.

\bibitem{liuzhama09}
L.~Liu, X.~Zhang, and H.~Ma,
\newblock ``Dynamic node collaboration for mobile target tracking in wireless
  camera sensor networks,''
\newblock in {\em Proceedings of IEEE INFOCOM 2009}, April 2009, pp.
  1188--1196.

\bibitem{linfarjov13}
F.~Lin, M.~Fardad, and M.~R. Jovanovi\'c,
\newblock ``Design of optimal sparse feedback gains via the alternating
  direction method of multipliers,''
\newblock {\em IEEE Transactions on Automatic Control}, vol. 58, pp.
  2426--2431, 2013.

\bibitem{boyvan04}
S.~Boyd and L.~Vandenberghe,
\newblock {\em Convex Optimization},
\newblock Cambridge University Press, Cambridge, 2004.

\bibitem{liuvemfarmasvar_14}
S.~Liu, A~Vempaty, M.~Fardad, E.~Masazade, and P.~K. Varshney,
\newblock ``Energy-aware sensor selection in field reconstruction,''
\newblock {\em IEEE Signal Processing Letters}, vol. 21, no. 12, pp.
  1476--1480, 2014.

\bibitem{panche99}
V.~Y. Pan and Z.~Q. Chen,
\newblock ``The complexity of the matrix eigenproblem,''
\newblock in {\em Proceedings of the Thirty-first Annual ACM Symposium on
  Theory of Computing}, 1999, pp. 507--516.

\bibitem{aspboy03report}
A.~d'Aspremont and S.~Boyd,
\newblock ``Relaxations and randomized methods for nonconvex {QCQP}s,''
\newblock Stanford, CA: Stanford Univ., Autumn 2003 [Online], Available:
  \emph{\url{http://web.stanford.edu/class/ee392o/relaxations.pdf}}.

\bibitem{nem12}
A.~Nemirovski,
\newblock ``Interior point polynomial time methods in convex programming,''
\newblock 2012 [Online], Available:
  {\emph{\url{http://www2.isye.gatech.edu/~nemirovs/Lect_IPM.pdf}}}.

\bibitem{yualiuye13}
L.~Yuan, J.~Liu, and J.~Ye,
\newblock ``Efficient methods for overlapping group lasso,''
\newblock {\em IEEE Transactions on Pattern Analysis and Machine Intelligence},
  vol. 35, no. 9, pp. 2104--2116, Sept 2013.

\bibitem{gold_bk}
A.~Goldsmith,
\newblock {\em Wireless Communications},
\newblock Cambridge University Press, New York, NY, USA, 2005.

\bibitem{gol73}
G.~H. Golub,
\newblock ``Some modified matrix eigenvalue problems,''
\newblock {\em SIAM Review}, vol. 15, no. 2, pp. 318--334, 1973.

\bibitem{parboy13}
N.~Parikh and S.~Boyd,
\newblock ``Proximal algorithms,''
\newblock {\em Foundations and Trends in Optimization}, vol. 1, no. 3, pp.
  123--231, 2013.

\end{thebibliography}

\end{document}